\documentclass[fleqn,usenatbib]{mnras}

\usepackage{newtxtext,newtxmath}

\usepackage[T1]{fontenc}

\DeclareRobustCommand{\VAN}[3]{#2}
\let\VANthebibliography\thebibliography
\def\thebibliography{\DeclareRobustCommand{\VAN}[3]{##3}\VANthebibliography}

\usepackage{booktabs}

\usepackage{graphicx}	
\usepackage{amsmath}	

\usepackage{rotating}

\title[MIGHTEE radio-loud AGN]{MIGHTEE: the nature of the radio-loud AGN population}

\author[I. H. Whittam et al.]{\parbox{\textwidth}
{I.~H.~Whittam,$^{1,2}$\thanks{E-mail: imogen.whittam@physics.ox.ac.uk}
M.~J.~Jarvis,$^{1,2}$
C.~L.~Hale,$^{3}$
M.~Prescott,$^{2,4}$ 
L.~K.~Morabito,$^{5,6}$
I.~Heywood,$^{1,7}$ 
N.~J.~Adams,$^{8}$
J.~Afonso,$^{9,10}$
Fangxia An,$^{4}$
Y.~Ao,$^{11,12}$
R.~A.~Bowler,$^{8}$
J.~D.~Collier,$^{13,14,15}$
R.~P.~Deane,$^{16,17}$
J.~Delhaize,$^{18}$
B.~Frank,$^{19,13,18}$
M.~Glowacki,$^{20,4}$
P.~W.~Hatfield,$^{1}$
N.~Maddox,$^{21}$
L.~Marchetti,$^{18,22}$
A.~M.~Matthews,$^{23}$
I.~Prandoni,$^{22}$
S.~Randriamampandry,$^{24,25,26}$
Z.~Randriamanakoto,$^{24,26}$
D.~J.~B.~Smith,$^{27}$
A.~R.~Taylor,$^{18,13,2}$
N.~L.~Thomas,$^{5,6}$
and M.~Vaccari$^{4,13,22}$
}\vspace{0.4cm}
\\
$^{1}$Astrophysics, University of Oxford, Denys Wilkinson Building, Keble Road, Oxford, OX1 3RH, UK\\
$^{2}$ Department of Physics and Astronomy, University of the Western Cape, Robert Sobukwe Road, Bellville 7535, South Africa\\
$^{3}$ School of Physics and Astronomy, Institute for Astronomy, University of Edinburgh, Royal Observatory, Blackford Hill, EH9 3HJ Edinburgh, UK\\
$^{4}$ Inter-University Institute for Data Astronomy, University of the Western Cape, Robert Sobukwe Road, 7535 Bellville, South Africa\\
$^{5}$  Centre for Extragalactic Astronomy, Department of Physics, Durham University, Durham, DH1 3LE, UK\\
$^{6}$ Institute for Computational Cosmology, Department of Physics, Durham University, Durham, DH1 3LE, UK\\
$^{7}$Department of Physics and Electronics, Rhodes University, PO Box 94, Grahamstown, 6140, South Africa\\
$^{8}$ Jodrell Bank Centre for Astrophysics, University of Manchester, Oxford Road, Manchester, UK\\
$^{9}$Instituto de Astrof\'{i}sica e Ci\^{e}ncias do Espa\c co, Universidade de Lisboa, OAL, Tapada da Ajuda, PT1349-018 Lisboa, Portugal\\
$^{10}$Departamento de F\'{i}sica, Faculdade de Ci\^{e}ncias, Universidade de Lisboa, Edif\'{i}cio C8, Campo Grande, PT1749-016 Lisbon, Portugal\\
$^{11}$ Purple Mountain Observatory and Key Laboratory for Radio Astronomy, Chinese Academy of Sciences, Nanjing, China\\
$^{12}$ School of Astronomy and Space Science, University of Science and Technology of China, Hefei, China\\
$^{13}$ The Inter-University Institute for Data Intensive Astronomy, Department of Astronomy, University of Cape Town, Private Bag X3, Rondebosch, 7701, South Africa\\
$^{14}$ School of Science, Western Sydney University, Locked Bag 1797, Penrith, NSW 2751, Australia\\
$^{15}$ CSIRO Astronomy and Space Science, PO Box 1130, Bentley, WA, 6102, Australia\\
$^{16}$ Wits Centre for Astrophysics, School of Physics, University of the Witwatersrand, 1 Jan Smuts Avenue, 2000, Johannesburg, South Africa\\
$^{17}$Department of Physics, University of Pretoria, Hatfield, Pretoria, 0028, South Africa\\
$^{18}$ Department of Astronomy, University of Cape Town, Private Bag X3, Rondebosch 7701, South Africa\\
$^{19}$ South African Radio Astronomy Observatory, 2 Fir Street, Observatory, 7925, South Africa\\
$^{20}$ International Centre for Radio Astronomy Research, Curtin University, Bentley, WA 6102, Australia\\
$^{21}$ University Observatory, Faculty of Physics, Ludwig-Maximilians-Universit\"at, Scheinerstr. 1, 81679 Munich, Germany\\
$^{22}$ INAF - Istituto di Radioastronomia, via Gobetti 101, 40129 Bologna, Italy\\
$^{23}$ The Observatories of the Carnegie Institution for Science, 813 Santa Barbara Street, Pasadena, CA 91101, USA\\
$^{24}$ South African Astronomical Observatory, P.O. Box 9, Observatory 7935, South Africa\\
$^{25}$ Southern African Large Telescope, P.O. Box 9, Observatory 7935, Cape Town, South Africa\\
$^{26}$ Department of Physics, University of Antananarivo, P.O. Box 906, Antananarivo, Madagascar\\
$^{27}$ Centre for Astrophysics Research, University of Hertfordshire, College Lane, Hatfield, AL10 9AB, UK\\
}

\date{Accepted XXX. Received YYY; in original form ZZZ}

\pubyear{2022}

\begin{document}
\label{firstpage}
\pagerange{\pageref{firstpage}--\pageref{lastpage}}
\maketitle

\begin{abstract}
We study the nature of the faint radio source population detected in the MeerKAT International GHz Tiered Extragalactic Exploration (MIGHTEE) Early Science data in the COSMOS field, focusing on the properties of the radio-loud active galactic nuclei (AGN). Using the extensive multi-wavelength data available in the field, we are able to classify 88 per cent of the 5223 radio sources in the field with host galaxy identifications as AGN (35 per cent) or star-forming galaxies (54 per cent). 
We select a sample of radio-loud AGN with redshifts out to $z \sim 6$ and radio luminosities $10^{20} < \textrm{L}_{1.4~\textrm{GHz}} / \textrm{W Hz}^{-1} < 10^{27}$ and classify them as high-excitation and low-excitation radio galaxies (HERGs and LERGs). The classification catalogue is released with this work. We find no significant difference in the host galaxy properties of the HERGs and LERGs in our sample.  In contrast to previous work, we find that the HERGs and LERGs have very similar Eddington-scaled accretion rates; in particular we identify a population of very slowly accreting AGN that are formally classified as HERGs at these low radio luminosities, where separating into HERGs and LERGs possibly becomes redundant. We investigate how black hole mass affects jet power, and find that a black hole mass $\gtrsim 10^{7.8}~\textrm{M}_\odot$ is required to power a jet with mechanical power greater than the radiative luminosity of the AGN ($L_\textrm{mech}/L_\textrm{bol} > 1$). We discuss that both a high black hole mass and black hole spin may be necessary to launch and sustain a dominant radio jet.
\end{abstract}

\begin{keywords}
surveys -- catalogues -- Galaxies -- galaxies: active -- radio continuum: galaxies 
\end{keywords}



\section{Introduction}


Radio-loud Active Galactic Nuclei (RLAGN) are thought to play an important role in galaxy evolution, regulating star-formation in massive galaxies. Accretion of matter onto the supermassive black hole (SMBH) at the centre of the galaxy can power relativistic jets (highly collimated beams of plasma) which produce synchrotron emission visible with radio telescopes. RLAGN have been widely studied over several decades, and they display a range of different properties, with linear sizes ranging from pc to several Mpc and radio powers spanning many orders of magnitude from $\sim10^{19}$~W/Hz up to $\sim10^{29}$~W/Hz at 1.4~GHz (see reviews by \citealt{2010A&ARv..18....1D,2014ARA&A..52..589H,2020NewAR..8801539H}). 

There is mounting evidence that there are two fundamentally different classes of radio-loud AGN; one class which display the typical signatures of efficient accretion such as an accretion disk and a dusty torus, and a second class which appear to be lacking these structures and emit most of their energy as a powerful radio jet (e.g.\ \citealt{1994ASPC...54..201L,2007MNRAS.381..589M,2009MNRAS.396.1929H}). The first class are referred to as high-excitation radio galaxies (HERGs, due to the high-excitation lines visible in their optical spectra), cold-mode sources, quasar mode or radiatively efficient sources. The second class are called low-excitation radio galaxies (LERGs), hot mode sources, radio mode sources, or radiatively inefficient sources.

In studies conducted to date, LERGs tend to be hosted by massive galaxies, often at the centre of a group or cluster \citep{1990AJ.....99...14B,2007MNRAS.379..894B}, which mostly contain older stars (e.g.\ \citealt{2009ApJ...699L..43S,2010MNRAS.406.1841H}). HERGs, however, are more likely to be found in less massive galaxies with a younger stellar population \citep{2012MNRAS.421.1569B}. There is evidence that HERGs show rapid cosmological evolution in their space density, while LERGs show little or no evolution out to $z \sim 1$ \citep{2004MNRAS.352..909C,2014MNRAS.445..955B,2016MNRAS.460....2P,2016MNRAS.457..730P} and may display strong negative evolution at higher redshifts \citep{2018MNRAS.475.3429W}.  

Initially, it was believed that there was a direct link between the HERG and LERG classes and the Fanaroff and Riley morphological classifications (FRI and II classes, \citealt{1974MNRAS.167P..31F}), with most HERGs displaying edge-brightened FRII morphologies and most LERGs appearing as centre-brightened FRI sources (e.g.\ \citealt{1997MNRAS.286..241J}). However, the existence of FRI HERGs \citep{2001ApJ...562L...5B} and FRII LERGs (e.g.\ \citealt{1979MNRAS.188..111H,1999MNRAS.309.1017W,2018MNRAS.476.1614C} ) contradicts this scenario, and work by \citet{2013MNRAS.430.3086G} and \citet{2022MNRAS.511.3250M} shows that the link between accretion mode and morphology is very indirect.
There is also evidence that HERGs seem to be more dominated by radio emission from their cores than LERGs \citep{2016MNRAS.462.2122W}.

It had been proposed that there is a direct relationship between the accretion mode and the source of fuel available, with HERGs accreting efficiently from sources of cold gas, while LERGs accrete inefficiently from the hot gas halo \citep{2007MNRAS.376.1849H}. More recent work, however, has argued that Eddington-scaled accretion rate is the key driver of the difference between HERGs and LERGs, rather than the source of the material being accreted (e.g.\ \citealt{2012MNRAS.421.1569B,2014MNRAS.445..955B,2018NatAs...2..273H,2020NewAR..8801539H}). It is therefore important that we fully understand the Eddington-scaled accretion rates of different AGN classes as we probe lower radio powers and higher redshifts.

In the scenario building up in the literature, there has been thought to be a dichotomy in accretion rates between the HERG and LERG classes; HERGs accrete efficiently at $\gtrsim 1$ per cent of their Eddington accretion rate, while LERGs accrete much more slowly at $\lesssim 1$ per cent of Eddington, with almost no overlap in the accretion rates of the two classes \citep{2012MNRAS.421.1569B,2014MNRAS.440..269M}. The current understanding is that this is because the two classes are a result of two fundamentally distinct accretion modes \citep{2005MNRAS.362...25B,2006MNRAS.370.1893H}. HERGs are thought to accrete cold gas efficiently via an optically thick, geometrically thin accretion disc (e.g.\ \citealt{1973A&A....24..337S}), while LERGs accrete from a hot gas reservoir (e.g.\ \citealt{2012A&A...541A..62J,2014ARA&A..52..529Y,2007MNRAS.376.1849H}) relatively slowly via an advection-dominated accretion flow \citep{1995ApJ...452..710N,2003ANS...324..435Q}. 

However, recent work by \citet{2018MNRAS.480..358W} using radio galaxies selected from a Karl G. Jansky Very Large Array (VLA) survey of Stripe-82 \citep{2016MNRAS.460.4433H} has suggested that the dichotomy in accretion rates of HERGs and LERGs may not be as clear cut as previously thought; \citeauthor{2018MNRAS.480..358W} find a significant overlap in the accretion rates of the two classes. The \citeauthor{2018MNRAS.480..358W} study probes lower radio luminosities than the \citet{2012MNRAS.421.1569B} work (see Fig.~\ref{fig:LzS82}) and the \citet{2014MNRAS.440..269M} study; this may be the reason for the larger overlap in accretion rates, but \citeauthor{2018MNRAS.480..358W} do not have the statistics at lower luminosities to confirm this trend. 

This suggests that the current leading model where there are two distinct accretion modes, which equate to different feedback processes (see review by \citealt{2012ARA&A..50..455F}), does not tell the full story, and instead galaxies may display a more continuous range of accretion rates and associated properties.  This has important implications for our understanding of galaxy evolution and how AGN processes affect star-formation in a galaxy (e.g.\ \citealt{2009Natur.460..213C,2020NewAR..8801539H}). However, further research at lower radio luminosities is required to confirm this, which is the aim of this work.

The MeerKAT International Tiered GHz Extragalactic Exploration (MIGHTEE, \citealt{2016mks..confE...6J}) survey is a large survey project currently underway with the MeerKAT radio telescope \citep{2009IEEEP..97.1522J}. When complete, MIGHTEE will cover 20 square degrees across four different fields (COSMOS, XMM-LSS, ELAIS-S1 and E-CDFS) to a depth of $\sim2~\muup$Jy/beam rms at 1.28~GHz. The unique combination of depth over a significant area combined with excellent multi-wavelength coverage means that the MIGHTEE survey has the potential to provide a significant step forward in our understanding of galaxy evolution. In particular, the MIGHTEE survey allows us to study the accretion rates of a large sample of AGN across a range of radio powers and redshifts. In this paper we use MIGHTEE Early Science observations in the COSMOS field \citep{2022MNRAS.509.2150H} to probe the nature of the apparently faint radio source population, focusing on the properties of radio-loud AGN.

This paper is laid out as follows: in Section~\ref{section:radio_data} we describe the MIGHTEE radio data used in this work, and the ancillary multi-wavelength data used is outlined in Section~\ref{section:multi_data}. In Section~\ref{section:classification} we explain the scheme used to classify the MIGHTEE radio sources. We then use this sample to explore the properties of radio-loud AGN in Section~\ref{section:properties}; first we discuss the host galaxy properties of different types of radio-loud AGN, next we investigate the accretion rates of these classes, then we explore the host galaxy properties as a function of accretion rate and finally we discuss the relationship between AGN power and black hole mass. The implications of these results are discussed in Section~\ref{section:discussion}, and our conclusions are presented in Section~\ref{section:conclusions}.

Throughout this paper the following values for the cosmological parameters are used: $H_0 = 70~{\rm km \, s}^{-1} ~\rm Mpc^{-1}$, $\Omegaup_{\rm M} = 0.3$ and $\Omegaup_{\Lambdaup} = 0.7$. Unless stated all magnitudes are AB magnitudes. We use the following convention for radio spectral index, $\alpha$: $S \propto \nu^{-\alpha}$, for a source with flux density $S$ and frequency $\nu$.

\section{Radio data - the MIGHTEE survey}\label{section:radio_data}

The MIGHTEE Early Science radio continuum data release consists of one pointing in the COSMOS field, covering $\sim1.6~\textrm{deg}^2$, and three overlapping pointings in the XMM-LSS field, covering $\sim3.5~\textrm{deg}^2$. In this work, we restrict our analysis to the central part of the COSMOS Early Science image, with a diameter of 1.04 degrees, as this is the region over which multi-wavelength counterparts for the radio sources have been identified (see Section~\ref{section:cross-matching} and Prescott et al., in prep). These observations consist of a single field of view with the MeerKAT telescope centred on RA 10h00m28.6s, Dec +02d12m21s. The observations were taken with the L-band receiver (bandwidth 900 - 1670 MHz) between 2018 and 2020 and include 17.45 hours on source. For full details of the observations and data reduction we refer the reader to \citet{2022MNRAS.509.2150H}. 

The MIGHTEE Early Science data contains two versions of the data processed with different \citet{1995AAS...18711202B} robust weighting values; the first uses Briggs' robustness parameter = 0.0 and is optimised for sensitivity, but has lower resolution. The second image uses robust = -1.2 which down-weights the short baselines in the core resulting in a higher resolution, but this comes at the expense of sensitivity. In this paper we use the maximum-sensitivity (robust = 0.0) image, which has a circular synthesised beam full-width half maximum (FWHM) diameter of 8.6~arcsec and a thermal noise of 1.7~$\muup$Jy/beam. However, due to confusion noise the effective rms noise in the centre of the image is $\sim4~\muup$Jy/beam \citep{2022MNRAS.509.2150H}. 

Source finding was carried out using the Python Blob Detector and Source Finder (\textsc{PyBDSF}; \citealt{2015ascl.soft02007M}) using the default source extract parameters, see \citet{2022MNRAS.509.2150H} for further details. Our initial sample contains 6263 radio sources with $S_{1.28~\textrm{GHz}} > 20~\muup$Jy in the central part of the COSMOS field\footnote{Note that components which make up part of an extended radio source are grouped together as part of the cross-matching process described in Section~\ref{section:cross-matching}}. As mentioned, the full MIGHTEE Early Science data released by \citeauthor{2022MNRAS.509.2150H} covers a larger area of the COSMOS field as well as the XMM-LSS field, so contains a much larger number of sources, but here we are limited to the region over which the multi-wavelength counterparts for the radio sources have been identified to date.

Due to the wide bandwidth of the MeerKAT L-band receivers used for the MIGHTEE observations ($900 - 1670$~MHz) and the varying response of the primary beam with frequency (together with other factors such as flagging of the raw data), the effective frequency of the MIGHTEE data varies across the image. This is discussed in detail in \citet{2022MNRAS.509.2150H}. In order to have measurements at a constant frequency and to aid comparisons with other work, we scale the MIGHTEE flux densities and radio luminosities to 1.4 GHz using the effective frequency map released with \citeauthor{2022MNRAS.509.2150H}, assuming a spectral index of 0.7. All radio luminosities are k-corrected assuming the same spectral index.

\begin{figure}
    \centering
    \includegraphics[width=\columnwidth]{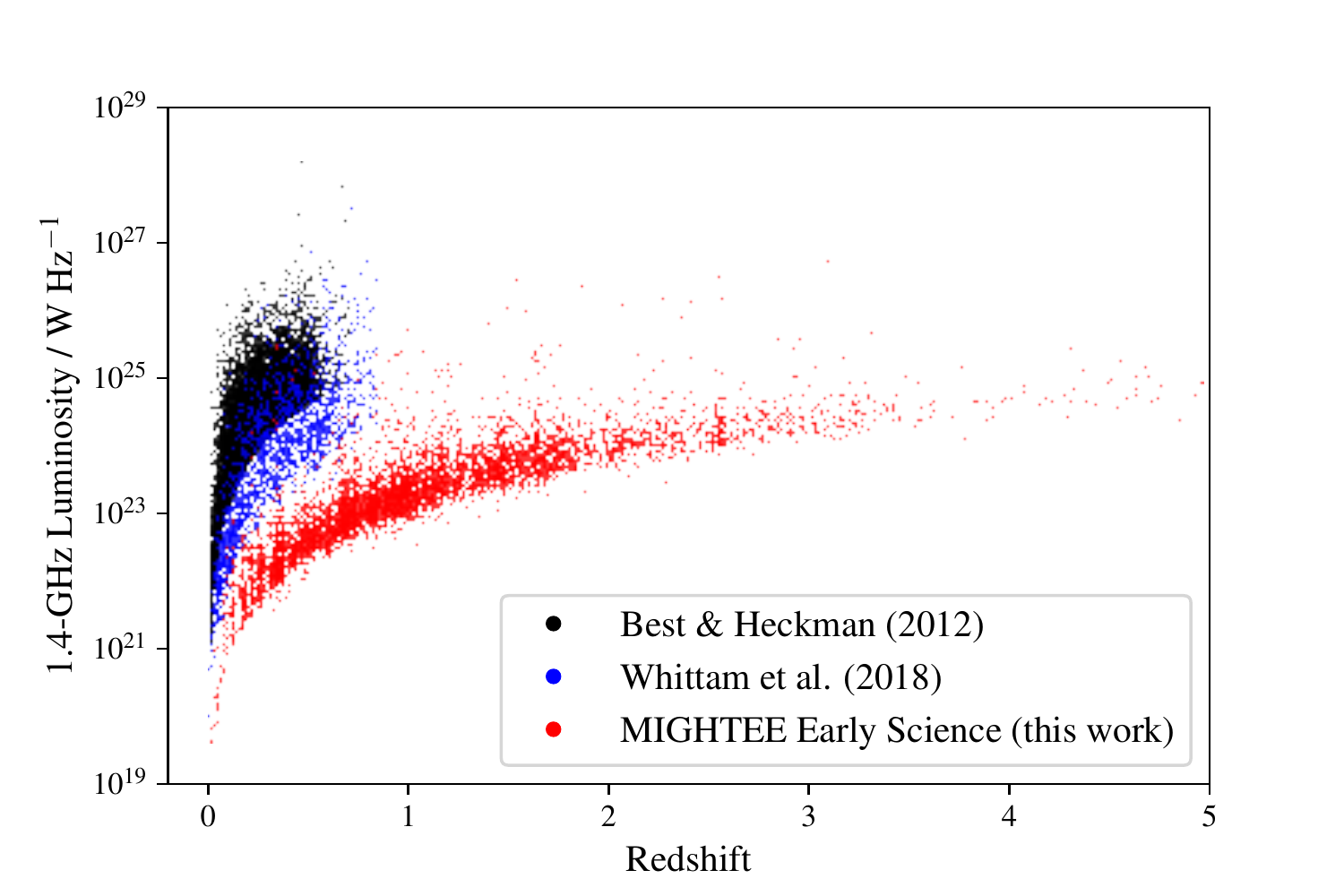}
    \caption{1.4-GHz radio luminosity and redshift of the MIGHTEE Early Science multi-wavelength cross-matched sample presented by Prescott et al. (in prep) and studied in this work (red), along with the samples studied by \citet{2012MNRAS.421.1569B} (black) and \citet{2018MNRAS.480..358W} (blue).}
    \label{fig:LzS82}
\end{figure}

Due to the depth of the radio data used to select the sample studied in this paper, together with the quality of the ancillary data (see Section~\ref{section:multi_data}) we are able to probe lower radio luminosities and higher redshifts than previous studies of the accretion rates of radio-loud AGN, such as \citet{2012MNRAS.421.1569B} and \citet{2018MNRAS.480..358W}. This is illustrated by Fig.~\ref{fig:LzS82} which shows the radio luminosity and redshift distribution probed by this work compared to those of two previous studies. As can been seen in the figure, there is almost no overlap in the parameter space probed by the MIGHTEE sample and the \citet{2012MNRAS.421.1569B} study. This demonstrates the unique power of the MIGHTEE survey to probe low-powered radio galaxies across cosmic time, which therefore has potential to provide a new insight into AGN activity.

\section{Multi-wavelength data}\label{section:multi_data}

\subsection{Optical and near-infrared}\label{section:optNIR-data}

This work makes use of the extensive optical and near-infrared data in the COSMOS field compiled by \citet{2020MNRAS.493.2059B}, (see also \citealt{2020MNRAS.494.1771A,2021MNRAS.506.4933A}). Briefly, this contains near-infrared imaging in the $YJHK_s$ band from the fourth data release (DR4) of the UltraVISTA survey \citep{2012A&A...544A.156M}, optical data in $grizy$ filters from Hyper Suprime-Cam Subaru Strategic Program DR1 (HSC SSP;  \citealt{2017arXiv170600566T}) along with data from two HSC narrow-band filters at 8160 and 9210~\AA, deeper optical imaging in the $u^*griz$ filters from the Canada-France-Hawaii Telescope Legacy Survey (CFHTLS), and deep $z'$-band imaging from Subaru/Suprime-Cam \citep{2016ApJ...822...46F}. Additionally, mid-infrared data from the \emph{Spitzer} Infrared Array Camera (IRAC) at 3.6 and 4.5~$\muup$m is included. This combines shallower imaging from the \emph{Spitzer} Large-Area Survey with the HSC (SPLASH; \citealt{2014ApJ...791L..25S}) with deeper data from the \emph{Spitzer} Matching Survey of the UltraVISTA ultra-deep Stripes survey (SMUVS; \citealt{2018ApJS..237...39A}) and the \emph{Spitzer} Extended Deep Survey (SEDS; \citealt{2013ApJ...769...80A}).  Full details of the catalogue creation are given in \citet{2020MNRAS.493.2059B}.

Additionally, this work makes use of the high-resolution \emph{Hubble Space Telesope (HST)} Advanced Camera for Surveys (ACS) $I$-band image, further details of these data are given in \citet{2007ApJS..172...38S}.

\subsection{Cross-matching}\label{section:cross-matching}

The procedure used to identify the host galaxy of each radio source is described in detail in Prescott et al. (in prep) and summarised briefly here. Overlays displaying the MIGHTEE radio contours and the higher-resolution \citet{2017A&A...602A...1S} VLA-COSMOS 3~GHz radio contours on top of the UltraVISTA $K_s$-band images were produced for each MIGHTEE radio component in the early science low resolution Level 0 catalogue (described in \citeauthor{2022MNRAS.509.2150H}). Although less sensitive for a typical radio source with a spectral index of 0.7 (median rms is $S_{3~\textrm{GHz}} \sim 2.3~\muup$Jy/beam), the higher-resolution (0.78~arcsec) VLA-COSMOS 3~GHz data are useful when identifying the correct host galaxy for the radio sources. 

Prescott et al. (in prep) use an updated version of  the \textsc{Xmatchit} code (described in \citealt{2018MNRAS.480..707P}) to quickly display the overlays for each radio component. These were examined by eye by three separate people to identify the most likely host galaxy for each radio component.  Any sources where the three classifiers did not agree were examined again by a committee.  In total, we have identified the host galaxy for 5223 out of 6262 (83 per cent) radio sources. The 1039 unmatched radio sources are a combination of sources which lie in masked regions close to bright sources in the $K_s$-band image (208 sources), sources which are too confused for us to be able to identify the correct host ID (i.e.\ where two or more individual sources with separate host galaxies are blended together in the radio image; 693 sources), sources where there is no host galaxy visible in the UltraVISTA $K_s$-band image (126 sources), and sources which appear to be artefacts in the radio image (12 sources; note that these do not appear in the MIGHTEE Early Science Level 1 catalogue released by \citeauthor{2022MNRAS.509.2150H} where artefacts have been removed). This is discussed further in Prescott et al. (in prep). This sample of 5223 sources is the focus of the remainder of this paper.

\subsection{Mid and far-infrared}\label{section:FIR-data}

In addition to the data from the two shorter-wavelength \emph{Spitzer} IRAC bands included in the \citet{2020MNRAS.493.2059B} compilation described in Section~\ref{section:optNIR-data}, we use data from the 5.8 and 8.0~$\muup$m bands from the SPLASH survey, accessed from \citet{2016ApJS..224...24L}.

Far-infrared data were obtained from the \emph{Herschel} Extra-galactic Legacy Project (HELP; \citealt{2021MNRAS.507..129S}). We use data at 24~$\muup$m from the Multiband Imaging Photometer (MIPS, \citealt{2004ApJS..154...25R}) instrument on the \emph{Spitzer} Space Telescope, 100 and 160~$\muup$m from the Photodetector Array Camera and Spectrometer (PACS, \citealt{2010A&A...518L...2P}) on \emph{Herschel} and 250, 350 and 500~$\muup$m from the Spectral and Photometric Imaging Receiver (SPIRE, \citealt{2010A&A...518L...3G}), also on \emph{Herschel}. Photometry was obtained using \textsc{xid}+ \citep{2017MNRAS.464..885H}, a probabilistic de-blender developed for the HELP project. This utilises prior information from the four \emph{Spitzer} IRAC bands, which have a higher resolution than the \emph{Herschel} data, and applies a Bayesian approach to extract photometry from the \emph{Herschel} maps.

The HELP far-infrared catalogues were combined with the cross-matched catalogue (Section~\ref{section:cross-matching}) by matching to the position of the host galaxy using a match radius of 1 arcsec. 4540 out of 5223 radio sources are detected in the MIPS and PACS data, and 4957 in the SPIRE data.

\subsection{X-ray}\label{section:xray}

Deep X-ray imaging is available across the COSMOS field from the \emph{Chandra} COSMOS-Legacy project, which combines 1.8 Ms of data from the C-COSMOS survey \citep{2009ApJS..184..158E} with 2.8 Ms of more recent \emph{Chandra} ACIS-I observations \citep{2015ApJ...808..185C} in the 0.5 - 10 keV energy band.
The optical and infrared counterparts to the \emph{Chandra} COSMOS-Legacy survey are presented in \citet{2016ApJ...817...34M}; we use this catalogue to identify X-ray counterparts to the MIGHTEE radio sources via the positions of the optical host galaxies. A total of 572 out of 5223 radio sources are detected in the X-ray observations.

\subsection{VLBI}\label{section:VLBI-data}

Very long baseline interferometry (VLBI) observations of the COSMOS field were carried out by \citet{2017A&A...607A.132H} using the Very Long Baseline Array (VLBA). The observations have a median resolution of $16.2 \times 7.3~\textrm{mas}^2$ and a central frequency of 1.54~GHz. We matched the VLBA catalogue to the MIGHTEE catalogue described in this work using the positions of the optical host galaxies and a match radius of 1~arcsec. A total of 255 MIGHTEE sources are detected in the VLBA catalogue.

\subsection{Redshifts}\label{section:redshifts}

Spectroscopic redshifts are available for a number of sources in the field from the following observing campaigns; the Deep Imaging Multi-Object Spectrograph (DEIMOS) 10K survey \citep{2018ApJ...858...77H}, the Fiber Multi-Object Spectrograph (FMOS) survey \citep{2015ApJS..220...12S}, the PRIsm MUlti-object Survey (PRIMUS, \citealt{2011ApJ...741....8C,2013ApJ...767..118C}), the zCOSMOS survey \citep{2009ApJS..184..218L}, the 3D-HST survey \citep{2014ApJS..214...24S,2016ApJS..225...27M}\footnote{The redshifts from the latter four surveys were compiled by the HSC team and are available here: \url{https://hsc-release.mtk.nao.ac.jp/doc/index.php/dr1_specz/}.}, the Sloan Digital Sky Survey 14th data release (SDSS DR14; \citealt{2018ApJS..235...42A}) and the Large Early Galaxy Astrophysics Census (LEGA-C) program \citep{2016ApJS..223...29V,2018ApJS..239...27S}. In total 2427 (46 per cent) of the sources in our sample have a spectroscopic redshift available.

For the remaining sources we use photometric redshifts derived from the excellent optical and near-infrared photometry available in the COSMOS field (see Section~\ref{section:optNIR-data}). Briefly, the photometric redshifts are calculated using a hierarchical Bayesian combination (as per \citealt{2018MNRAS.477.5177D}) of two different approaches; the more traditional template fitting, and a machine learning approach. The template fitting is carried out using the \textsc{LePhare} Spectral Energy Distribution (SED) fitting code \citep{2011ascl.soft08009A} set up as described in \citet{2021MNRAS.506.4933A}, and the machine learning using the \textsc{GPz} algorithm \citep{2016MNRAS.455.2387A,2016MNRAS.462..726A} set up as described in \citet{2020MNRAS.498.5498H}. For further details of this process we refer the reader to Hatfield et al. (submitted).

Optical spectral line measurements are available for the 92 sources with SDSS spectra and the 80 sources with spectra from LEGA-C, and are used in this work. The procedures used to obtain the emission line fluxes for the SDSS sources are described in \citet{2013MNRAS.431.1383T}.

\subsection{AGNfitter}\label{section:AGNfitter}

We use the SED fitting code \textsc{AGNfitter} to investigate the properties of the AGN in the sample. \textsc{AGNfitter} is a fully Bayesian SED-fitted code and is described in \citet{2016ApJ...833...98C}. It uses a library of theoretical, empirical and semi-empirical models to characterise the nuclear and host galaxy emission simultaneously. The SED is separated into four components; two arise from the AGN and two from the host galaxy. Following the nomenclature of \citeauthor{2016ApJ...833...98C}, the AGN emission comprises of a UV/optical accretion disk component (BB) and a hot dust torus component (TO). The host galaxy emission components are stellar emission (GA) and reprocessed light from dust (SB).

\begin{figure}
    \centering
    \includegraphics[width=\columnwidth]{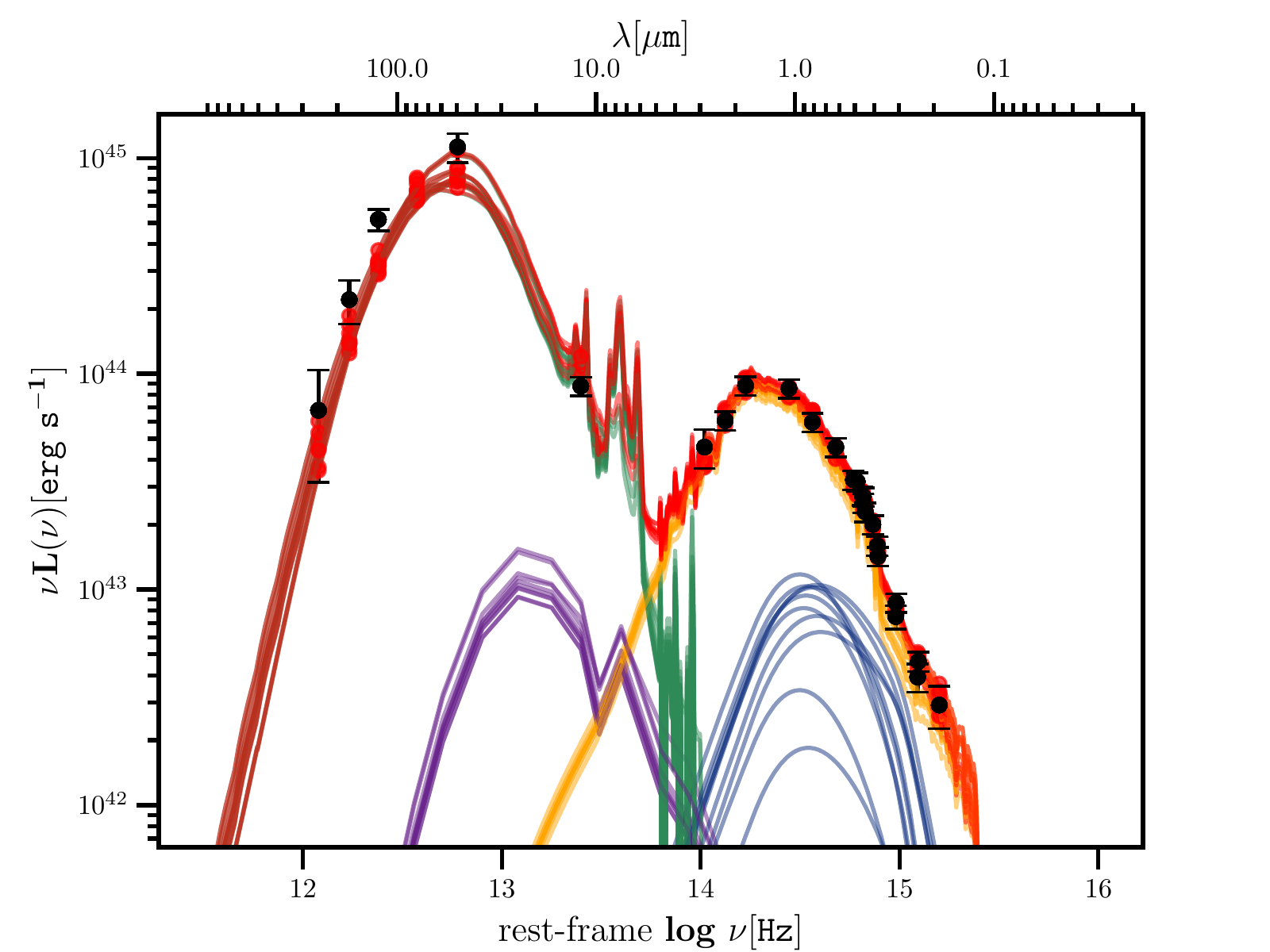}
    \caption{Example of multiple SEDs taken from the posterior from \textsc{AGNfitter}. The black points show the photometry and associated uncertainties. The purple lines show different realisations of the AGN torus component (TO), the blue lines are the AGN accretion disk component (BB), the yellow lines are the stellar emission (GA) and the green lines are the reprocessed stellar light from dust (SB). The red lines are the total SED. The red circles are the simulated measurements derived by convolving the filter profile with the fit SED.}
    \label{fig:AGNfitter-SED}
\end{figure}

We used a total of 24 bands ranging from the far-infrared to the optical as described in Sections~\ref{section:optNIR-data} and \ref{section:FIR-data}. Due to the well-known problem with SED-fitting using different wavebands with (potentially) different systematic uncertainties, we set any uncertainties which were $< 10$~per cent to 10 per cent. The redshift for each source is fixed to the best available value, as described in Section~\ref{section:redshifts}. We achieved a good fit (quantified by log likelihood values $>-100$) for 4969 out of 5223 sources. 
An example SED is shown in Fig.~\ref{fig:AGNfitter-SED}.

We use the results of the \textsc{AGNfitter} SED-fitting to estimate the accretion rate of the AGN in our sample, as described in Section~\ref{section:accretion-rates}. The code also produces estimates of the host galaxy's stellar mass and star-formation rate, which are used in the analysis. We compared the stellar masses derived by \textsc{AGNfitter} to those estimated by the \textsc{LePhare} SED-fitting code (see \citealt{2021MNRAS.506.4933A}), and found that the stellar masses from the two SED-fitting codes are generally in good agreement; we therefore use those from \textsc{AGNfitter} in this work, but using the values from \textsc{LePhare} instead does not change any of the conclusions in this paper.

\section{Source classification}\label{section:classification}

The classifications in this work are based on five different criteria, which are described in the following sub-sections. These are: radio excess (Section~\ref{section:RLAGN}), mid-infrared colour cuts (Section~\ref{section:midIRAGN}), optical morphology (Section~\ref{section:optAGN}), X-ray luminosity (Section~\ref{section:XAGN_class}) and VLBI detection (Section~\ref{section:VLBA_AGN}).  These different selection criteria identify different AGN properties; the mid-infrared colour cut identifies objects with a dusty AGN torus, the optical morphology criterion selects optical quasars and the X-ray cut identifies AGN with accretion-related X-ray emission. These are all signatures of quasar-like AGN. The two radio criteria, however, identify sources with radio-AGN activity. This is discussed further in Section~\ref{section:overall_class}, where the five criteria are combined to give the overall classification scheme.

These classifications are released publicly with this work, details of the catalogue are given in Appendix~\ref{appendix:level3-cat}. The classifications are compared to classifications from optical spectra in Section~\ref{section:opt_spectra_comparison} and to classifications from the VLA-COSMOS 3-GHz Large Project \citep{2017A&A...602A...1S,2017A&A...602A...2S} which covers the same field in Section~\ref{section:vla}. 

We note that \citet{2018MNRAS.475.3429W} use information from \textsc{AGNfitter} SED fitting to classify the sources in the LOFAR-Boot\"{e}s sample as HERGs and LERGs. We tested using this classification scheme and found that it was not appropriate for the fainter sources in our sample, for details see Appendix~\ref{appendix:williams}.

\subsection{Radio-excess AGN}\label{section:RLAGN}

We make use of the infrared -- radio correlation to identify sources with significantly more radio emission than would be expected from star-formation alone. The infrared -- radio correlation can be quantified by the parameter $q_\text{IR}$, which is defined as the logarithmic ratio of the infrared and radio luminosities:

\begin{equation}
    q_\text{IR} = \text{log}_{10} \frac{L_\text{IR}~[\text{W}] \, / \, 3.75 \times 10^{12} ~[\text{Hz}]} {L_{1.4~\text{GHz}}~[\text{W Hz}^{-1}]}
\end{equation}

\noindent where $L_\text{IR}$ is the total infrared luminosity between 8 - 1000 $\muup$m, estimated by \textsc{AGNfitter}. This is divided by the central frequency of $3.75 \times 10^{12} ~\text{Hz}$ (80 $\muup$m) so that $q_\text{IR}$ is a dimensionless quantity.

We use the stellar mass and redshift dependent infrared-radio correlation (IRRC) from \citet{2021A&A...647A.123D} to identify sources which display a radio excess above what would be expected from star-formation. \citet{2021A&A...647A.123D} use the MIGHTEE Early Science data, as well as data from the VLA-COSMOS 3 GHz Large Project \citep{2017A&A...602A...1S} to investigate how the infrared-radio correlation evolves with stellar mass and redshift simultaneously.  Following \citeauthor{2021A&A...647A.123D} and \citet{2018MNRAS.475.3429W}, we consider any source which lies more than 0.43 dex below the best fit correlation given by equation 5 in their paper as having a radio excess (this is equivalent to 2 $\sigma$, where $\sigma$ is the intrinsic scatter in the relation). We note that adopting this cut may result in a small number of SFG being misclassified as radio-excess AGN, but using a less-aggressive cut would result in the opposite problem, where radio-excess AGN are missed and classified as radio quiet. \citeauthor{2021A&A...647A.123D} tested a number of different cuts to separate radio-AGN and star-forming populations, and found the $2\sigma$ threshold used here to be the best compromise, with contamination from SFG only around 3 - 4 per cent. 

\begin{figure}
    \centering
    \includegraphics[width=\columnwidth]{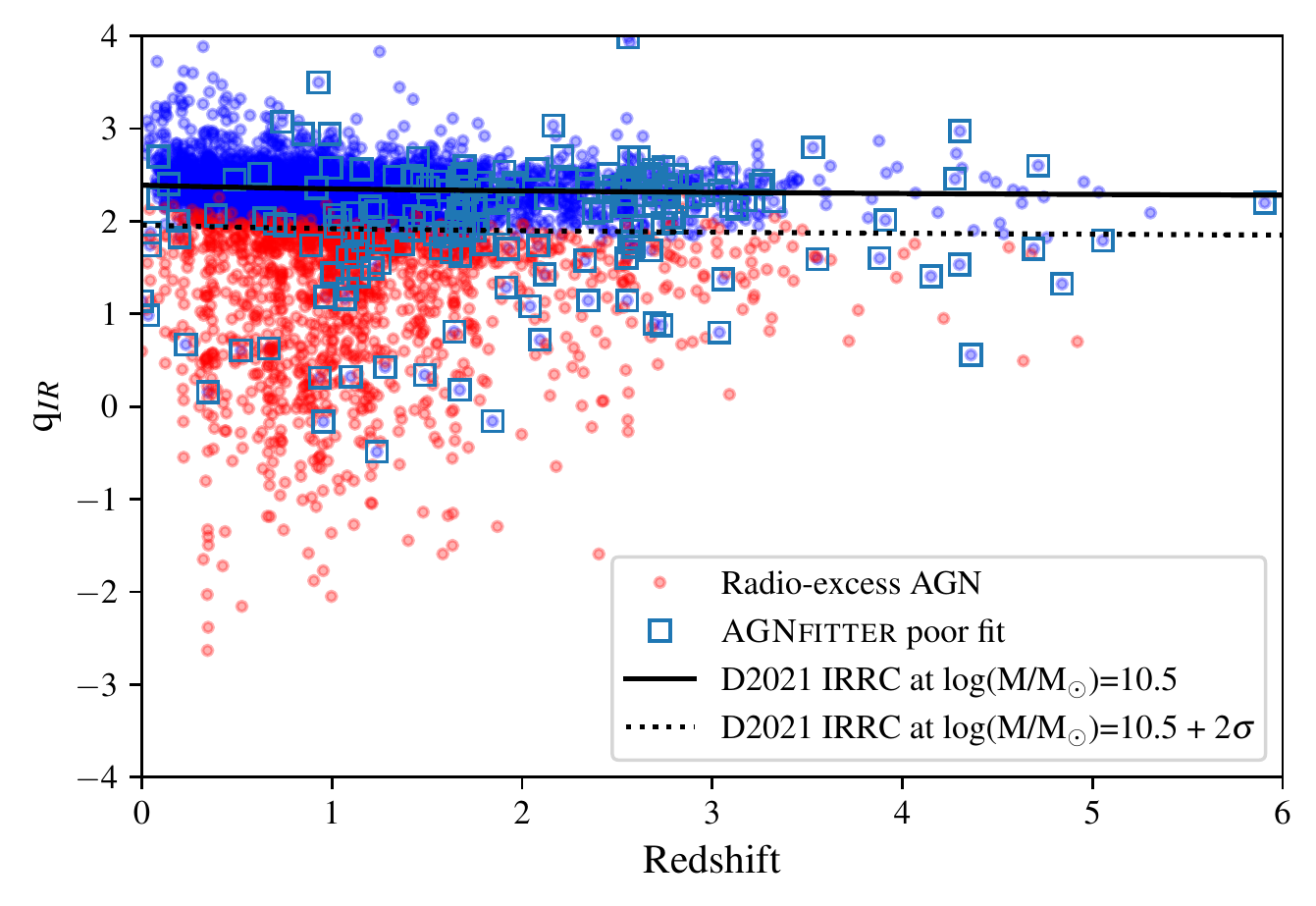}
    \includegraphics[width=\columnwidth]{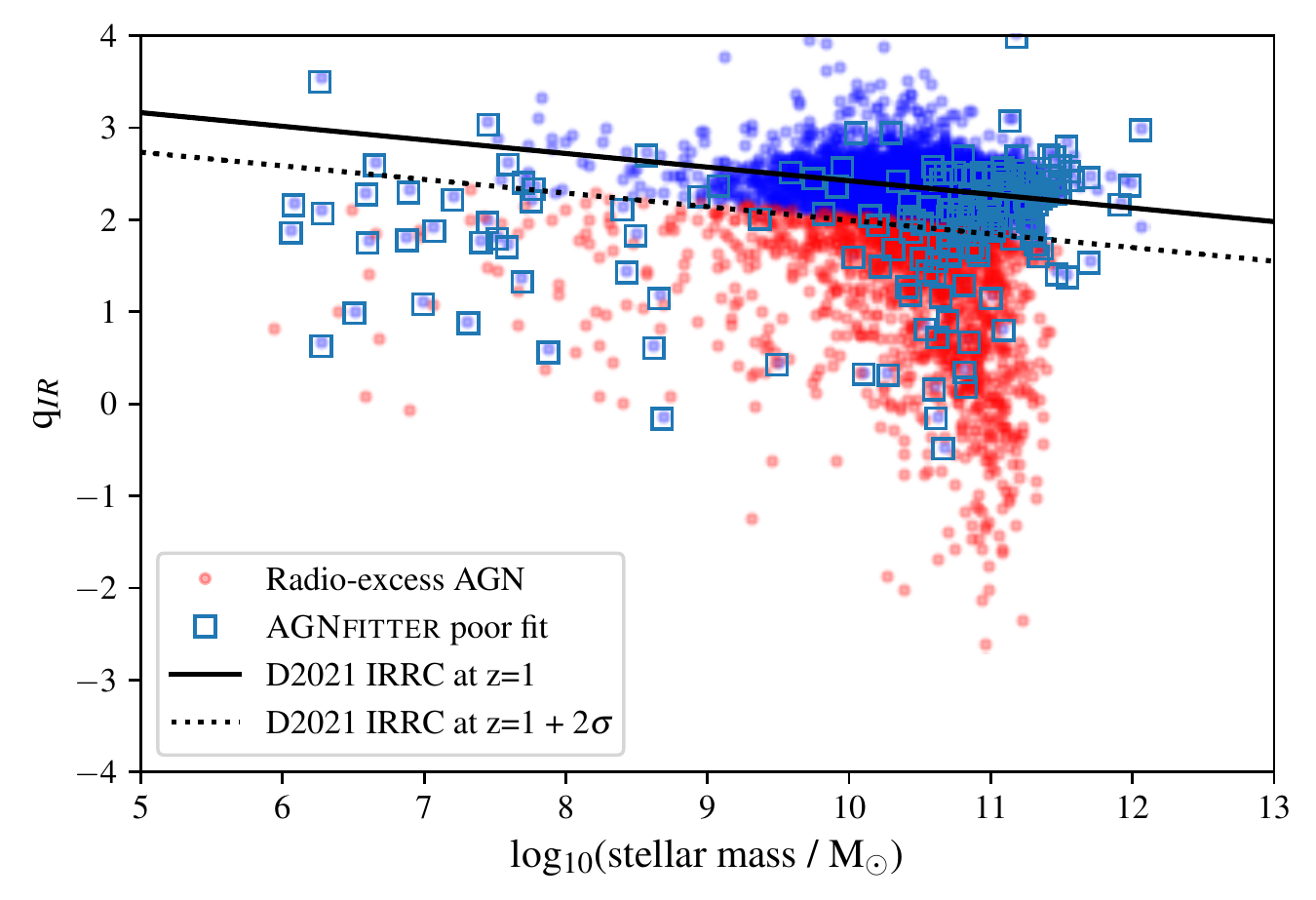}
    \caption{$q_\text{IR}$ as a function of redshift (top) and stellar mass (bottom). We use the mass- and redshift-dependent IRRC from \citet{2021A&A...647A.123D} to identify sources with a radio excess. This correlation (black solid line), along with the cut used (black dotted line), are shown at a fixed stellar mass (log$_{10}(\textrm{M/M}_\odot) = 10.5 $) and redshift ($z = 1$) in the top and bottom panels respectively. Sources classified as radio-excess AGN are shown as red points, while the rest of the sample studied in this paper are shown as blue points. As the full mass- and redshift-dependent relation is used to classify the sources, several of the classified sources appear scattered above or below the classification line; this is because they are at a different mass or redshift to the one for which the relation is shown.  Sources with a poor fit from \textsc{AGNfitter} are marked by squares.}
    \label{fig:qir}
\end{figure}

Fig.~\ref{fig:qir} shows $q_\text{IR}$ as a function of redshift and stellar mass, with the \citeauthor{2021A&A...647A.123D} IRRC relation and radio excess cut shown. Using this criterion, 1332 MIGHTEE sources are classified as having a radio excess, and 3590 are classified as not having a radio excess. A further 301 sources are unable to be classified, as \textsc{AGNfitter} gives a poor fit (with log likelihood $<-100$) so we are unable to place a reliable constraint on their total infrared luminosity.

\subsection{Mid-infrared AGN}\label{section:midIRAGN}

Galaxies which display power-law emission from an AGN torus can be identified using a mid-infrared colour-colour diagram. We use the region on a $S_{8.0~\muup\textrm{m}}/S_{4.5~\muup\textrm{m}}$ vs $S_{5.8~\muup\textrm{m}}/S_{3.6~\muup\textrm{m}}$ diagram defined by \citet{2012ApJ...748..142D}, and classify any galaxy lying in this region as a mid-infrared AGN. 

\begin{figure}
    \centering
    \includegraphics[width=\columnwidth]{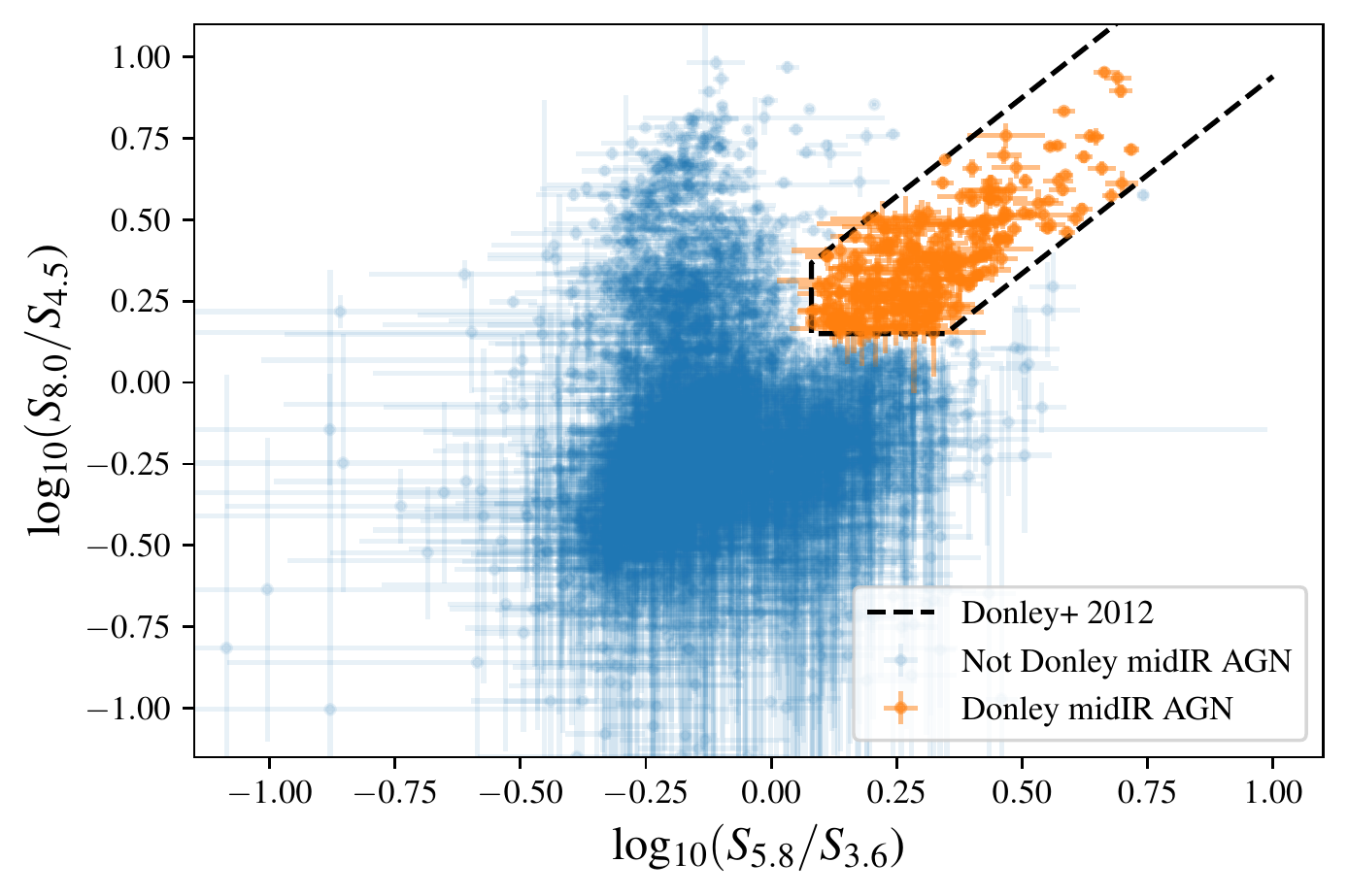}
    \caption{Mid-infrared colour-colour diagram showing sources with detections in all four bands (subscripts on axis labels are band wavelengths in $\muup$m). Sources lying inside the \citet{2012ApJ...748..142D} AGN region, shown as a black dashed line, are classified as Donley mid-infrared AGN (orange points). Sources which cannot lie inside this region, taking into account the uncertainties on the flux densities, are classified as `not Donley mid-infrared AGN' (blue points).}
    \label{fig:IRcolour1}
\end{figure}

Sources which are found outside the \citeauthor{2012ApJ...748..142D} AGN region, and which, taking into account the quoted $1 \sigma$ uncertainties on the flux density measurements in all four bands, are not within $1 \sigma$ of the boundary of the AGN region, are classified as `not Donley mid-infrared AGN'. Additionally, 3$\sigma$ upper limits are calculated for sources which are not detected at 5.8 and/or 8.0 $\muup$m. Taking into account these limits, sources which could not lie inside the \citeauthor{2012ApJ...748..142D} AGN region are also classified as `not Donley mid-infrared AGN'. These sources are shown on a mid-infrared colour-colour diagram in Fig.~\ref{fig:IRcolour1}. We note that the \citeauthor{2012ApJ...748..142D} region was designed to select a clean AGN sample at the expense of completeness, so lying outside this region does not guarantee that a source is not an AGN. While using a larger selection region would improve completeness of our mid-infrared AGN sample, it would also contaminate the sample with a number of misclassified star-forming galaxies.

In total, 273 sources are classified as Donley mid-infrared AGN, and 4542 are `not Donley mid-infrared AGN'. A further 409 sources are unable to be classified, either because they are undetected in one or more bands and the relevant flux density limits make it uncertain whether or not they lie inside the \citeauthor{2012ApJ...748..142D} AGN region, or because they lie close to the boundary and the flux error bars could take them inside the region. Of the 273 sources identified as AGN in the mid-infrared, 61 also have a radio excess.

\subsection{Optical point-like AGN}\label{section:optAGN}

Sources which display a point-like morphology in optical imaging can be classified as AGN, since emission from the nucleus outshines the emission from the host galaxy. To investigate the optical morphology of the MIGHTEE sources, we use the \emph{HST} ACS $I$-band image \citep{2007ApJS..172...38S}, with a resolution of 0.09" FWHM. We classify any source with Class\_star$>=0.9$, which is therefore point-like in the image, as an optical point-like AGN. Sources detected in the ACS $I$-band image with Class\_star$<0.9$ are classified as `not optical point-like AGN'. This gives 157 optical point-like AGN and 4540 not optical point-like AGN. A further 526 sources are unable to be classified as they are not detected in the ACS image. Of the 157 sources which are point-like in the optical image, 71 are identified as mid-infrared AGN in Section~\ref{section:midIRAGN} and 36 are classified as radio-excess AGN in Section~\ref{section:RLAGN}.

\subsection{X-ray AGN}\label{section:XAGN_class}

X-ray observations can be used to identify AGN, as some of the brightest AGN display characteristic accretion related X-ray emission. Using the X-ray data described in Section~\ref{section:xray}, we consider a source to be an X-ray AGN if its rest-frame (0.5 -- 10 keV) X-ray luminosity $L_{X} > 10^{42}~\text{erg s}^{-1}$ \citep{2004ApJS..155..271S}. The X-ray luminosities are K-corrected assuming an X-ray spectral index of $\Gamma = 1.4$, and no obscuration correction is applied (see \citealt{2016ApJ...817...34M} for details). If a source is not detected in the X-ray and has $z < 0.5$, we are able to place an upper limit on its X-ray luminosity such that $L_{X} < 10^{42}~\text{erg  s}^{-1}$, so these sources are classified as `not X-ray AGN'. Undetected sources with redshifts $>0.5$ could have X-ray luminosities greater than this cutoff, so they are not classified. This gives 519 X-ray AGN, 1084 `not X-ray AGN' and 3620 unclassified sources which lie at $z>0.5$. Of the 519 X-ray AGN, 213 are identified as radio-excess AGN in Section~\ref{section:RLAGN}, 154 are classified as mid-infrared AGN in Section~\ref{section:midIRAGN} and 94 display a point-like morphology in the optical (Section~\ref{section:optAGN}).

\subsection{VLBI AGN}\label{section:VLBA_AGN}

We use the VLBI observations of the COSMOS field described in Section~\ref{section:VLBI-data} to identify AGN. In order to be detectable in the VLBI observations, a source must have a brightness temperature above $10^6$~K \citep{2017A&A...607A.132H}. However, at frequencies $\sim1$~GHz the brightness temperature of normal galaxies does not exceed $10^5$~K \citep{1992ARA&A..30..575C}, so in order to have a brightness temperature high enough to be detected by the VLBA observations, a source must have AGN activity. We therefore classify the 255 MIGHTEE sources detected in the VLBA observations as VLBA AGN. Of these, 216 are classified as radio-excess AGN using the infrared -- radio correlation as described above, 26 do not display a radio excess and 13 are unable to be classified using the IRRC (due to poor constraints on their infrared luminosity).

\subsection{Overall classifications}\label{section:overall_class}

\begin{figure*}
    \centering
    \includegraphics[width=\columnwidth]{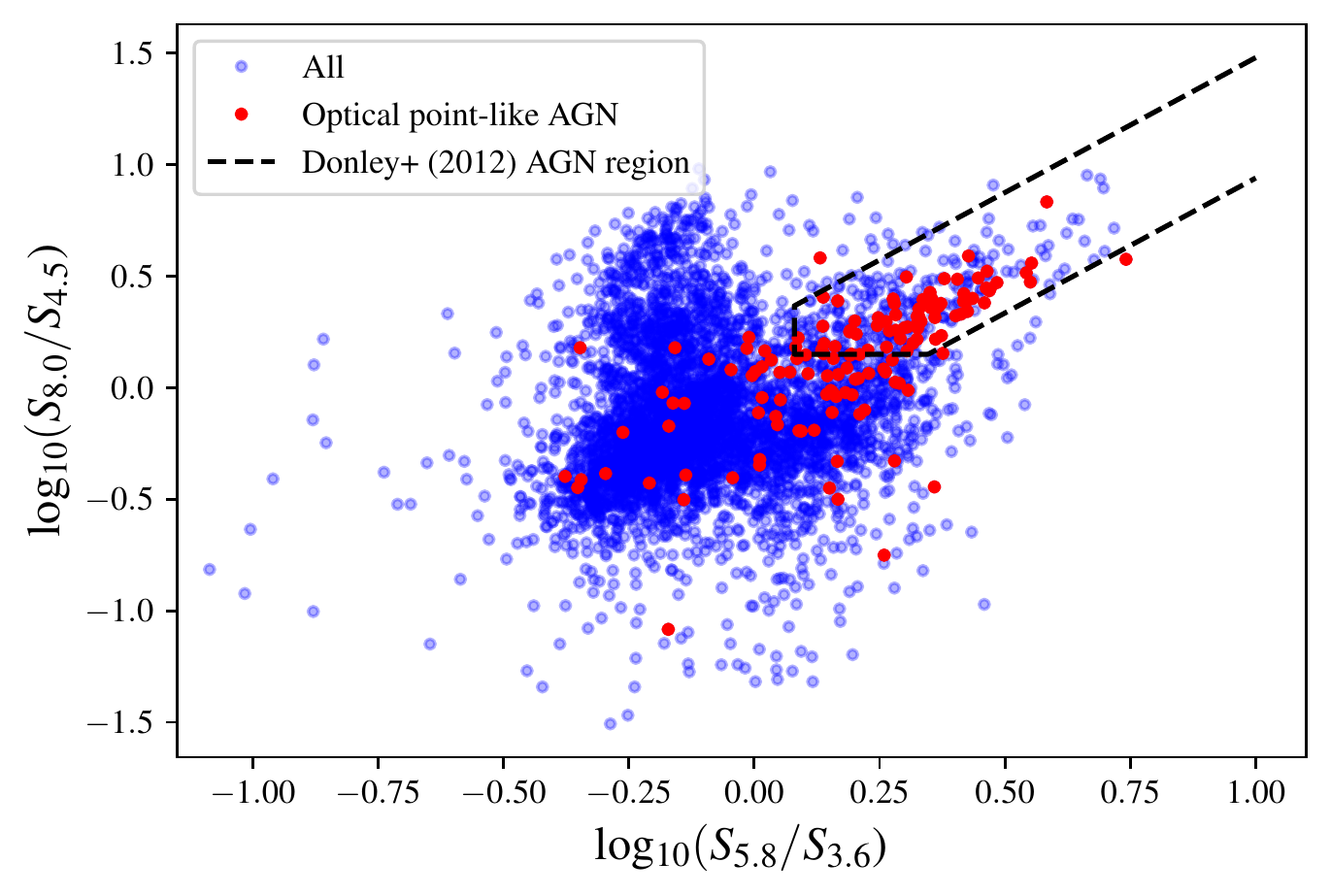}
    \includegraphics[width=\columnwidth]{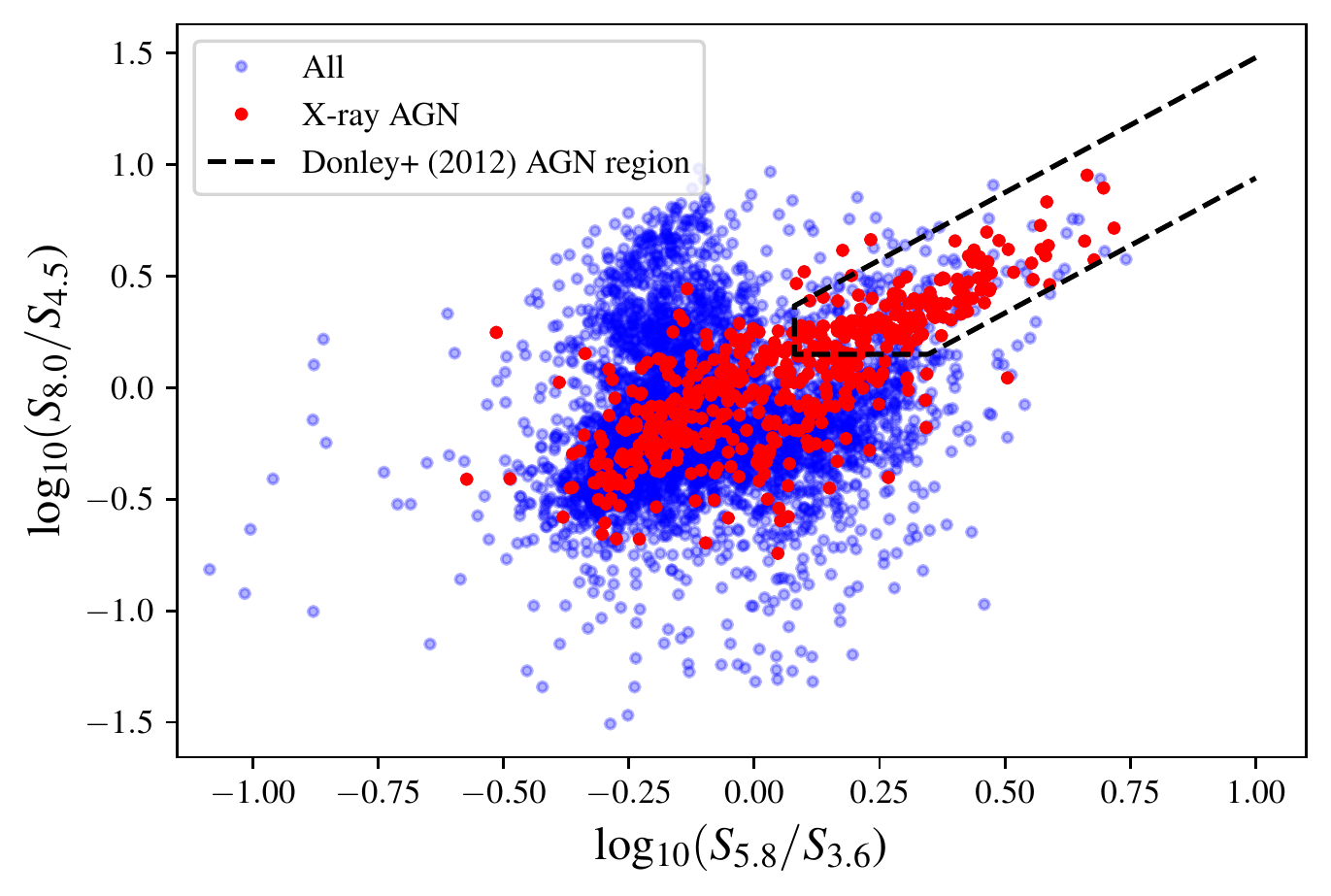}
    \includegraphics[width=\columnwidth]{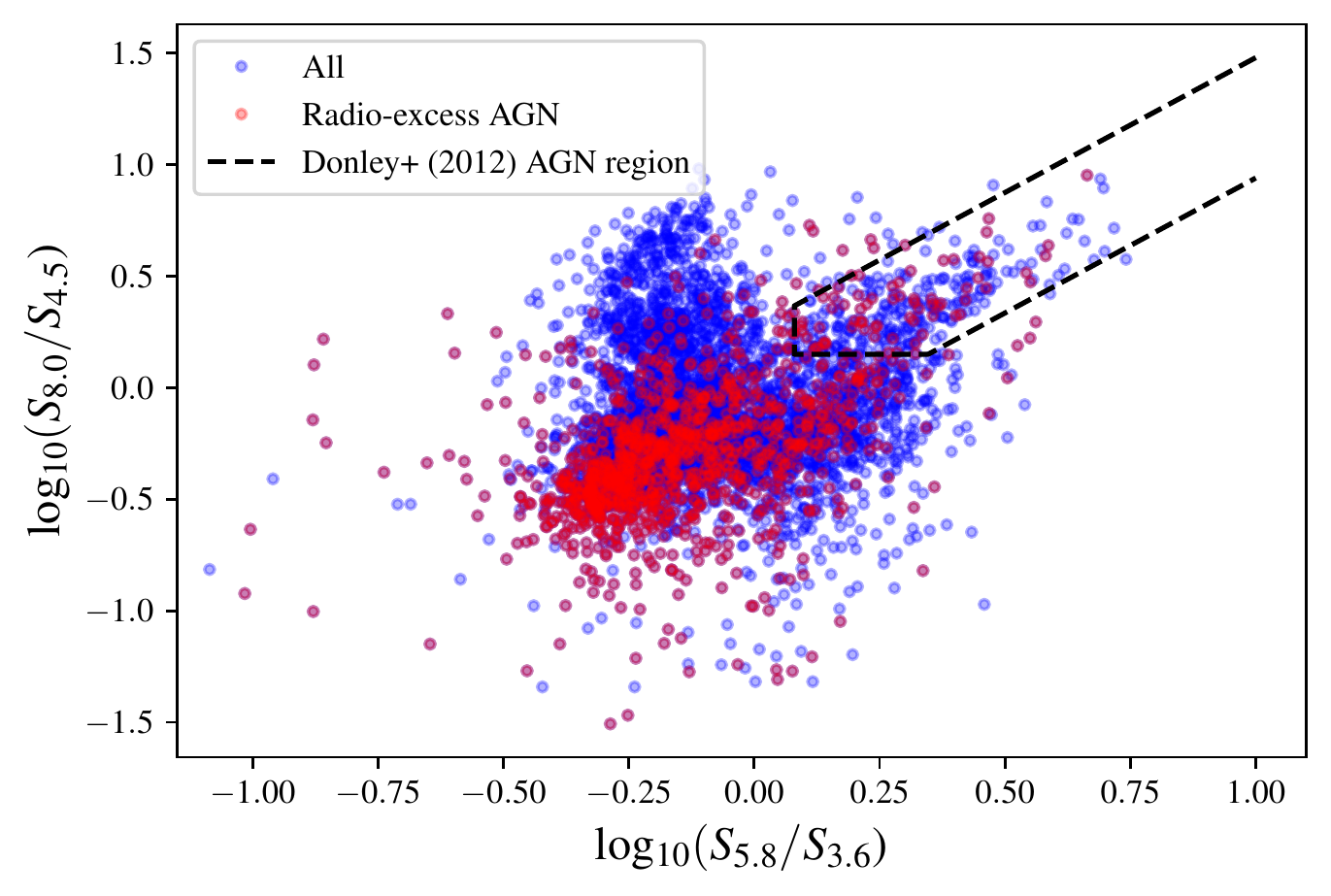}
    \includegraphics[width=\columnwidth]{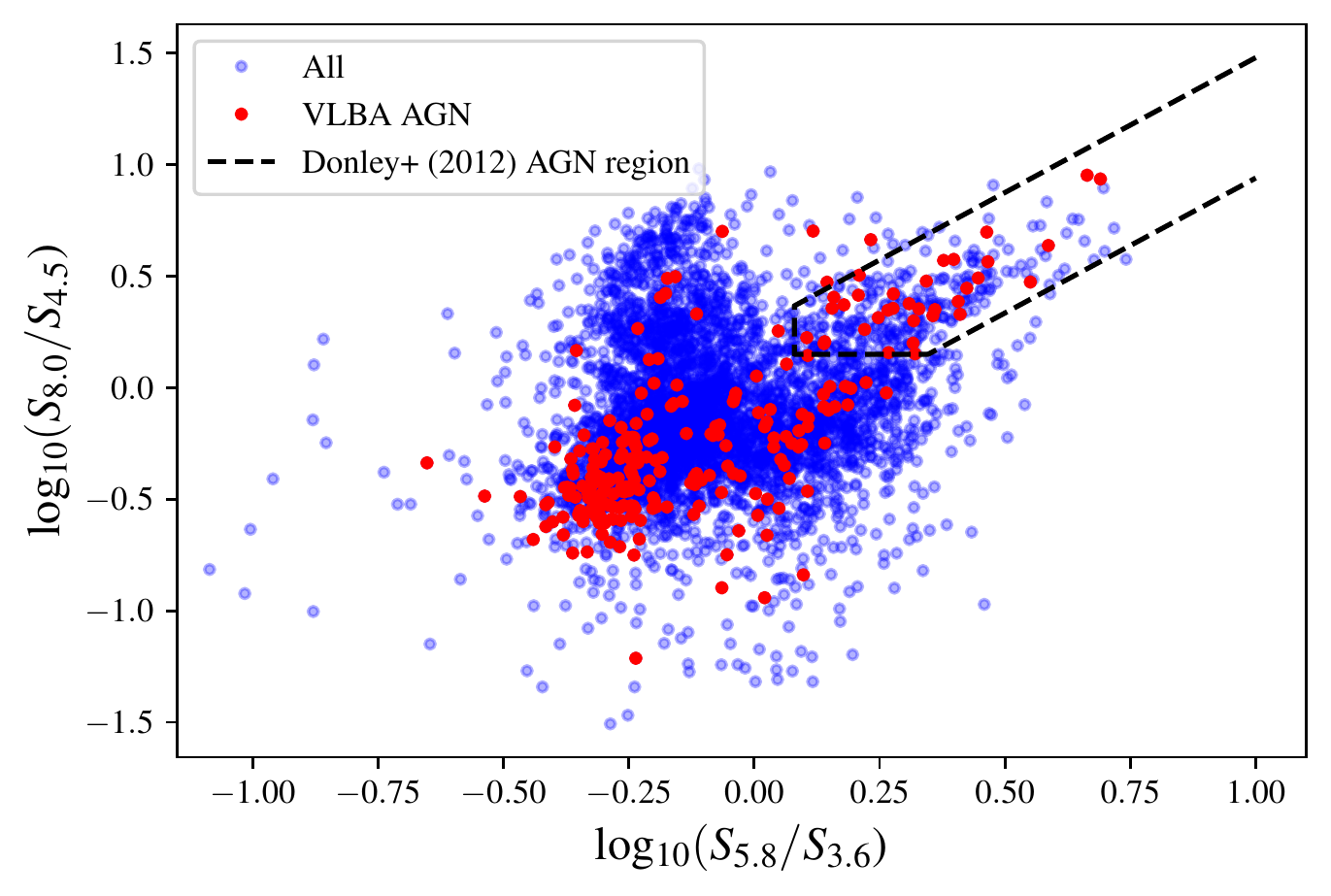}
    \caption{A mid-infrared colour-colour diagram (subscripts on axis labels are band wavelengths in $\muup$m). Sources lying inside the \citet{2012ApJ...748..142D} AGN region, shown as a black dashed line, are classified as mid-infrared AGN. Sources classified as AGN using other criteria are marked in red. Top left: optical point-like AGN (point-like in the optical; Section~\ref{section:optAGN}), top right: X-ray AGN (Section~\ref{section:XAGN_class}), bottom left: radio-excess AGN (Section~\ref{section:RLAGN}), bottom right: VLBA AGN (Section~\ref{section:VLBA_AGN}).}
    \label{fig:IRcolour2}
\end{figure*}

We have five indicators of AGN emission (radio excess, mid-infrared colours, optical morphology, X-ray power and compact radio emission detected by the VLBA), as described in the previous sections and summarised in Table~\ref{tab:classes}. These criteria are compared in Fig.~\ref{fig:IRcolour2}, which shows mid-infrared colour-colour diagrams with sources classified as AGN using the four other AGN criteria plotted separately. This shows that the X-ray, radio, optical point-like and VLBA-detected AGN generally lie in the expected regions on this diagram; the optical point-like AGN are generally found along the power-law wedge defined by \citet{2012ApJ...748..142D} (shown by the black dashed line in Fig.~\ref{fig:IRcolour2}), although there is considerable scatter outside this region. Many of the X-ray AGN are found in the \citeauthor{2012ApJ...748..142D} AGN region, although a large number of them lie along the extension of this power law to the bottom left. The radio-excess AGN are concentrated in the bottom left part of the distribution, in the region where `red and dead' host galaxies are likely to be found, as expected. Reassuringly, there are very few AGN of any type in the region generally occupied by SFGs (i.e.\ the left `bunny ear' region with log$_{10}(S_{8.0} / S_{4.5}) \gtrsim 0.2$ and $-0.5 \lesssim \textrm{log}_{10}(S_{8.0} / S_{4.5}) \lesssim 0.0 $). The cluster of radio-excess AGN found in the bottom left part of the diagram are mainly LERGs, as expected. Note that we do not necessarily expect agreement between the different AGN diagnostics, as any given AGN may only display one or two indicators of AGN activity. 

 For the overall classifications, we classify a source as an AGN if it meets any one (or more) of the five AGN criteria. Sources which we can securely classify as not being an AGN using all five criteria are classified as star-forming galaxies. Because the X-ray criterion limits us to only classifying sources with redshifts $<0.5$ in this way, we introduce the additional classification of `probable SFG' for sources which have redshifts $>0.5$ so are unable fulfil the `not X-ray AGN' criteria, but which are classified as `not AGN' using the other four criteria.

\begin{figure}
    \centering
    \includegraphics[width=\columnwidth]{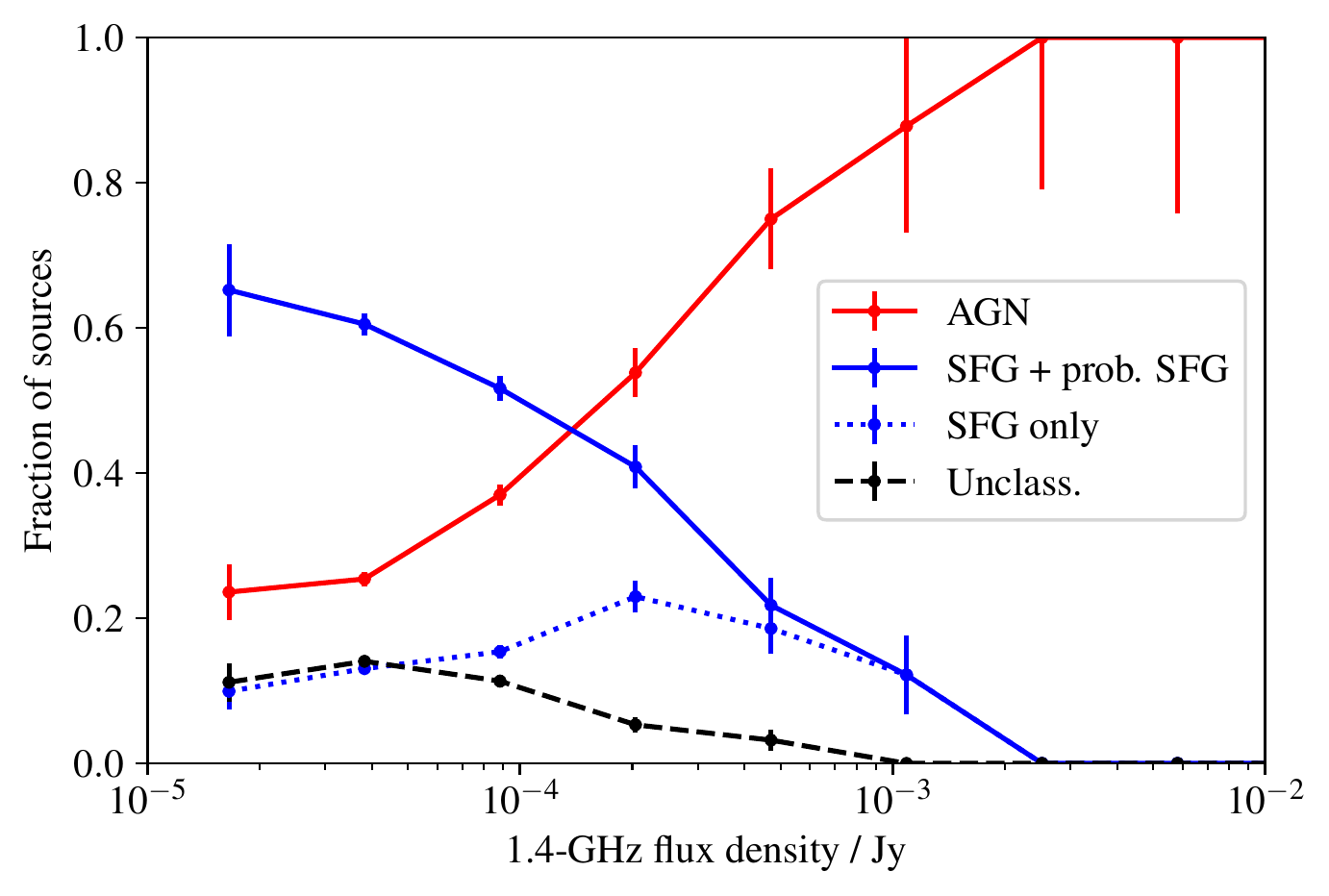}
    \includegraphics[width=\columnwidth]{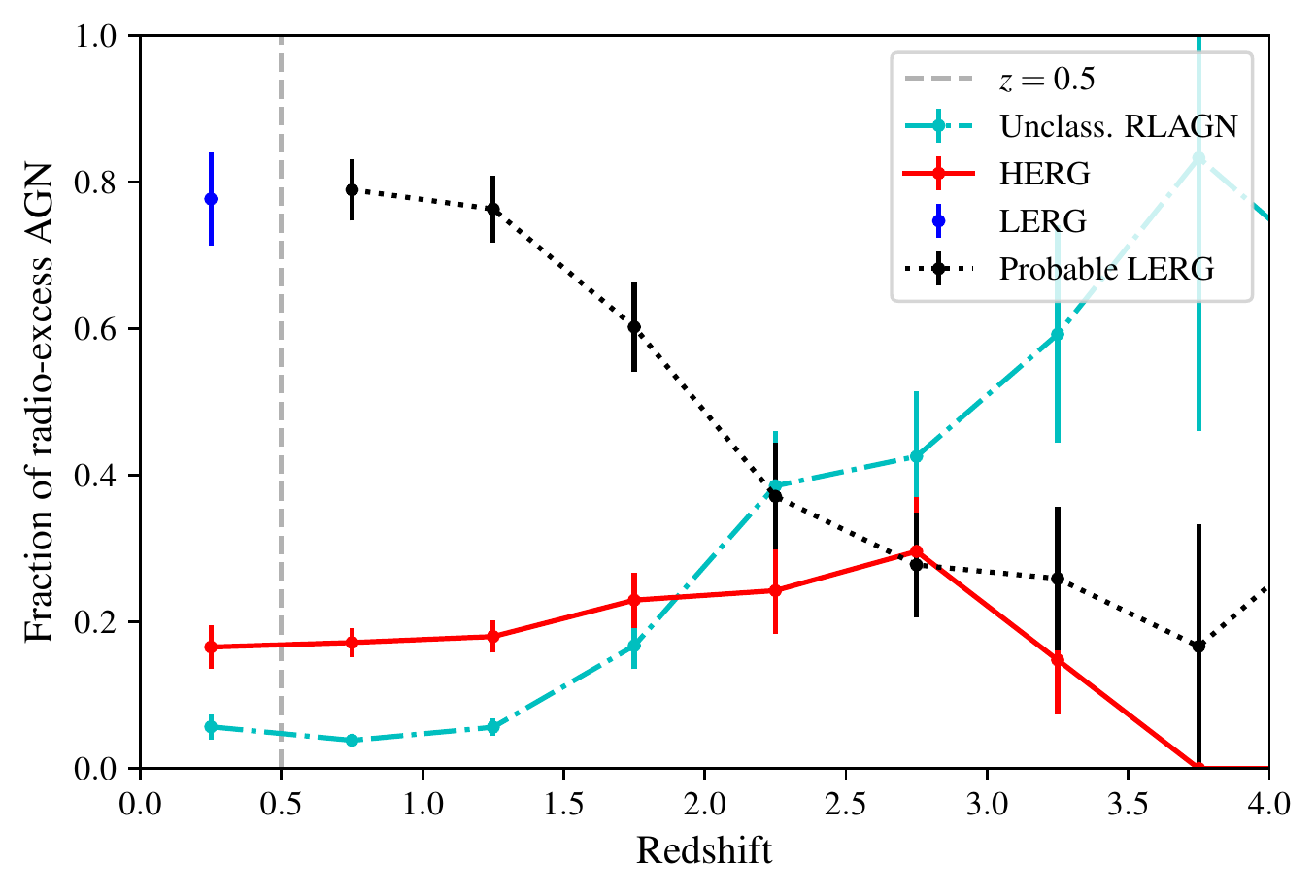}
    \caption{Top: proportion of sources with a host galaxy identification that are classified as AGN and SFG as a function of radio flux density. The red solid line shows the AGN, the solid blue line shows the SFG and probable SFG combined, the dotted blue line shows the sources securely classified as SFG only, and the dashed black line shows the unclassified sources. Bottom: proportion of radio-excess AGN which are HERGs (red solid line), LERGs (blue dashed line), probable LERGs (black dotted line) and unclassified RLAGN (cyan dot-dashed line) as a function of redshift. The grey dashed line is at $z=0.5$, the redshift beyond which we cannot securely classify LERGs so adopt the `probable LERG' classification instead.}
    \label{fig:fracHERG}
\end{figure}

Additionally, we classify the radio excess AGN as high-excitation and low-excitation radio galaxies (HERGs and LERGs) in a similar way to \citet{2016MNRAS.462.2122W}. HERGs generally display typical AGN signatures across the electromagnetic spectrum, while LERGs typically lack the usual AGN apparatus (e.g. accretion disk, dusty torus). This means that while LERGs are radio loud, they do not display AGN characteristics in other parts of the electromagnetic spectrum. Therefore, we classify radio excess sources which are classified as an AGN using at least one of the following other diagnostics as HERGs: optical morphology, mid-infrared colours or X-ray power. Sources which have a radio excess and are securely classified as `not AGN' using the remaining three criteria are classified as LERGs. Due to the X-ray limit, this restricts us to sources with $z<0.5$. Therefore, in a similar way as for the SFGs, we define a `probable LERG' class - any radio excess source with $z>0.5$ which is securely classified as not being an AGN in the optical and mid-infrared is considered a `probable LERG'. Radio-loud AGN which are not classified as a HERG, LERG or probable LERG are referred to as `unclassified RLAGN'; these are radio-loud AGN which do not meet the HERG criteria (i.e.\ they are not identified as AGN using the X-ray, mid-infrared or optical morphology data) but are not securely classified as `not AGN' using all the relevant criteria so are not able to be classified as LERGs or probable LERGs. Sources which do not have a radio excess but which fulfil one of the other three AGN criteria are classified as radio-quiet AGN; 417 sources fit these criteria. Note that in order to be classified as either radio loud or radio quiet a source must not have a poor fit with \textsc{AGNfitter}  (quantified by log likelihood $<-100$) so that we are able to place a reliable constraint on its total infrared luminosity. This means that a small number of AGN (57) are not classified as either radio loud or radio quiet. These classifications are summarised in Table~\ref{tab:overallclasses}. Fig.~\ref{fig:fracHERG} shows the proportion of different source types as a function of radio flux density and redshift.

\begin{table}
    \centering
    \caption{Number of sources classified as AGN, not robust AGN, and unclassified using the criteria outlined in Section~\ref{section:classification}. The `not robust AGN' column refers to sources which are securely identified as not having radio excess, have an X-ray luminosity $<10^{42} \textrm{erg \, s}^{-1}$, lie outside the Donley AGN selection region on a mid-infrared colour-colour diagram, are extended in the optical or are not detected by the VLBA observations respectively.}
    \begin{tabular}{l|ccc}\hline
    AGN criteria & AGN & Not robust AGN & Unclassified\\\hline
    Radio excess AGN & 1332 & 3590 & 301 \\
    Donley mid-infrared AGN & 273 & 4541 & 409 \\
    Optical point-like AGN & 157 & 4540 & 526 \\
    X-ray AGN & 519 & 1084$^1$ & 3620 \\
    VLBA AGN & 255 & 4968 & \\\hline
    \end{tabular}
    \\$^1$ Limited to sources with $z<0.5$ (see text).
    \label{tab:classes}
\end{table}

\begin{table}
    \centering
    \caption{Summary of the number of sources in each of the classes described in Section ~\ref{section:overall_class}.}
    \begin{tabular}{l|c}\hline
    Overall class & Number of sources \\\hline
    AGN & 1806 \\
    SFG & 766 \\
    Probable SFG & 2040 \\
    Unclassified & 611 \\\hline
    Total RLAGN & 1332 \\
    HERG & 249 \\
    LERG & 150 \\
    Probable LERG & 782 \\
    Unclassified RLAGN & 151 \\\hline
    Radio quiet AGN & 417\\\hline
    Total number of sources & 5223 \\\hline
    \end{tabular}
    \label{tab:overallclasses}
\end{table}

The catalogue containing these classifications is known as the Level 3 catalogue and is released with this work; the columns in this catalogue are described in Appendix~\ref{appendix:level3-cat} and the full catalogue is available as supplementary material.

\subsection{Comparison with optical spectra}\label{section:opt_spectra_comparison}

\begin{table}
    \centering
    \caption{Classifications using the scheme outlined in Section~\ref{section:overall_class} for the 82 sources in our sample with SDSS spectra.}
    \begin{tabular}{l|c|cccc}\hline
      BPT class     &       & \multicolumn{4}{|c|}{Classifications from this work}\\
      \cmidrule(lr){3-6}
      (from SDSS)   &  Total & AGN & SFG & Prob. SFG & Unclass. \\\hline
      Star-forming  & 28 & 5 & 22 & 0 & 1 \\
      Composite     & 22 & 9  & 13 & 0 & 0\\
      Seyfert       & 10 & 9  & 1  & 0 & 0\\
      Seyfert/LINER & 1  & 1  & 0  & 0 & 0\\
      LINER         & 21 & 15 & 5  & 0 & 1\\\hline
    \end{tabular}
    \label{tab:spec_comparison}
\end{table}

In our sample, 82 sources have a Baldwin, Philips and Terlevich (BPT; \citealt{1981PASP...93....5B}) diagram classification from SDSS. These classifications are compared to those using the scheme described in Section~\ref{section:overall_class} in Table~\ref{tab:spec_comparison}. The classifications are broadly in agreement, with a few discrepancies. Notably, 5 sources which are found in the star-forming region of the BPT diagram are classified as AGN using the scheme described in this paper. All 5 of these sources have a radio excess but do not display any other signatures of AGN activity. This demonstrates that classifications are not always clear-cut when dealing with faint populations (in both the radio and multi-wavelength data). 

\begin{table}
    \centering
    \caption{Comparison between the classification of HERGs and LERGs used in this paper and those using [O\textsc{iii}] equivalent width measurements from optical spectra.}
    \begin{tabular}{c|ccccc}\hline
         &  Total & HERG & LERG & Prob. LERG & Unclass.\\\hline
     Spec. HERG  & 6 & 3 & 0 & 3 & 0\\
     Spec. LERG & 33 & 6 & 15 & 12 & 0\\\hline
    \end{tabular}
    \label{tab:specHERG}
\end{table}

\begin{figure}
    \centering
    \includegraphics[width=\columnwidth]{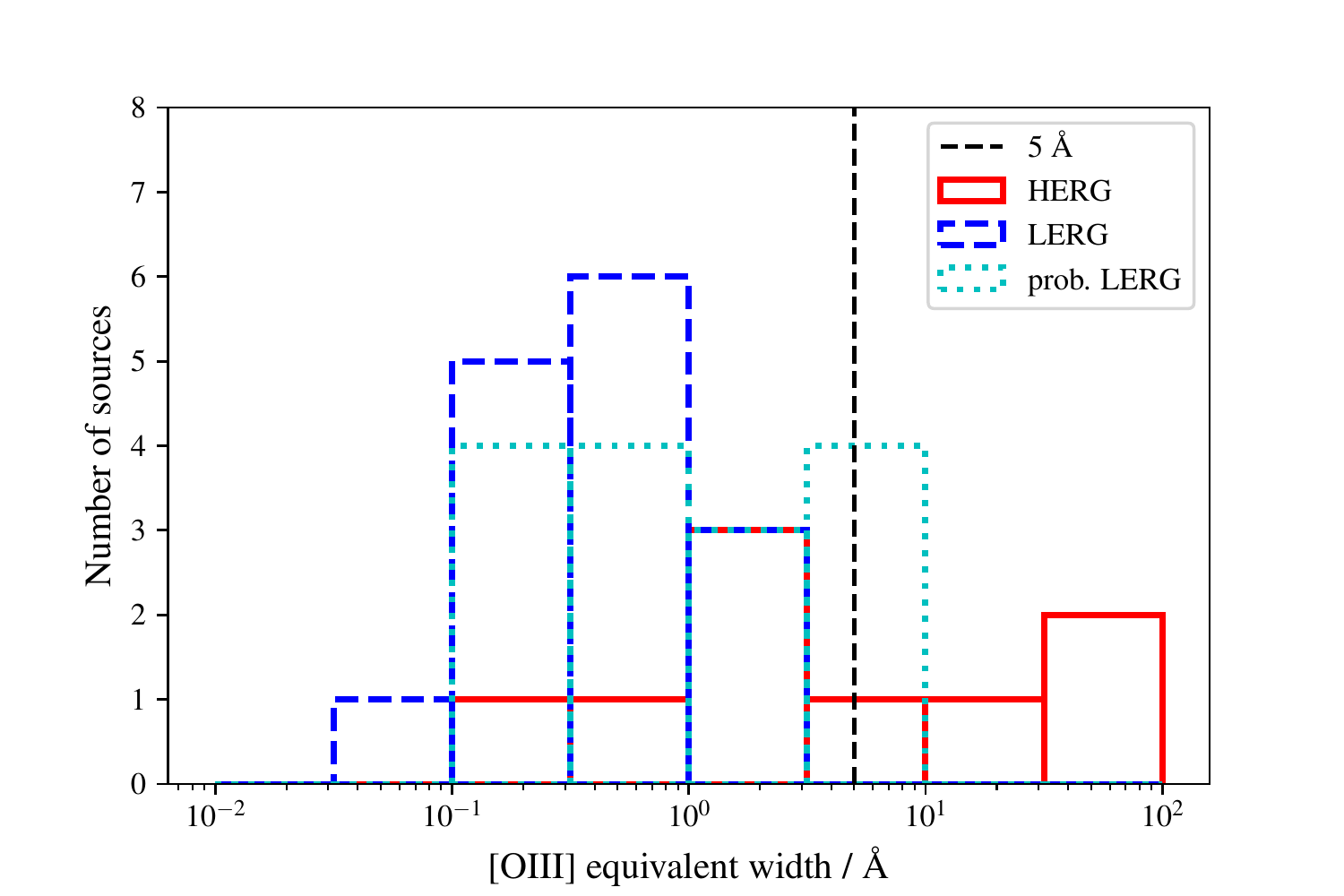}
    \caption{[O\textsc{iii}] equivalent width for sources in our sample with line measurements available from the SDSS or LEGA-C surveys. Sources classified as HERGs (red solid line), LERGs (blue dashed line) and probable LERGs (cyan dotted line) using the criteria in Section~\ref{section:overall_class} shown separately.}
    \label{fig:OIII}
\end{figure}

The equivalent width of the [O\textsc{iii}] line at 5007~\AA~is often used in the literature to separate HERGs and LERGs, for example by \citet{1994ASPC...54..201L,1997MNRAS.286..241J,1998MNRAS.298.1035T,2017MNRAS.469.4584C,2018MNRAS.480..707P}. Where [O\textsc{iii}] line measurements are available, we classify sources with [O\textsc{iii}] equivalent width $>5$~\AA~as HERGs, and those with values $<5$~\AA~as LERGs. 42 radio excess sources in our sample have an [O\textsc{iii}] equivalent width measurement available, from either SDSS or LEGA-C, so are able to be classified in this way.

The classifications using [O\textsc{iii}] equivalent width are compared to those derived from the criteria in Section~\ref{section:overall_class} in Table~\ref{tab:specHERG} and Fig.~\ref{fig:OIII}. While the classifications using [O\textsc{iii}] equivalent width are broadly in agreement with those used in this work, six sources with [O\textsc{iii}] equivalent width $< 5$~\AA~are classified as HERGs. All six sources have X-ray luminosities $>10^{42}$ erg/s, indicating that there is AGN-accretion related X-ray emission, which is characteristic of HERGs. Again, this shows that source classifications are not always clear-cut, and depend on the criteria used. In another example, \citet{2014MNRAS.440..269M} found seven sources in their sample of very bright radio galaxies ($S_{2.7~\textrm{GHz}} > 2$~Jy) that were classified as LERGs based on their optical spectra but which showed signs of being radiatively efficient at other wavelengths.

\subsection{Comparison with VLA-COSMOS 3~GHz Large Project}\label{section:vla}

The VLA-COSMOS 3 GHz Large Project (\citealt{2017A&A...602A...1S}; hereafter VLA-COSMOS) surveyed 2~deg$^2$ of the COSMOS field at 3~GHz with median rms of 2.3~$\muup$Jy/beam and a resolution of 0.75~arcsec. They use a combination of mid-infrared colours, rest-frame optical colours, X-ray emission, radio excess, SED fitting and \emph{Herschel} detections to classify the radio sources detected in their observations (full details are given in \citealt{2017A&A...602A...2S}). As many of these sources also appear in our catalogue, in this section we compare their classifications to the classification scheme described in this work. While there are some similarities in the classification schemes employed by the two studies, there are some notable differences; for example, our work utilises the high-resolution \emph{HST} data while the VLA-COSMOS study uses rest-frame optical colours, and the two studies use different SED-fitting codes and different radio-excess definitions.

\begin{figure}
    \centering
    \includegraphics[width=\columnwidth]{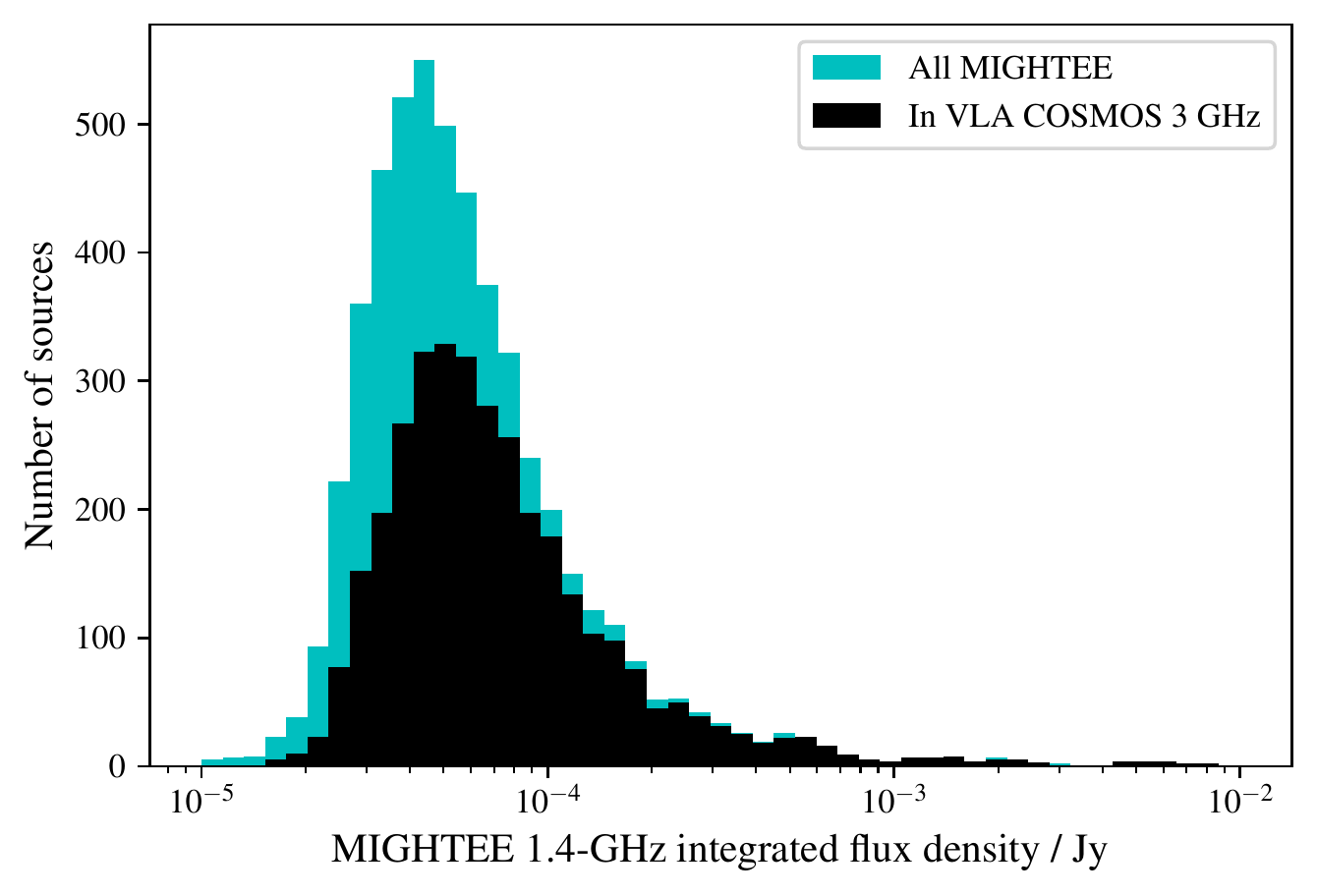}
    \caption{1.4-GHz integrated flux density from MIGHTEE for the MIGHTEE Early Science cross-matched sample considered in this work (cyan), and the sub-sample of these sources which are also in the VLA-COSMOS 3~GHz multi-wavelength sample (black). Measured flux densities are scaled to 1.4-GHz using the effective frequency map and assuming a spectral index of 0.7.}
    \label{fig:flux_vla}
\end{figure}

There are 3386 sources with multi-wavelength identifications in both the MIGHTEE catalogue and the VLA-COSMOS catalogue (matched using optical positions and match radius of 1~arcsec). The total MIGHTEE radio flux densities of these sources are shown in Fig.~\ref{fig:flux_vla}. There are 2975 sources with source type classifications available in both the VLA-COSMOS catalogue and this work, the classifications of these sources are compared in Table~\ref{tab:vla_comparison}. Here we have combined the sources classified as SFG and probable SFG in this work in one column, as this is approximately comparable to the VLA-COSMOS `SFG' class, and listed the sources classified as SFG using our criteria in another column. We have also combined our LERG and probable LERG classes in this table to ease comparison. The redshift values used by the two projects are not necessarily the same as we utilise new spectroscopic redshifts available for some sources, and have used updated photometry and a different method to estimate the photometric redshifts (see Section~\ref{section:redshifts} for details). Although the redshift estimates for many of the sources are consistent, the values for 320 (12 per cent) of sources are outliers, where an outlier is defined as being when $|z_\textrm{MIGHTEE} - z_\textrm{VLA-COSMOS}| / (1 + z_\textrm{MIGHTEE}) > 0.15$. The number of sources in each category where the redshifts used by the two projects do not significantly disagree (i.e.\ when $|z_\textrm{MIGHTEE} - z_\textrm{VLA-COSMOS}| / (1 + z_\textrm{MIGHTEE}) < 0.15$) are shown in brackets in Table~\ref{tab:vla_comparison}. We find no trend with the type of source and discrepant redshift.

76 per cent of the sources classified as AGN in this work are also classified as AGN in VLA-COSMOS, while 89 percent of our SFG are classified as clean SFG\footnote{The VLA-COSMOS team define a SFG class, which includes some sources with a radio excess, and a `clean SFG' class where these radio-excess sources are removed. Here we compare to their `clean SFG' class, as this is more consistent with the definition of SFG used in this work, which excludes all radio-excess sources.} in their work. This demonstrates that there is broad agreement between the two classification schemes. However, despite the excellent photometric data utilised by both groups, there are some differences evident in the classifications, highlighting the dependence on the exact classification method used. For example, 288 sources which are classified as AGN in our catalogue are identified as clean SFGs in the VLA-COSMOS catalogue. The vast majority of the sources (260 out of 288) are identified as being AGN in the radio only. While both classification schemes use the far-infrared-radio correlation to identify sources with a radio excess, we use different versions of
the correlation. \citet{2017A&A...602A...2S} use a redshift-dependent relation based on \citet{2017A&A...602A...3D} and \citet{2017A&A...602A...4D}. In this work, we use the updated \citet{2021A&A...647A.123D} correlation, which is dependent on stellar mass as well as redshift. We also employ a different cut to identify radio-excess AGN; following \citet{2021A&A...647A.123D} and  \citet{2018MNRAS.475.3429W}, we classify sources which lie more than $2\sigma$ from the IRRC as having a radio-excess, where $\sigma$ is the instrinsic scatter on the relation. \citeauthor{2017A&A...602A...2S}, however, use $3\sigma$ instead. This may well be partly responsible for the higher number of radio AGN identified in our work. Additionally, the compact configuration of MeerKAT means the telescope has excellent sensitivity to low-surface brightness emission, meaning that it picks up such extended emission from sources which may be missed by the VLA-COSMOS observations. This issue is discussed in briefly Hale et al. (in prep) and could be responsible for some radio excess AGN being identified in the MIGHTEE observations and missed in the VLA survey.

This highlights the difficulties of classifying sources based on photometry alone, and the sensitivity to different approaches.

\begin{table*}
    \centering
    \caption{Comparison of sources classified in this work and the classifications by the VLA-COSMOS team for sources which feature in both catalogues. The numbers in brackets are the number of sources where the redshift values used by each project agree within $|z_\textrm{MIGHTEE} - z_\textrm{VLA-COSMOS}| / (1 + z_\textrm{MIGHTEE}) > 0.15$. HLAGN and MLAGN are high and medium luminosity AGN respectively, as classified by the VLA-COSMOS team. These can be very approximately related to our HERG and LERG classes respectively. The last row shows the percentage of sources in each MIGHTEE class classified as the approximately equivalent class using the VLA classification scheme. For these purposes, MIGHTEE AGN is compared to VLA AGN, MIGHTEE SFG \& prob. SFG is compared to VLA clean SFG in the second column, MIGHTEE SFG is also compared to VLA clean SFG in the third column, MIGHTEE HERG is compared to VLA HLAGN, and MIGHTEE LERG \& prob. LERG is compared to VLA MLAGN. Note that the classification methods are not identical so these classes are not directly equivalent}. Sources classified as AGN and SFG by both classification schemes are shown in bold to aid the reader.
    \begin{tabular}{c|cccccc}\hline
    VLA classes   & \multicolumn{5}{c}{Classes in this work} & Total \\
    \cmidrule(lr){2-6}
               & AGN           & SFG \& prob. SFG  &   SFG & HERG    & LERG \& prob. LERG &  \\\hline
    AGN                 & \textbf{1075}  (936)    & 200 (180)          & 27 (26)       & 201 (175)   & 457 (431)  & 1275 (1116)   \\
    clean SFG             &  288 (239)          & \textbf{1288} (1196)         & 318 (315)      & 5 (3)    & 229 (202) & 1576 (1435)    \\
    HLAGN               & 598 (500)          & 102  (92)         & 18 (17)       & 198 (173)  & 51 (45) & 700 (592)      \\
    MLAGN               & 477 (436)          & 98 (88)           & 9 (9)       & 3 (2)    & 406 (386) & 575 (524)      \\\hline
    Total                & 1408 (1207)         & 1567 (1448)         & 357 (353)      & 207 (179)   & 717 (659) & 2975 (2655) \\
     \% agree with VLA class  & 76 (78) & 82 (83) & 89 (89)  & 96 (97) & 57 (59) & \\\hline
    \end{tabular}
    \label{tab:vla_comparison}
\end{table*}

\section{Properties of the radio-loud AGN}\label{section:properties}

\subsection{Host galaxies of HERGs and LERGs}\label{section:hosts}

\begin{figure}
    \centering
    \includegraphics[width=\columnwidth]{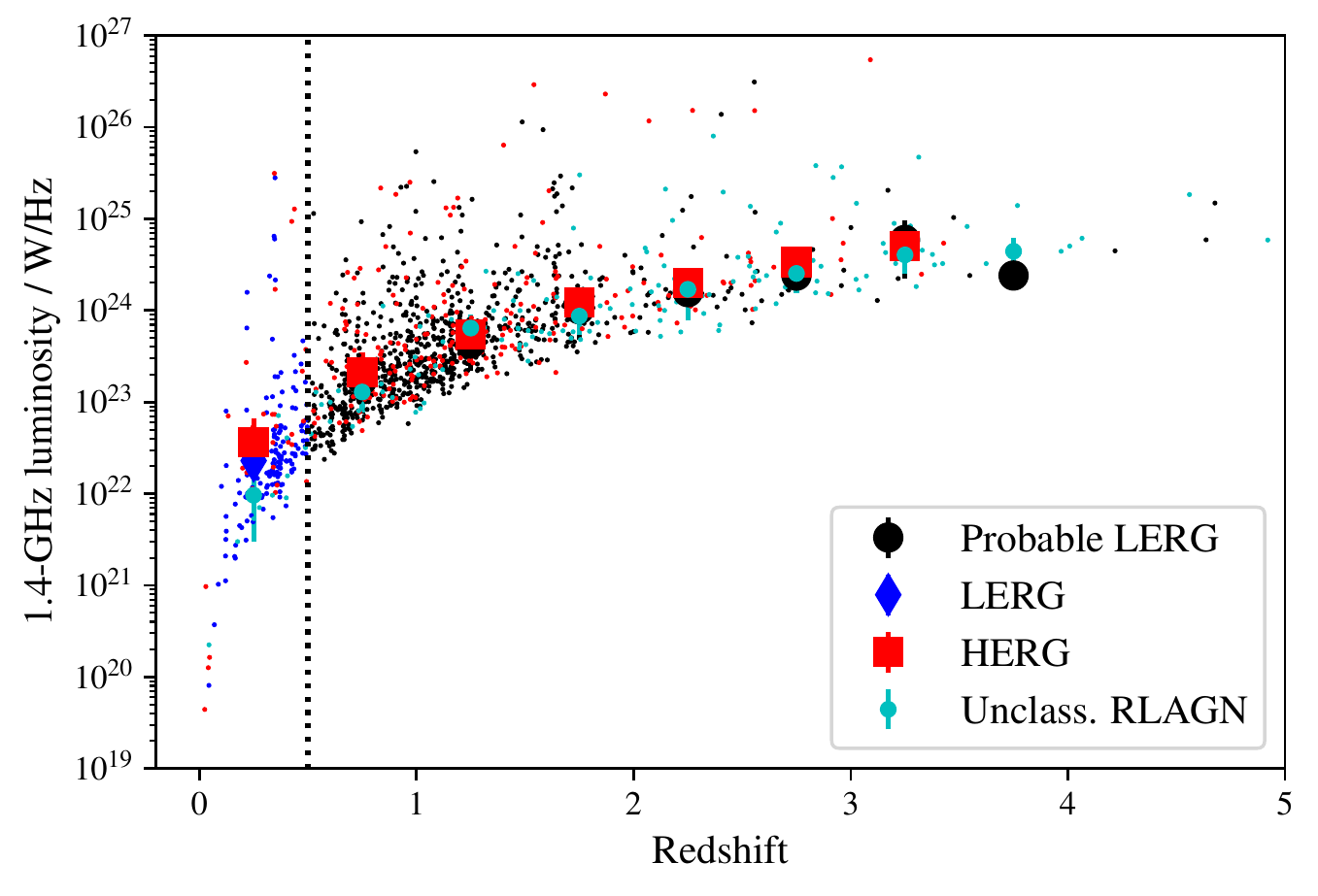}
    \caption{Radio luminosity as a function of redshift for the HERGs (red), LERGs (blue), probable LERGs (black) and unclassified RLAGN (cyan) in this sample. The large solid shapes show the median radio luminosity in each redshift bin. The dotted line is at $z=0.5$, the redshift beyond which we are unable to securely classify sources as LERGs due to the depth of the X-ray data (see text), and instead adopt the probable LERG classification.}
    \label{fig:lz-HERG}
\end{figure}

Fig.~\ref{fig:lz-HERG} shows the radio luminosity and redshift distribution of the HERGs and LERGs in this sample. The median radio luminosities of the two classes of radio galaxy are indistinguishable across the redshift range probed by this sample ($0 < z < 5$).

\begin{figure}
    \centering
    \includegraphics[width=\columnwidth]{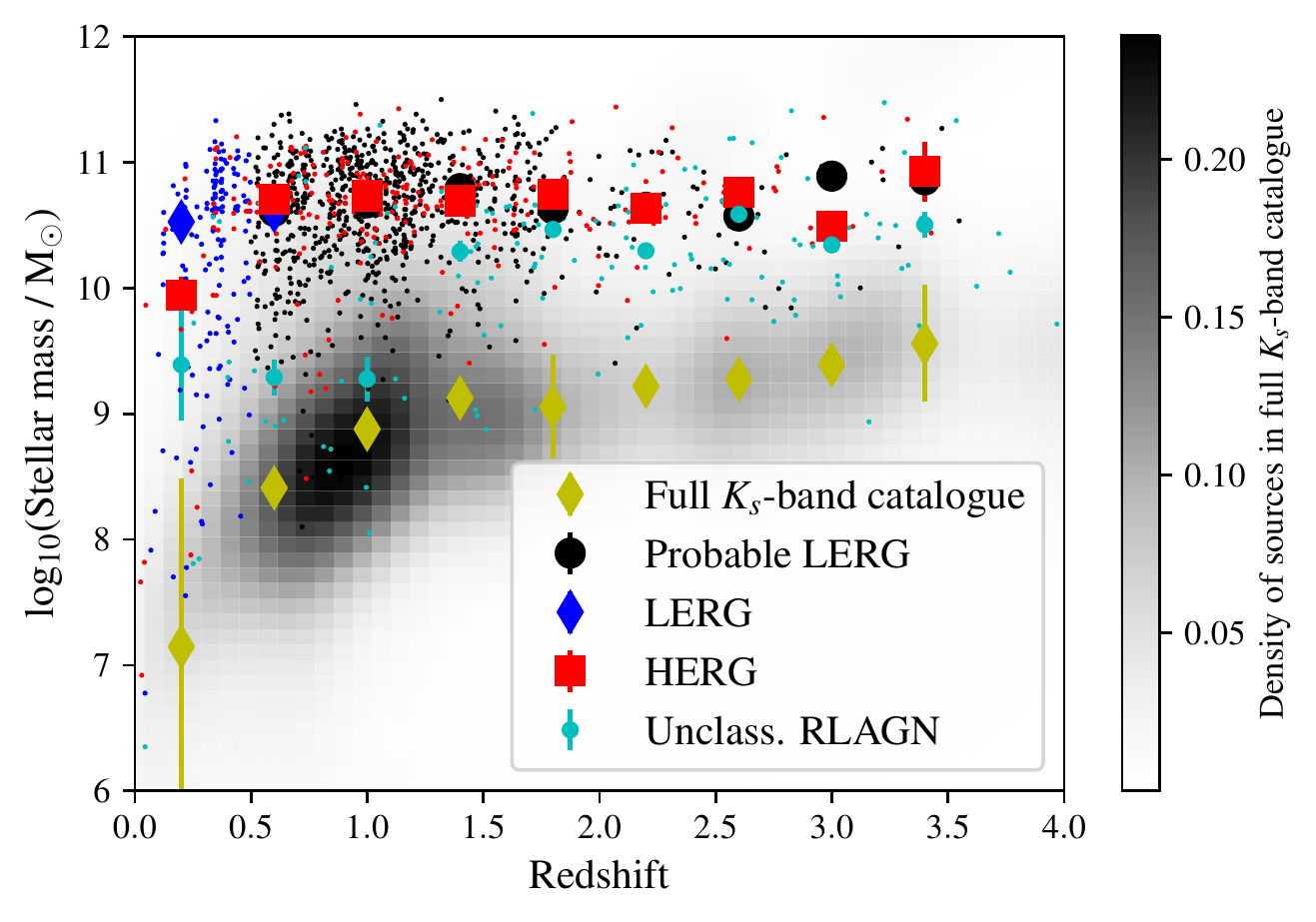}
    \includegraphics[width=\columnwidth]{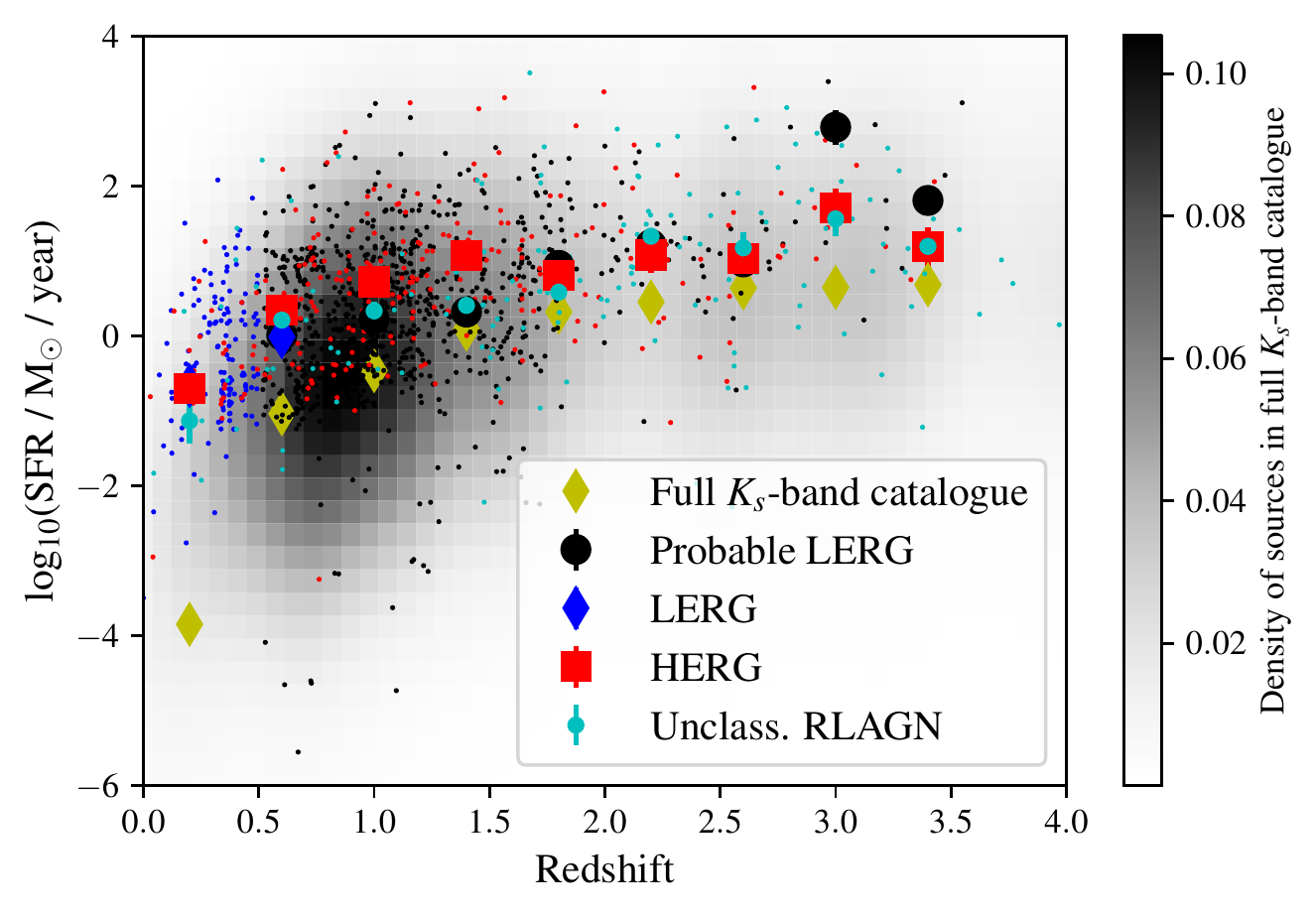}
    \caption{Stellar mass (top) and star formation rate (bottom) as a function of redshift for the HERGs (red), LERGs (blue), probable LERGs (black) and unclassified RLAGN (cyan) in this sample (small shapes). The greyscale shows the distribution of galaxies in the full $K_s$-band selected optical/near-infrared catalogue, the large yellow diamonds show the medians in each redshift bin for this sample. The other large solid shapes show the median stellar mass/star-formation rate for the different radio samples in each redshift bin.Stellar masses and star-formation rates shown in this plot are estimated from \textsc{LePhare}, rather than the \textsc{AGNfitter} values used in the rest of this paper, see text for details. }
    \label{fig:Mstar-HERG}
\end{figure}

The stellar masses and star-formation rates of HERGs and LERGs as a function of redshift are shown in Fig.~\ref{fig:Mstar-HERG}. These plots demonstrate that the HERGs and LERGs in our sample are hosted by galaxies with very similar host galaxy properties out to $z \sim 4$. The only exception is in the lowest redshift bin (z < 0.4), where LERGs seem to be hosted by galaxies with higher stellar masses than HERGs. This is in agreement with previous studies at higher radio powers which overlap with this redshift bin (e.g.\ \citealt{2012MNRAS.421.1569B,2018MNRAS.480..358W}), which find a significant difference in the properties of galaxies hosting these more powerful HERGs and LERGs. Our results demonstrate that at lower radio powers and higher redshifts the host galaxies of HERGs and LERGs are becoming indistinguishable.

While we are reasonably confident that all the sources in our HERG sample are indeed HERGs, it is possible that some of the sources in our LERG and `probable LERG' samples are HERGs where signs of efficient accretion have been missed.  In particular, the sources at $z>0.5$ classified as `probable LERGs' could have accretion-related X-ray emission which is too faint to be detected in the observations used in this work, and therefore should in reality be classified as HERGs. In addition, the Donley et al. mid-infrared selection used in this work is incomplete, so may have missed some mid-infrared HERGs. These potentially misclassified HERGs could cause the distributions of the host galaxy properties of the HERGs and LERGs to appear more similar than they actually are. However, as the stellar mass and star-formation distributions of the HERGs and probable LERGs are indistinguishable at $z>0.5$, adding a small number of potentially misclassified probable LERGs into the HERG sample should make little difference to the overall results.

The greyscale and the yellow points in Fig.~\ref{fig:Mstar-HERG} show the properties of the full optical catalogue, plotted here to highlight possible selection effects. At higher redshifts we can only detect more massive galaxies, but the median stellar mass of the full population is at least an order of magnitude below that for our radio galaxy sample across the full redshift range under consideration here, so this should not affect our results. The host galaxies of the radio galaxies are more massive than the general optically-selected population, as expected. The stellar masses and star-formation rates shown in this figure are estimated using the \textsc{LePhare} SED-fitting code by fixing the redshift at the best available value for each source (see Section~\ref{section:redshifts}). We show the \textsc{LePhare} galaxy properties in this figure, rather than those estimated by \textsc{AGNfitter} used in the remainder of the paper, in order to be directly comparable to the full optically-selected population also shown, for which only the \textsc{LePhare} values are available. We note that plotting the \textsc{AGNfitter} properties for our sample instead makes negligible difference to this plot, and would not change any of the conclusions drawn from this figure.  

Recent work by \citet{2022MNRAS.513.3742K} has uncovered a population of LERGs hosted by star-forming (rather than quiescent) galaxies at higher redshifts ($z>1$), in contrast to previous work at lower redshifts which have found LERGs are usually hosted by quiescent galaxies. This is consistent with the result discussed here that HERGs, LERGs and probable-LERGs have very similar star-formation rates. They also find that as the radio luminosity limit is lowered, their sample contains more LERGs with lower stellar masses (at all redshifts), compared to higher luminosities where LERGs tend to be hosted by only the most massive galaxies. This is again consistent with the work presented in this paper, where we find the host galaxy masses of HERGs and LERGs to be very similar at $z>0.5$.

This is in good agreement with what is predicted by the \textsc{Simba} cosmological hydrodynamical simulation \citep{2019MNRAS.486.2827D}. \citet{2021MNRAS.503.3492T} investigated the properties of radio-loud AGN in \textsc{Simba}, and found that HERG and LERG hosts became increasingly similar as they probed lower radio powers, which is supported by our results. However, although \citet{2021MNRAS.503.3492T} find HERG and LERG hosts in \textsc{Simba} have very similar stellar masses, they do find a difference when looking at the most massive galaxies; all galaxies with stellar masses $ > 2 \times 10^{11} M_\odot$ host LERGs. This effect is not evident in our MIGHTEE sample. Note however that the \citeauthor{2021MNRAS.503.3492T} study is at $z=0$, while the sample selected from MIGHTEE covers a range of redshifts out to $z=5$, which may well account for the differences seen. The properties of MIGHTEE radio galaxies will be compared to an equivalent sample selected from \textsc{Simba} in a future paper (Thomas et al., in prep).

There is tentative evidence that the HERGs in our sample are more likely to be detected by the VLBA observations than the LERGs, suggesting that they may have more compact radio emission: $25 \pm 3$~per cent of HERGs in the sample are detected in the VLBA catalogue, compared to $17 \pm 3$ per cent of LERGs and $15 \pm 1$~per cent of probable LERGs. This supports the findings of \citet{2016MNRAS.462.2122W} that HERGs are more dominated by emission from their cores than LERGs.

\subsection{Accretion Rates}\label{section:accretion-rates}

\citet{2018MNRAS.480..358W} studied $\sim1000$ radio galaxies selected from a VLA survey of Stripe 82 \citep{2016MNRAS.460.4433H}, and found significantly more overlap in the accretion rates of the HERGs and LERGs than has been found in previous studies \citep{2012MNRAS.421.1569B,2014MNRAS.440..269M} at higher radio luminosities. Here we extend this study to even lower radio luminosities using the MIGHTEE Early Science data. For this section we limit the study to sources with a good match only (flag $=$ 100 or 120, see Prescott et al. (in prep) for further details), and sources with a good fit in \textsc{AGNfitter} (characterised by log likelihood > -20) leaving a sample of 819 radio-loud AGN. We looked at the SEDs of all sources excluded for having a poor fit to ensure they do not represent e.g.\ a particular class of source. There are a variety of reasons for poor fits, such as low signal to noise or scatter in the photometry, but no pattern is evident.

Although we do not have line measurements from optical spectra for the majority of the sources in the MIGHTEE sample, we are able to estimate the AGN bolometric luminosity ($L_{\rm bol}$) from \textsc{AGNfitter} by integrating the best-fit rest-frame direct AGN emission (the BB component) from 0.1 $\muup$m to 1 keV, and adding in the torus component (TO component). 

The mechanical luminosity of the radio jet ($L_{\rm mech}$) was estimated from the 1.4~GHz luminosity using the relationship from \citet{2010ApJ...720.1066C}, $L_{\rm mech} = 7.3 \times 10^{36} (L_{1.4~\rm{GHz} }/ 10 ^{24} \rm{W~Hz^{-1}})^{0.7}$~W, assuming a spectral index of 0.7, and taking into account the variation in effective frequency across the MIGHTEE radio image to convert radio luminosities to rest frame 1.4 GHz. This scaling relation was derived by studying the energies associated with the cavities in hot X-ray gas evacuated by radio jets, and the scatter is found to be around 0.7~dex \citep{2010ApJ...720.1066C}. This scatter may be partly due to variations in the particle content of radio jets. \citet{2018MNRAS.476.1614C} studied the particle composition of radio jets and found no systematic difference between HERGs and LERGs, suggesting that any difference in radio luminosity observed between the two populations translates (on average) to a difference in mechanical powers.
The \citet{2010ApJ...720.1066C} relation used here is consistent with the \citet{1999MNRAS.309.1017W} scaling relation if the ratio between non-radiating particles and relativistic elections is assumed to be several tens of thousands (this is supported by results in the literature, e.g.\ \citealt{2005MNRAS.364.1343D,2008ApJ...686..859B}). We note that the \citeauthor{2010ApJ...720.1066C} relation was derived for powerful radio galaxies so may not be entirely applicable for the lower-powered radio galaxies studied in this work, however we choose to use it to be consistent with previous work (e.g.\ \citealt{2012MNRAS.421.1569B}). 

The black hole masses were estimated from stellar masses using the \citet{2004ApJ...604L..89H} relation: $M_\textrm{BH} \sim 0.0014M_{\ast}$, which has a scatter of $\sim0.3$~dex. The Eddington limit for each source was then calculated as follows: $L_{\textrm{Edd}} = 1.3 \times 10^{31} M_{\textrm{BH}} / M_\odot \textrm{W}$.  The Eddington-scaled accretion rate for each source was calculated as $\lambda = (L_{\rm bol} + L_{\rm mech}) / L_{\rm Edd}$. Although there is significant scatter on the scaling relations used to estimate the accretion rates, as discussed, we do not expect this to introduce a systematic effect. It is still informative to compare the accretion rates estimated in this way for the different samples discussed in this paper.

\begin{figure}
    \centering
    \includegraphics[width=\columnwidth]{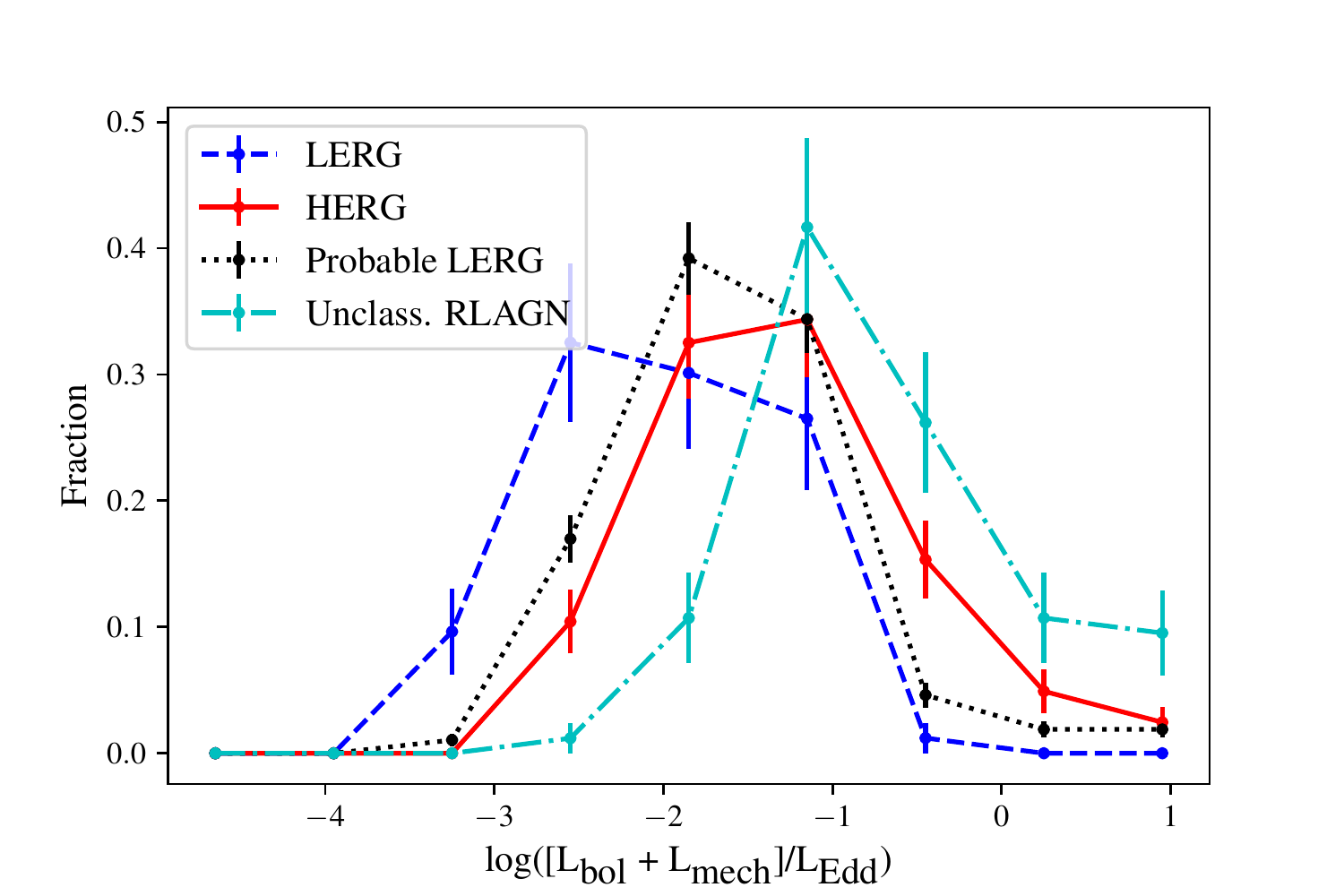}
    \includegraphics[width=\columnwidth]{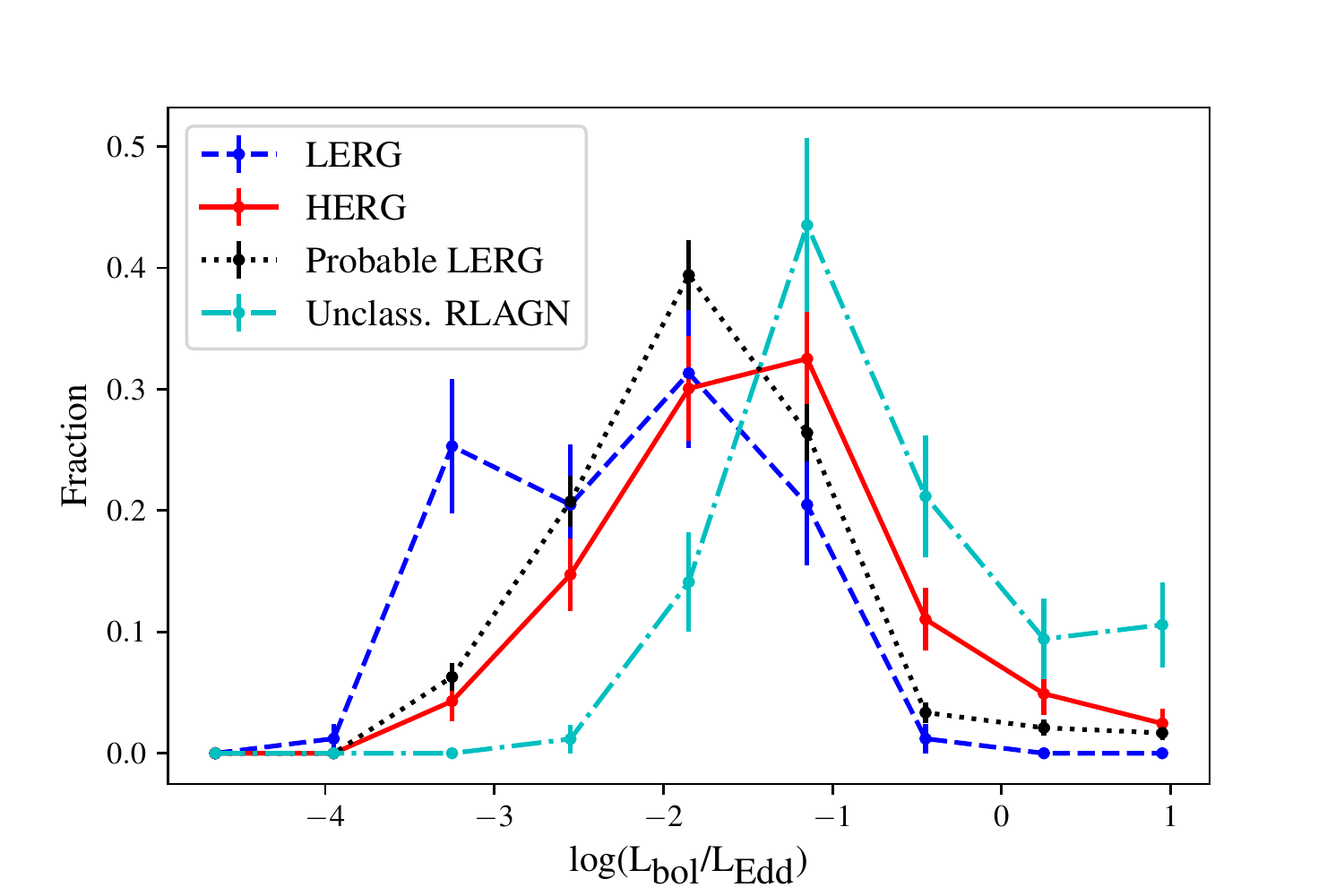}
    \caption{Distribution of Eddington-scaled accretion rates for HERGs (red solid line), LERGs (blue dashed), probable LERGs (black dotted) and unclassified radio-loud AGN (cyan dot-dashed). The top panel shows the combined accretion rates estimated from the radiative and mechanical luminosities, while the bottom panel shows the accretion rates considering only the radiative luminosity.}
    \label{fig:lbol}
\end{figure}

\begin{figure}
    \centering
    \includegraphics[width=\columnwidth]{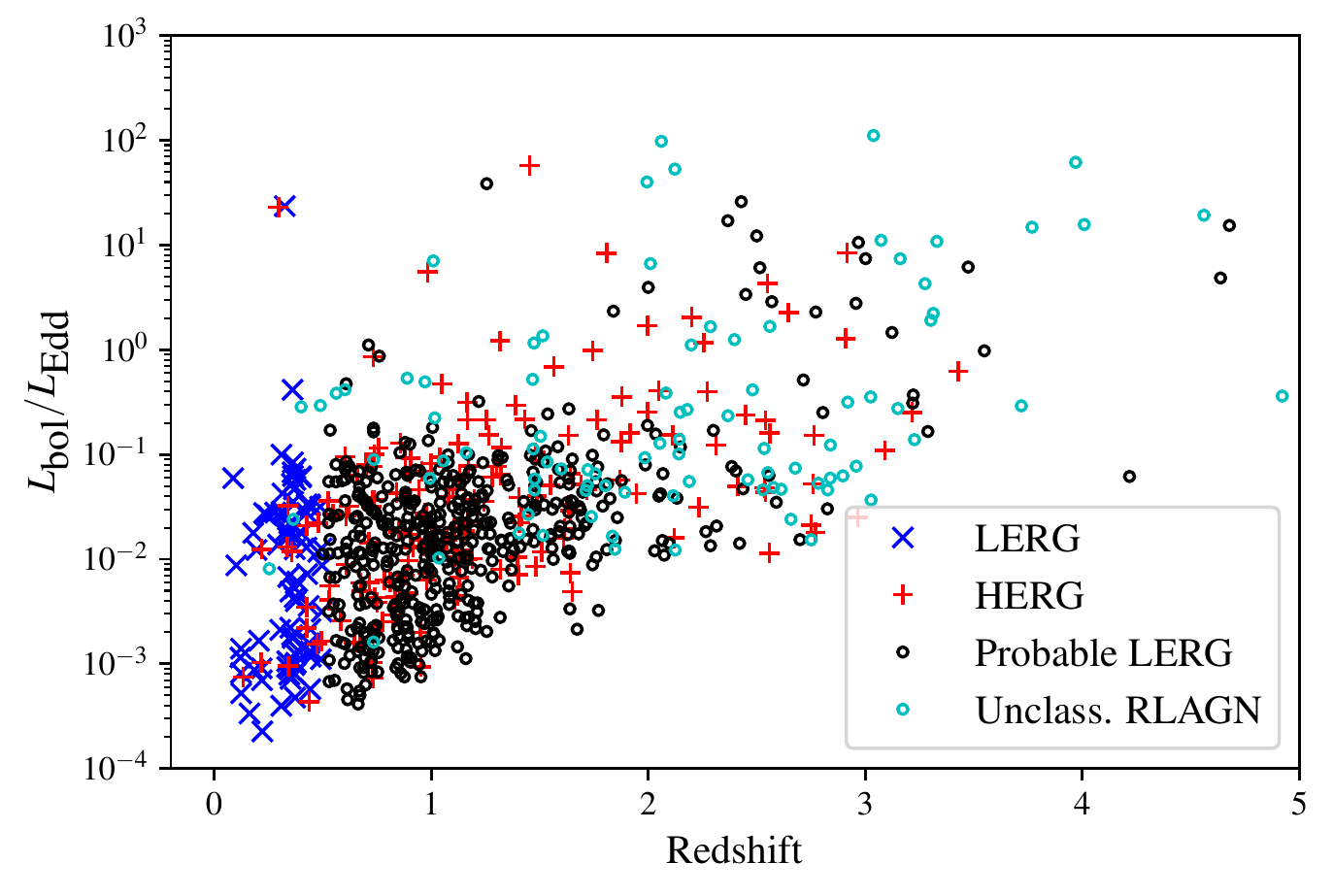}
    \includegraphics[width=\columnwidth]{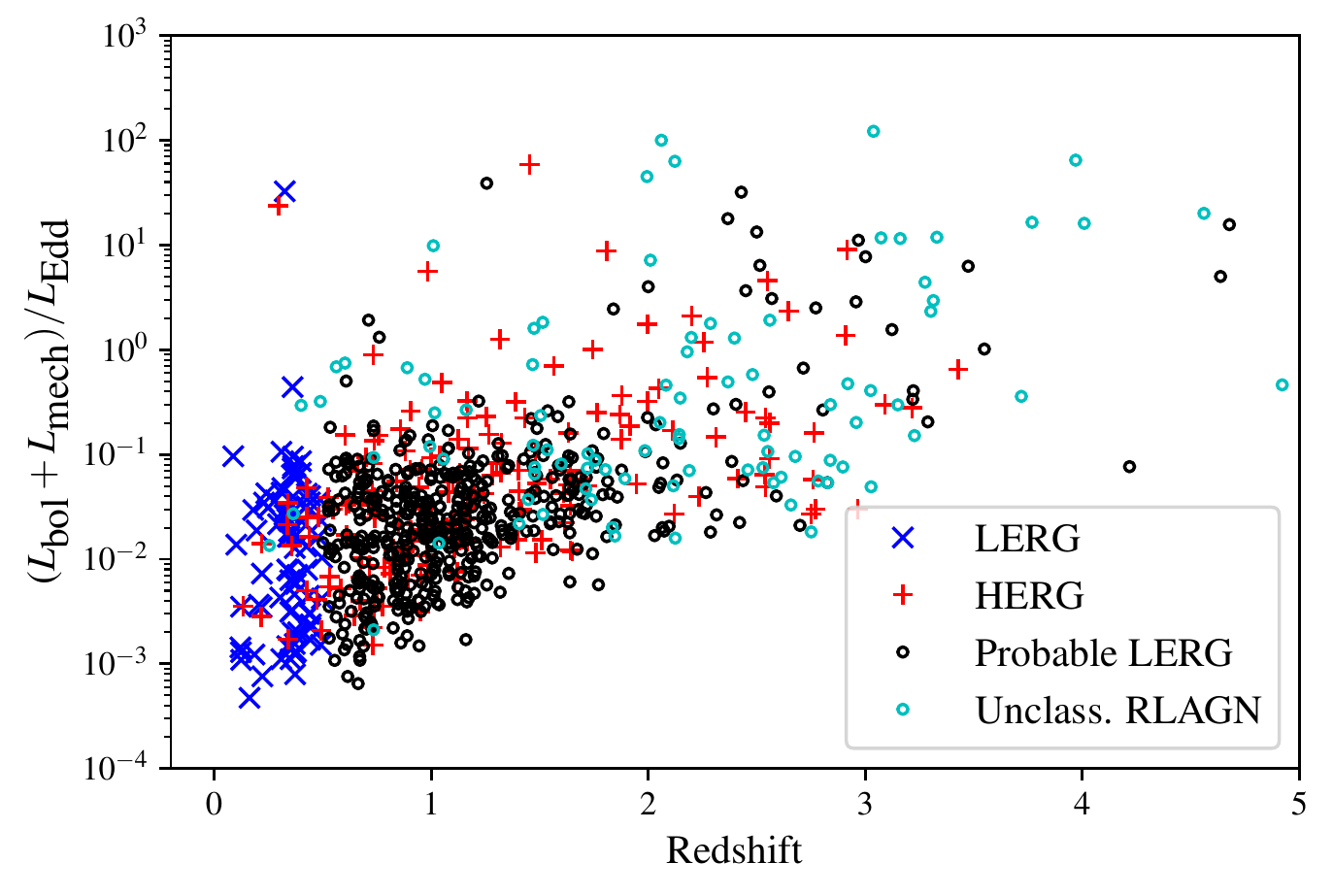}
    \caption{Eddington-scaled accretion rates as a function of redshift for the different source classifications (HERGs = red pluses, LERGs = blue crosses, probable LERGs = black circles and unclassified RLAGN = cyan circles). The top panel shows accretion rates considering only the radiative luminosity, while the bottom panel shows the combined radiative and mechanical luminosities.}
    \label{fig:lbol_z}
\end{figure}

The distribution of Eddington-scaled accretion rates for the HERGs, LERGs, probable LERGs and unclassified radio-loud AGN are shown in Fig.~\ref{fig:lbol}. It is clear that there is a significant overlap in the distribution of accretion rates of the HERGs and the LERGs in this sample. It appears that the unclassified radio-loud AGN have higher accretion rates than the sources classified as HERGs and LERGs, however this is a selection effect. Sources are unclassified when we do not have enough data to robustly classify them as HERGs or LERGs, which is more likely to be the case at higher redshifts, as can be seen from Fig.~\ref{fig:fracHERG}. At these higher redshifts (typically $z \gtrsim 2$), we are only able to detect AGN with high accretion rates, resulting in the unclassified sources we do detect having some of the highest accretion rates. This effect can be seen more clearly in Fig.~\ref{fig:lbol_z}, which shows the Eddington scaled accretion rates as a function of redshift.

Fig.~\ref{fig:lbol_z} shows that although there is significant overlap in the accretion rates of HERGs and LERGs, nearly all of the sources with high Eddington-scaled accretion rates ($\gtrsim 0.1$) are HERGs or unclassified sources (compared to the divide at $\sim 0.01$ found by past work). Previous studies of radio galaxy accretion rates by \citet{2012MNRAS.421.1569B} and \citet{2014MNRAS.440..269M} also found that all of the LERGs have accretion rates < 0.01; the difference with our sample is that as well as the efficiently-accreting HERGs, we are also detecting significant numbers of HERGs with Eddington-scaled accretion rates well below this value (between $10^{-3}$ and $10^{-2}$). This is due to the fact that MIGHTEE allows us to probe lower luminosities and higher redshifts than previous studies, as is illustrated by Fig.~\ref{fig:LzS82}. Having said that, the small number of sources with extremely low accretion rates, below $10^{-3}$ of their Eddington rate, are all LERGs or probable LERGs.

Due to the small volume probed at low redshifts ($z<0.7$), MIGHTEE detects very few of the rarer, powerful AGN which dominate both the \citet{2012MNRAS.421.1569B} and \citet{2018MNRAS.480..358W} samples at these redshifts. Our sample is therefore lacking the efficiently accreting HERGs which constitute the bulk of the HERG population in \citet{2012MNRAS.421.1569B} and \citet{2018MNRAS.480..358W}.

While the scaling relation used to estimate the mechanical luminosities in this work is the same as that used by \citet{2012MNRAS.421.1569B} and \citet{2018MNRAS.480..358W}, those two studies use L$_\textrm{[OIII]}$ to estimate the AGN bolometric luminosity, whereas we use the results from the SED fitting. While we do not expect this to affect the results presented here, we note that the bolometric luminosities may not be directly comparable with the work by \citet{2012MNRAS.421.1569B} and \citet{2018MNRAS.480..358W}.

It is possible that misclassified sources could be somewhat responsible for part of the overlap in accretion rates observed in our work. While we are confident that sources classified as HERGs are indeed HERGs, it is possible that some of the sources in our LERG and `probable LERG' samples are actually HERGs, where signs of efficient accretion have been missed. This is particularly true for the `probable LERG' class, where the limitations of the X-ray data available mean we are unable to rule out accretion-related X-ray emission. This means that some of the LERGs and probable-LERGs with high Eddington-scaled accretion rates could be misclassified HERGs, but this does not explain the slowly-accreting HERGs observed.

\citet{2021MNRAS.503.3492T} investigated the accretion rate of HERGs and LERGs in the \textsc{Simba} simulation, which probes a similar volume to MIGHTEE and is therefore comparable as it also does not contain the high luminosity sources found in larger-area surveys. They found that the simulation predicts no obvious difference in the Eddington fractions of the two classes at $z=0$. The results from MIGHTEE are in agreement with this prediction, although the small sample size at $z<0.1$ means that we cannot rule out other scenarios.

We have not included error bars on Figs.~\ref{fig:lbol_z} and \ref{fig:acc_lam} to avoid cluttering the figures, but note that the approximate size of the uncertainties can be seen by the error bars in Fig.~\ref{fig:rates-discussion}. In addition to the plotted uncertainties, there is significant scatter on the scaling relations used to estimate the mechanical luminosity (0.7~dex) and black hole mass (0.3~dex) in this work. There is no evidence that this scatter will introduce a systematic effect, so we do not expect this to affect the overall trends for the populations discussed here. We also note that the mechanical luminosities and black hole masses were estimated using the same scaling relations in \citet{2018MNRAS.480..358W}, where a difference in the accretion rates of the HERG and LERG samples was observed, suggesting that the scatter in the scaling relations is not responsible for the large overlap observed in this work.

\subsection{Host galaxy properties as a function of accretion rate}\label{section:properties-accrate}

\citet{2018MNRAS.480..358W} found that the radio-loud AGN host galaxy properties varied continuously with accretion rate, with the most slowly accreting sources being hosted by the most massive galaxies with the oldest stellar population, while the sources accreting matter more rapidly have lower-mass hosts and younger stellar populations. This is consistent with the idea that the accretion rate of a galaxy is linked to the supply of cold gas, with the sources which have a readily available gas supply having both high accretion rates and young stellar populations.

\begin{figure}
    \centering
    \includegraphics[width=\columnwidth]{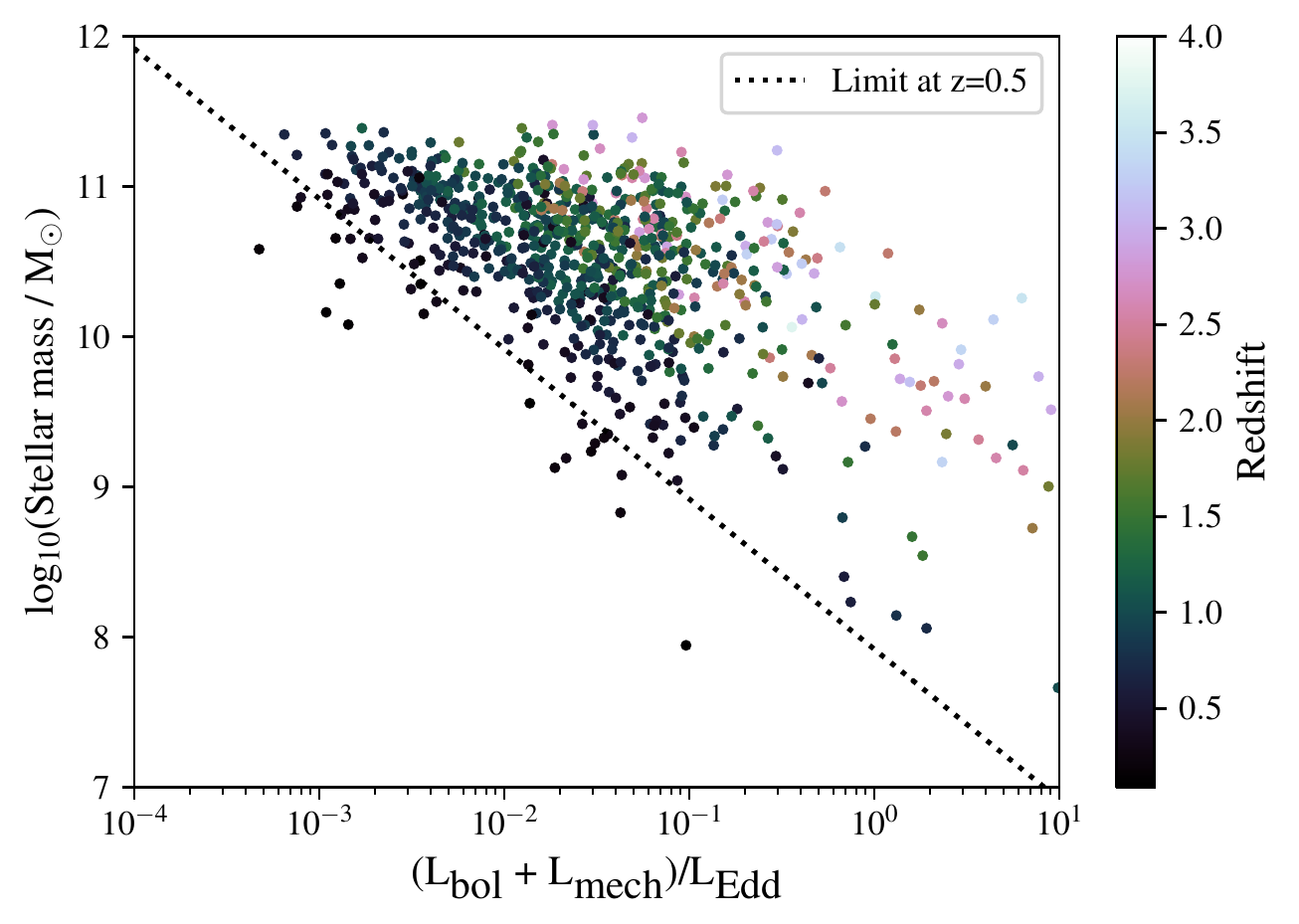}
    \includegraphics[width=\columnwidth]{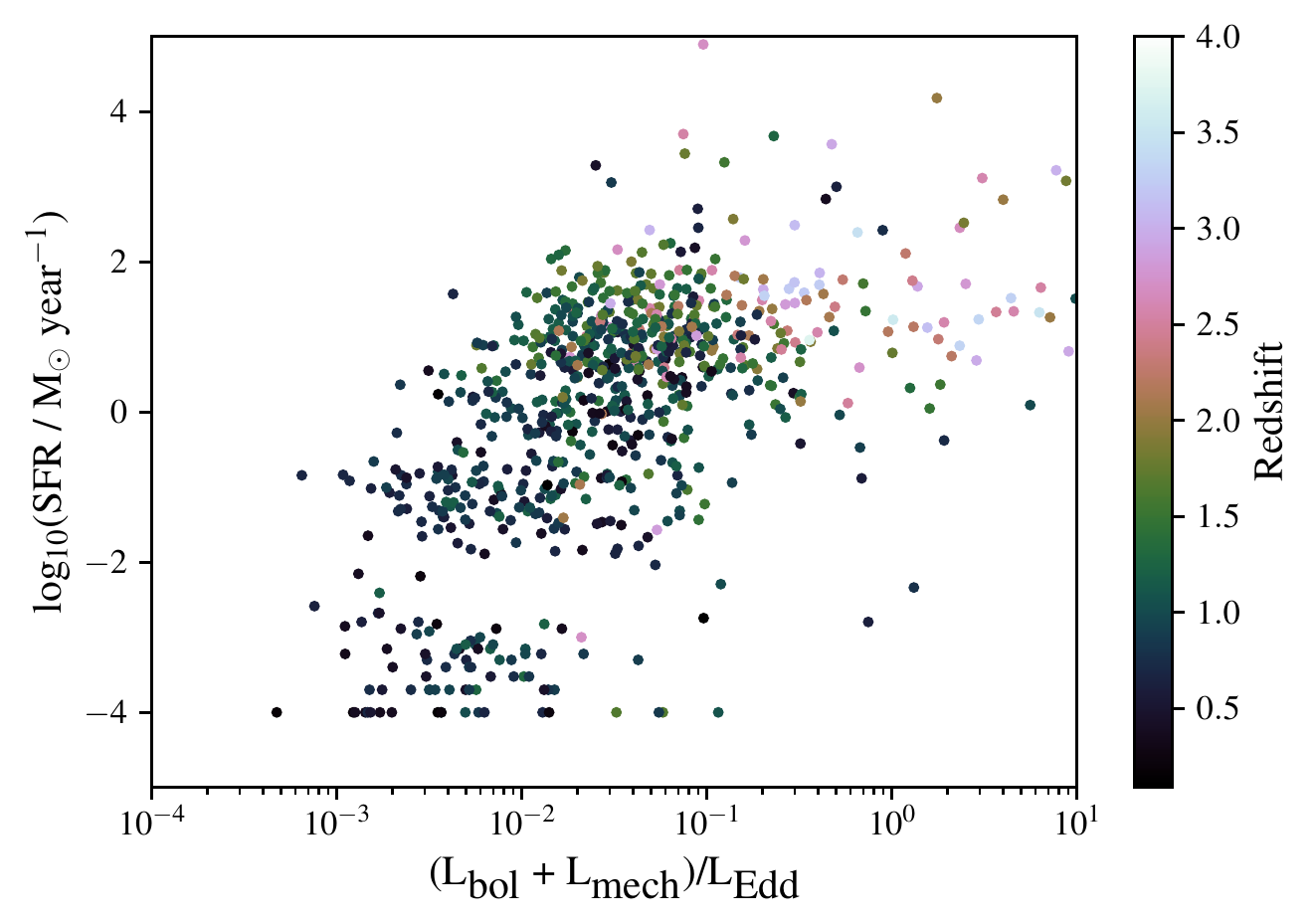}
    \caption{Host galaxy properties of radio-loud AGN as a function of Eddington-scaled accretion rate. The top panel shows stellar mass and the bottom panels shows star-formation rate. The black dotted line in the top panel shows the selection limit at $z=0.5$; at this redshift we can only detect sources in the region above this line. The colour bar shows the redshift of each source.}
    \label{fig:acc_lam}
\end{figure}

Fig.~\ref{fig:acc_lam} shows the host galaxy properties of the MIGHTEE radio-loud AGN as a function of Eddington-scaled accretion rate. Sources are coloured according to their redshift. This figure highlights a similar trend to that found by \citet{2018MNRAS.480..358W}; the sources accreting more efficiently (with higher Eddington-scaled accretion rates) have higher star-formation rates and lower stellar masses. Selection effects mean that we are only able to detect the most slowly accreting sources (with Eddington-scaled accretion rate fractions ($< 10^{-2}$) in the most massive galaxies (with stellar masses $\gtrsim 10^{10} \textrm{M}_\odot$ at $z\sim0.5$). However, source with higher accretion rates, which we are able to detect across the full range of stellar masses, tend to be found in host galaxies with lower stellar masses. For example, we find that none of  galaxies that lie beyond the knee in the stellar mass function at $\textrm{log}_{10}(\textrm{M} / \textrm{M}_\odot) \sim 10.8$ have Eddington-scaled accretion rates in excess of 1.

We have seen that in the MIGHTEE sample HERGs and LERGs are hosted by very similar galaxies (Section~\ref{section:hosts}) and accretion rate is not a good predictor of whether a source is a HERG or LERG (Section~\ref{section:accretion-rates}). It seems that the host galaxy properties are more linked to the Eddington-scaled accretion rate than HERG/LERG classification, raising the question as to whether or not the HERG/LERG class is meaningful as we extend out parameter space to the lower radio luminosity population.

\subsection{The relationship between AGN power and black hole mass}\label{section:power-mass}

\begin{sidewaysfigure*}
\centerline{\includegraphics[width=6cm]{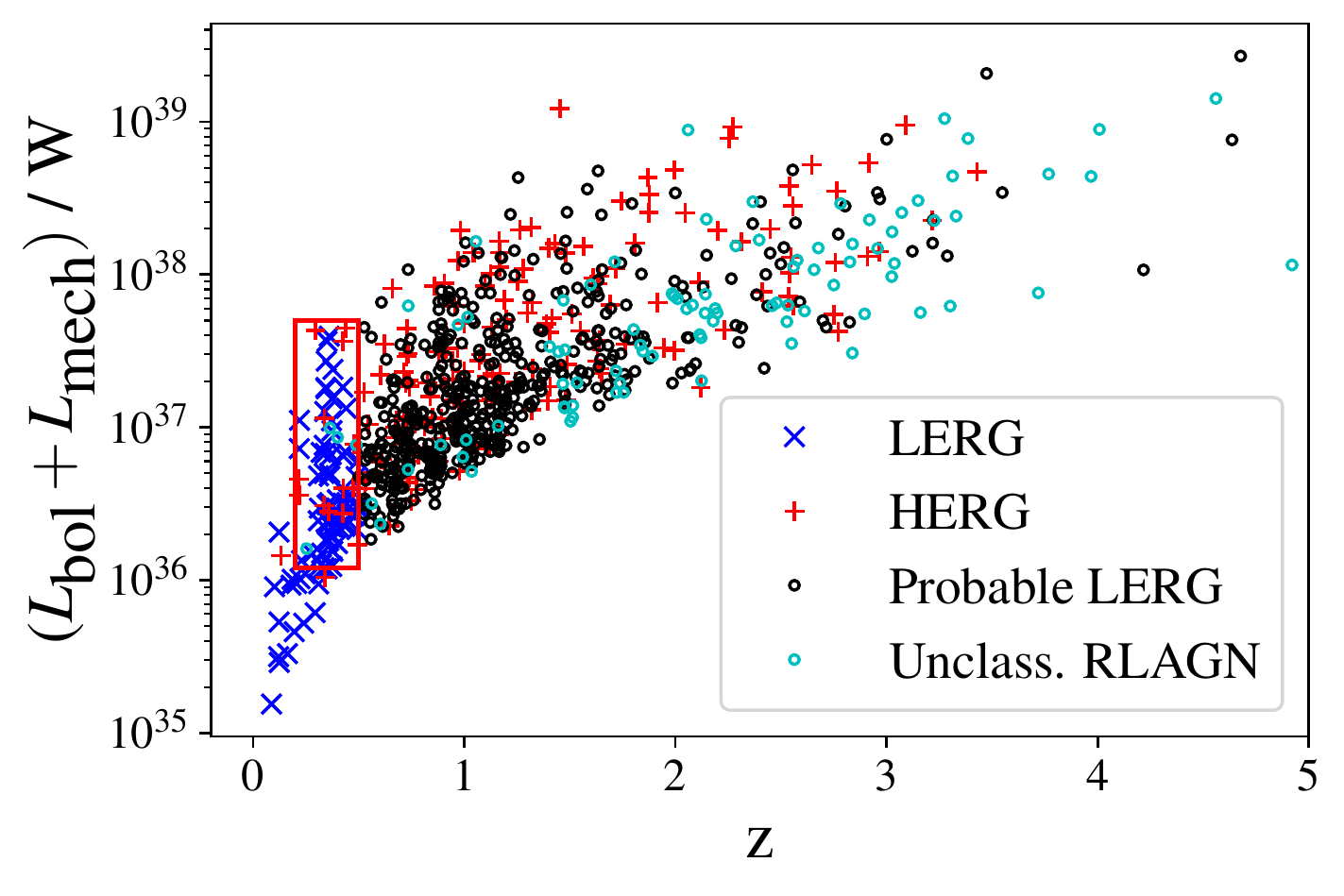}
            \includegraphics[width=6cm]{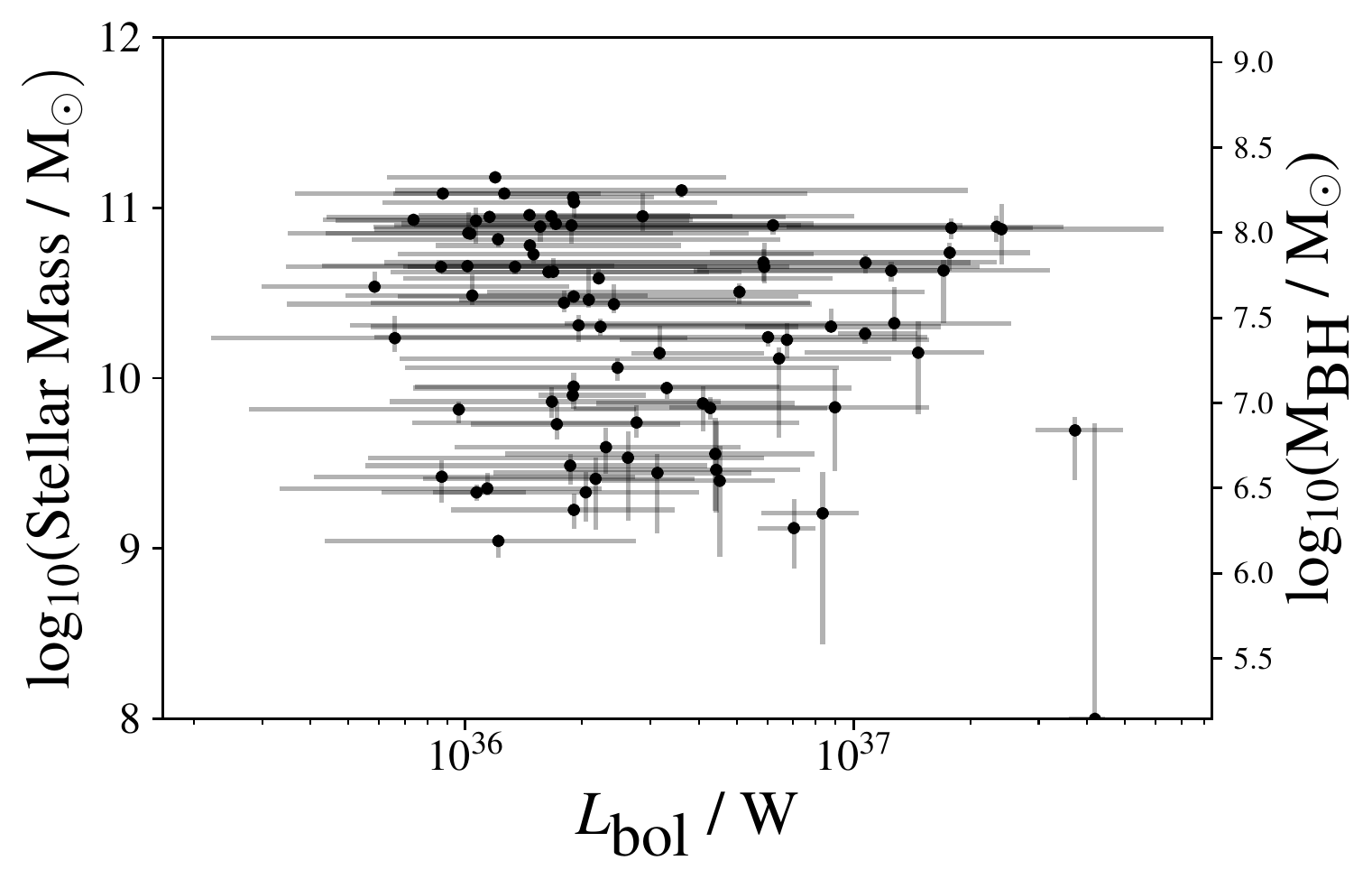}
            \includegraphics[width=6cm]{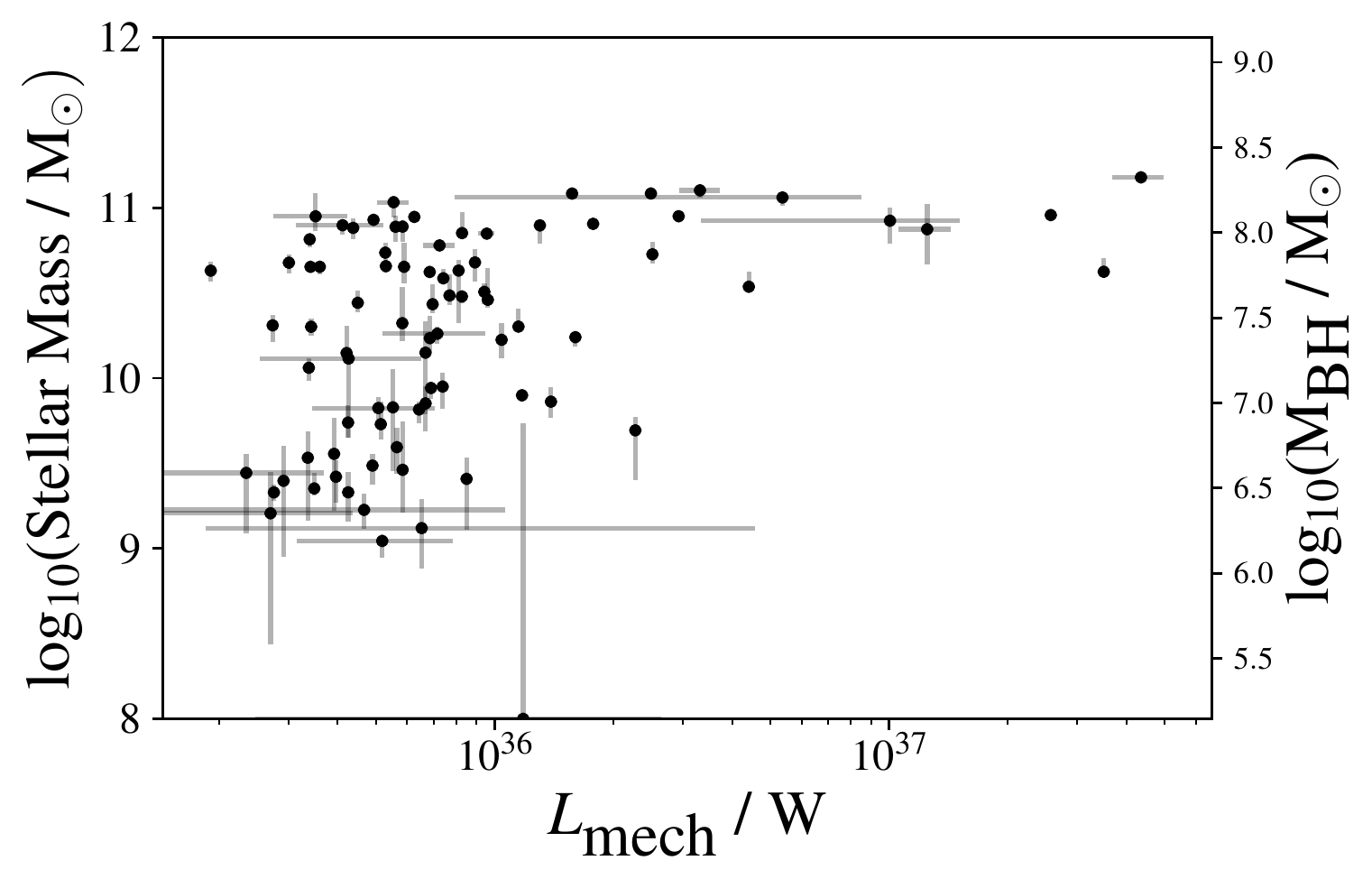}
            \includegraphics[width=6cm]{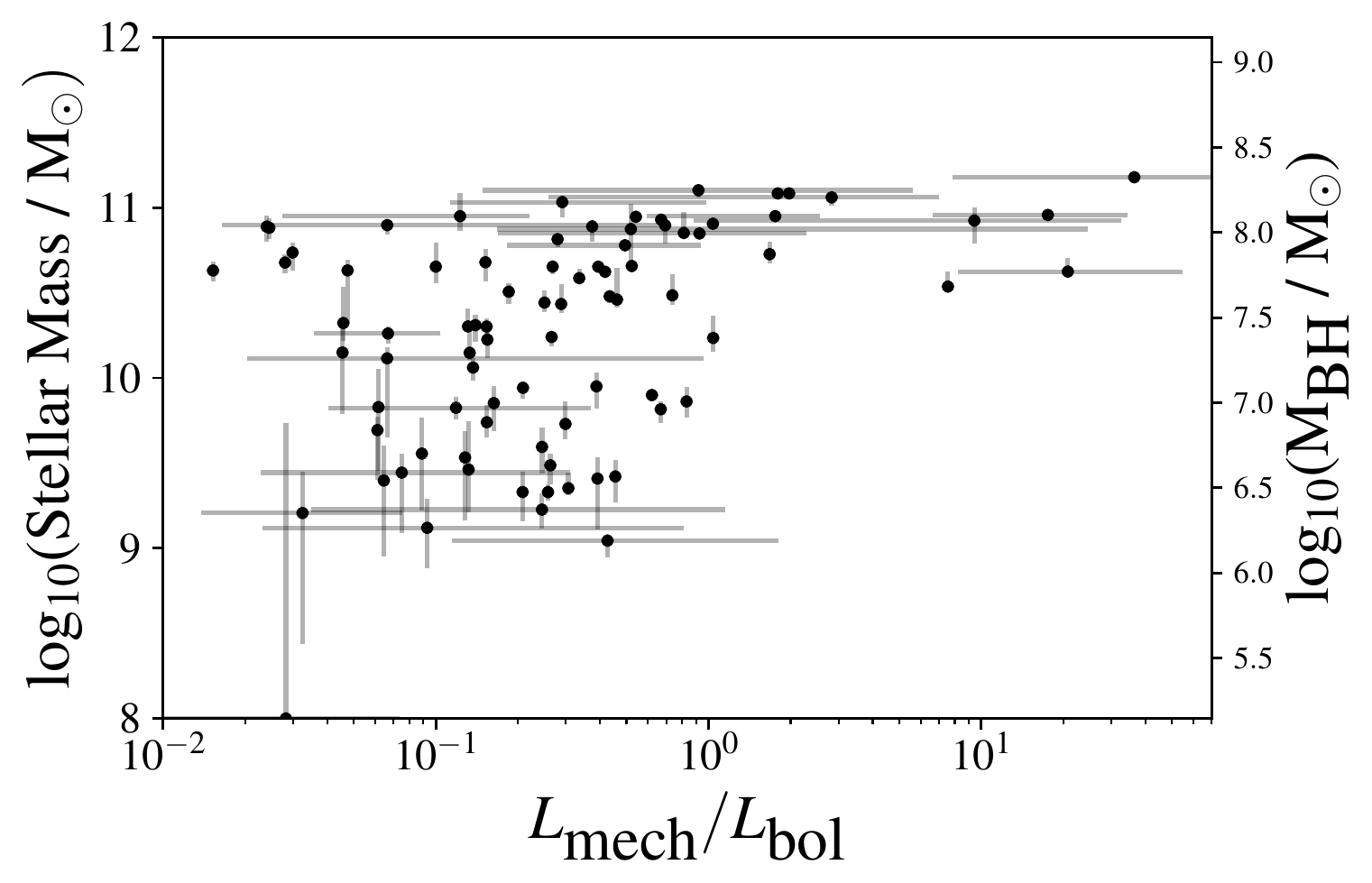}}
\centerline{\includegraphics[width=6cm]{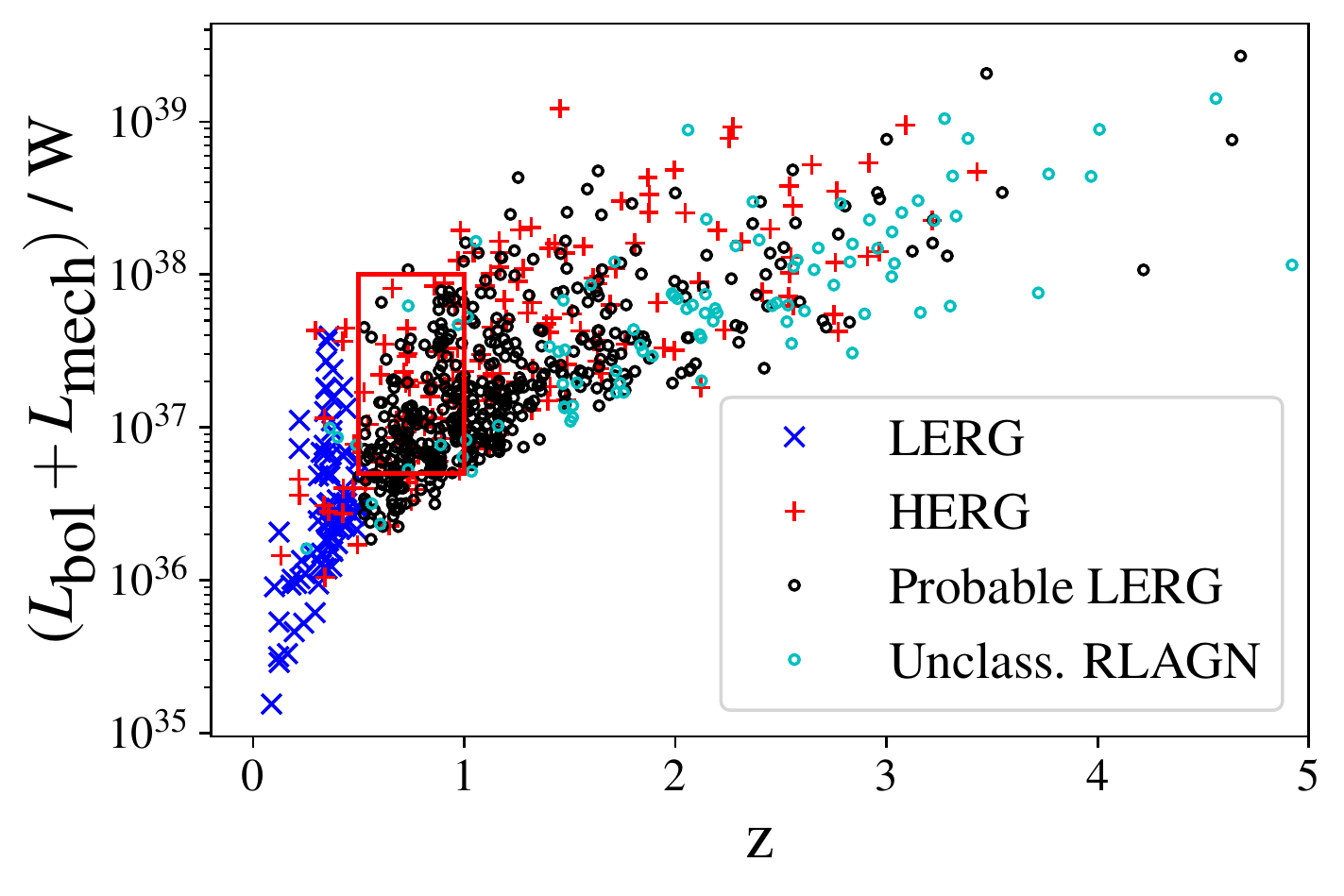}
            \includegraphics[width=6cm]{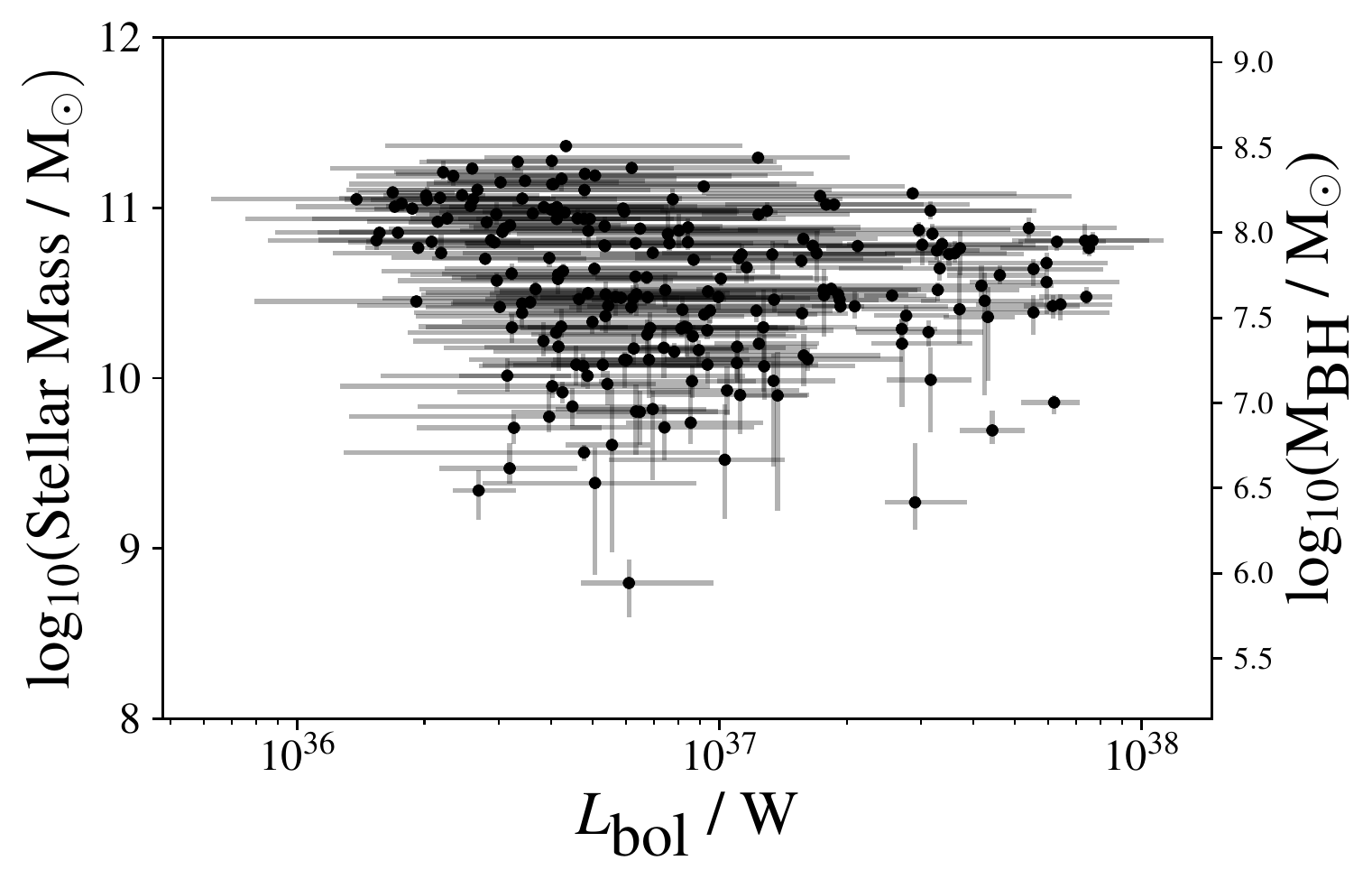}
            \includegraphics[width=6cm]{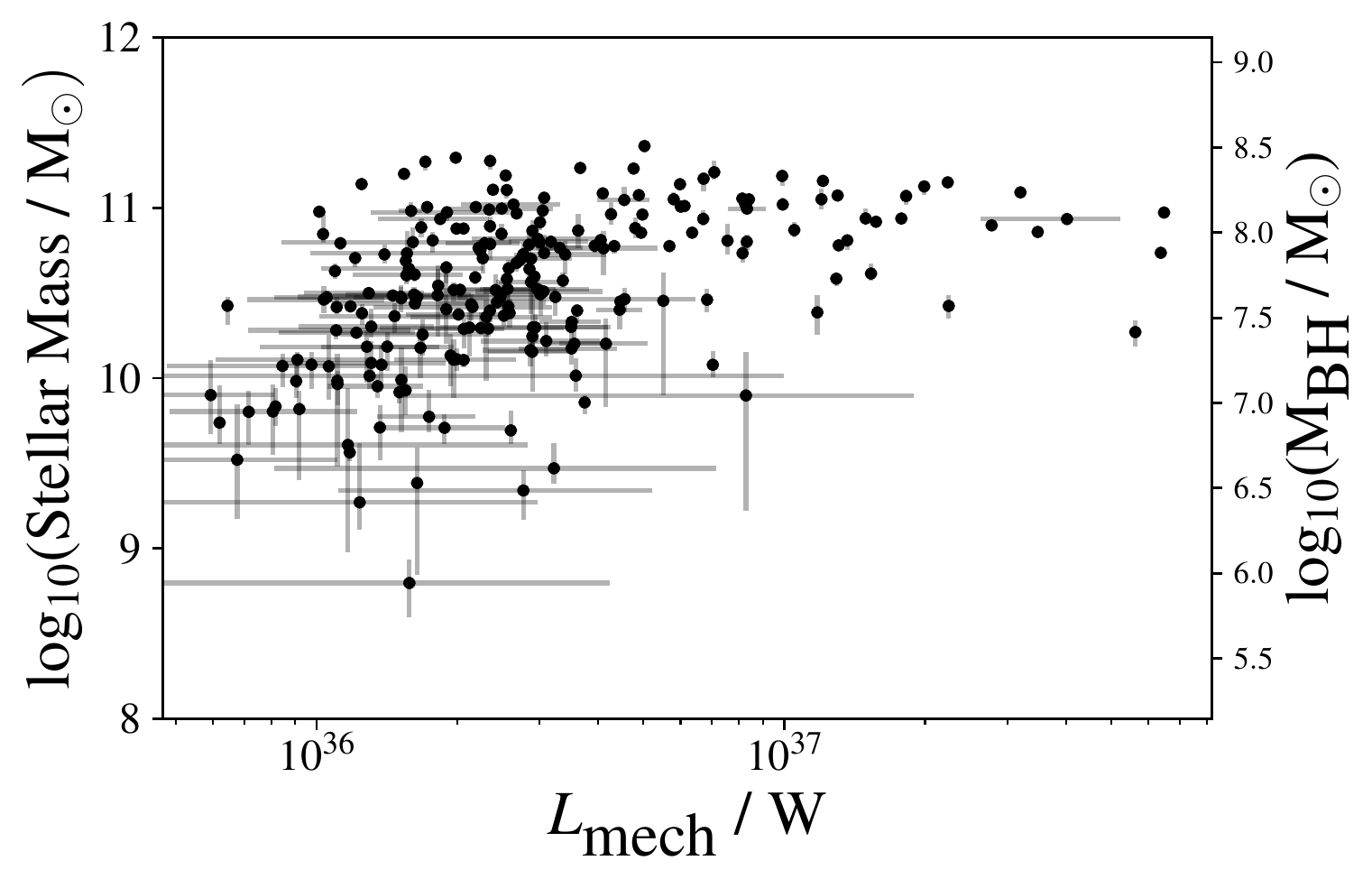}
            \includegraphics[width=6cm]{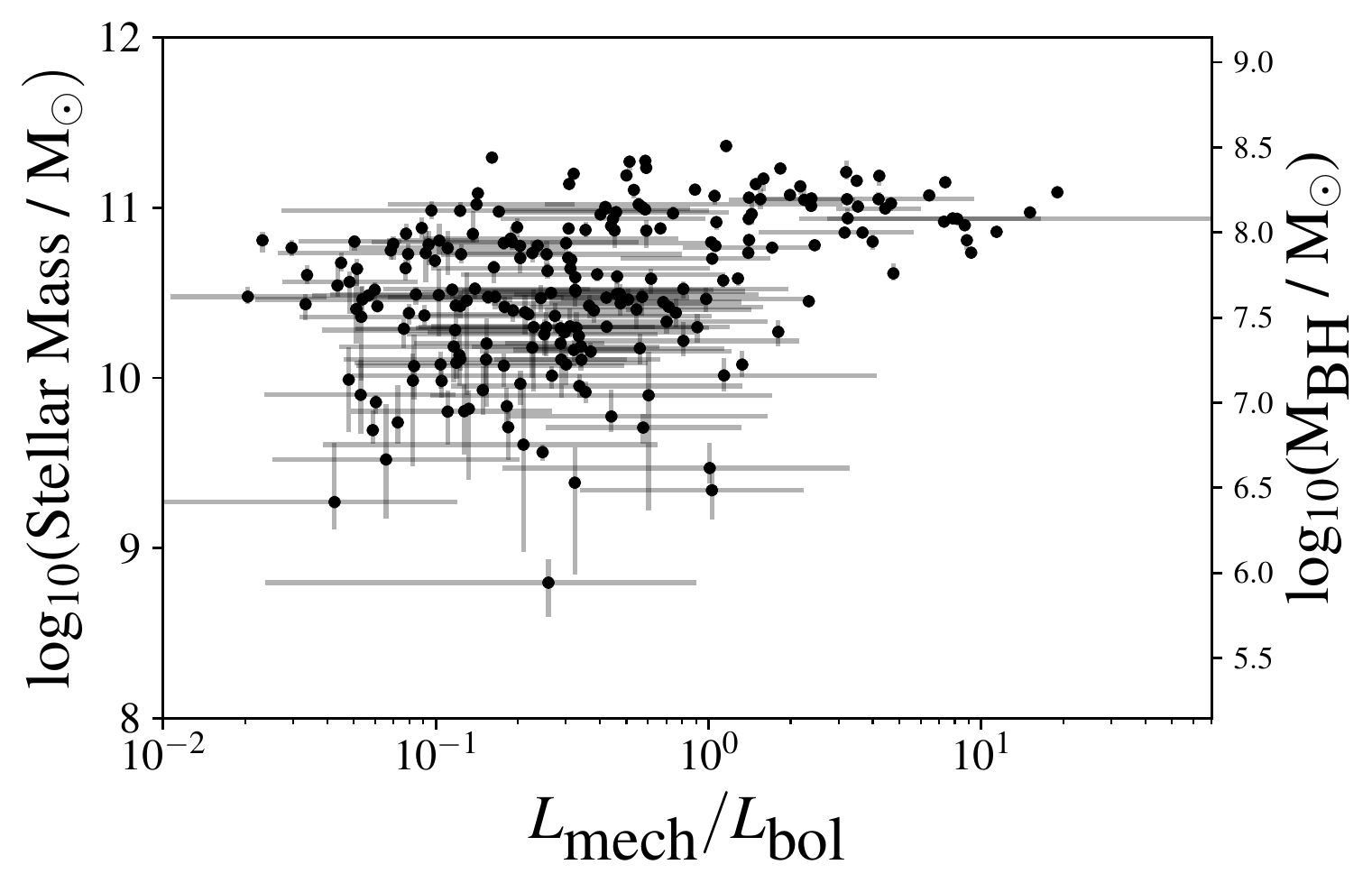}}
\centerline{\includegraphics[width=6cm]{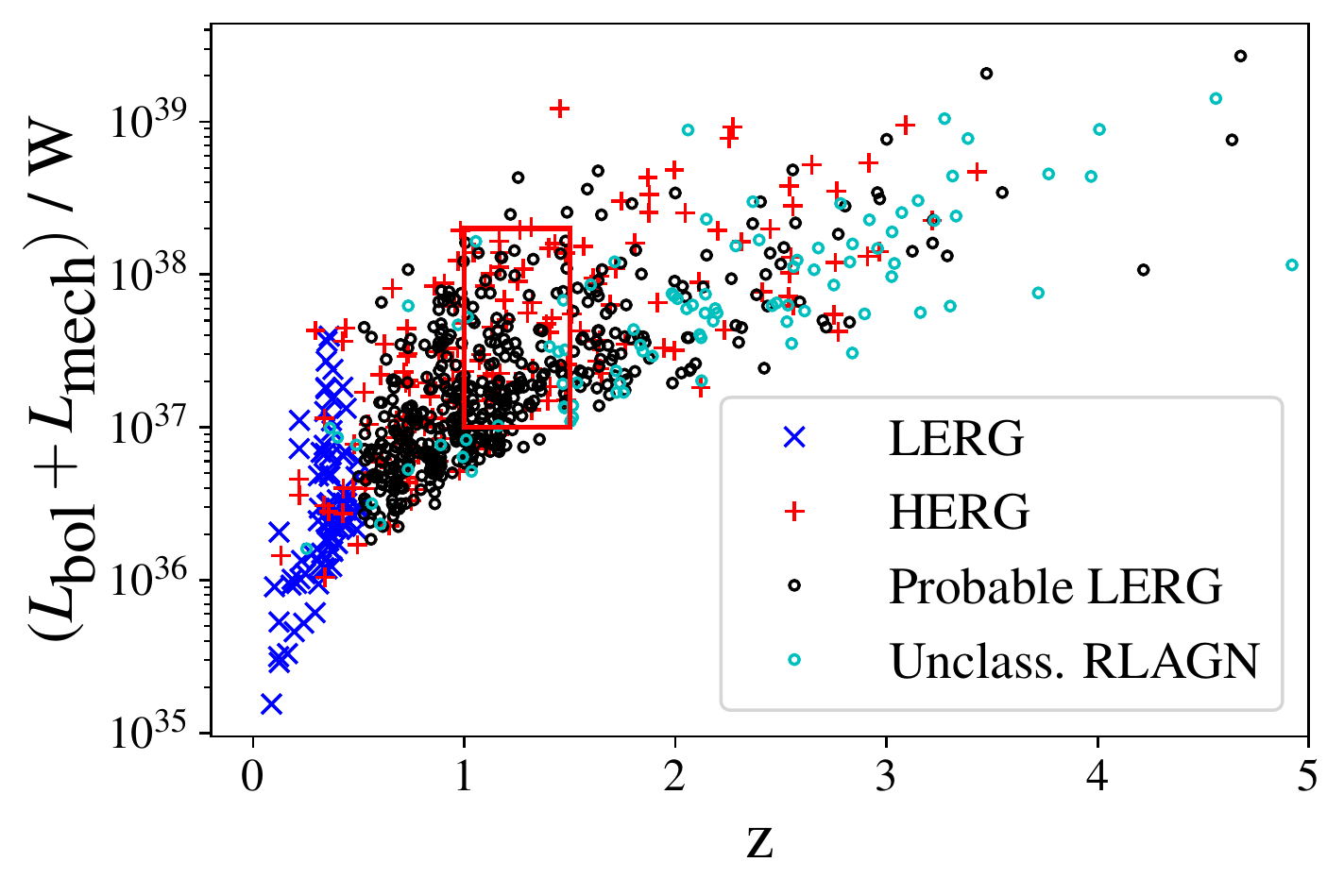}
            \includegraphics[width=6cm]{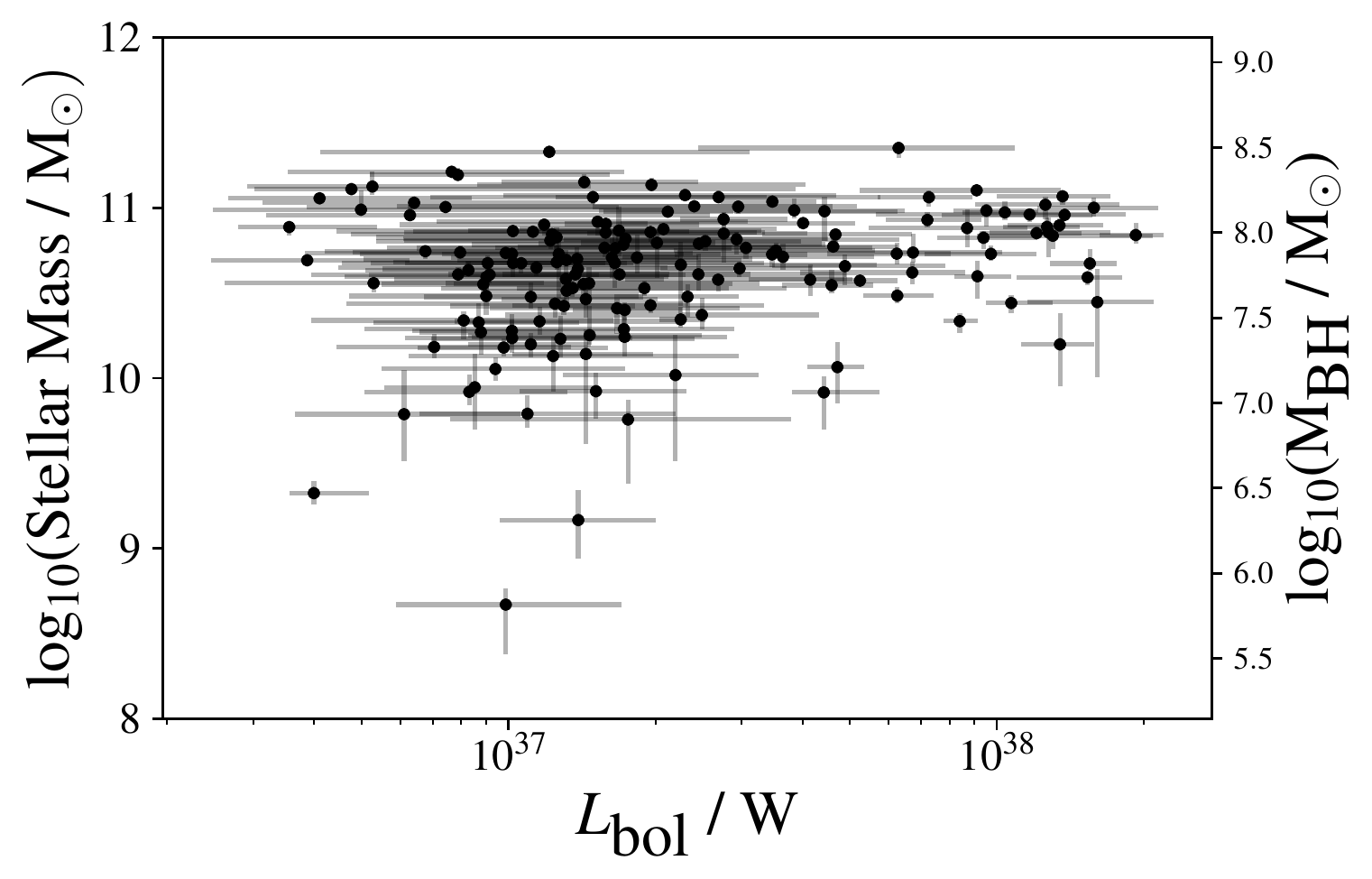}
            \includegraphics[width=6cm]{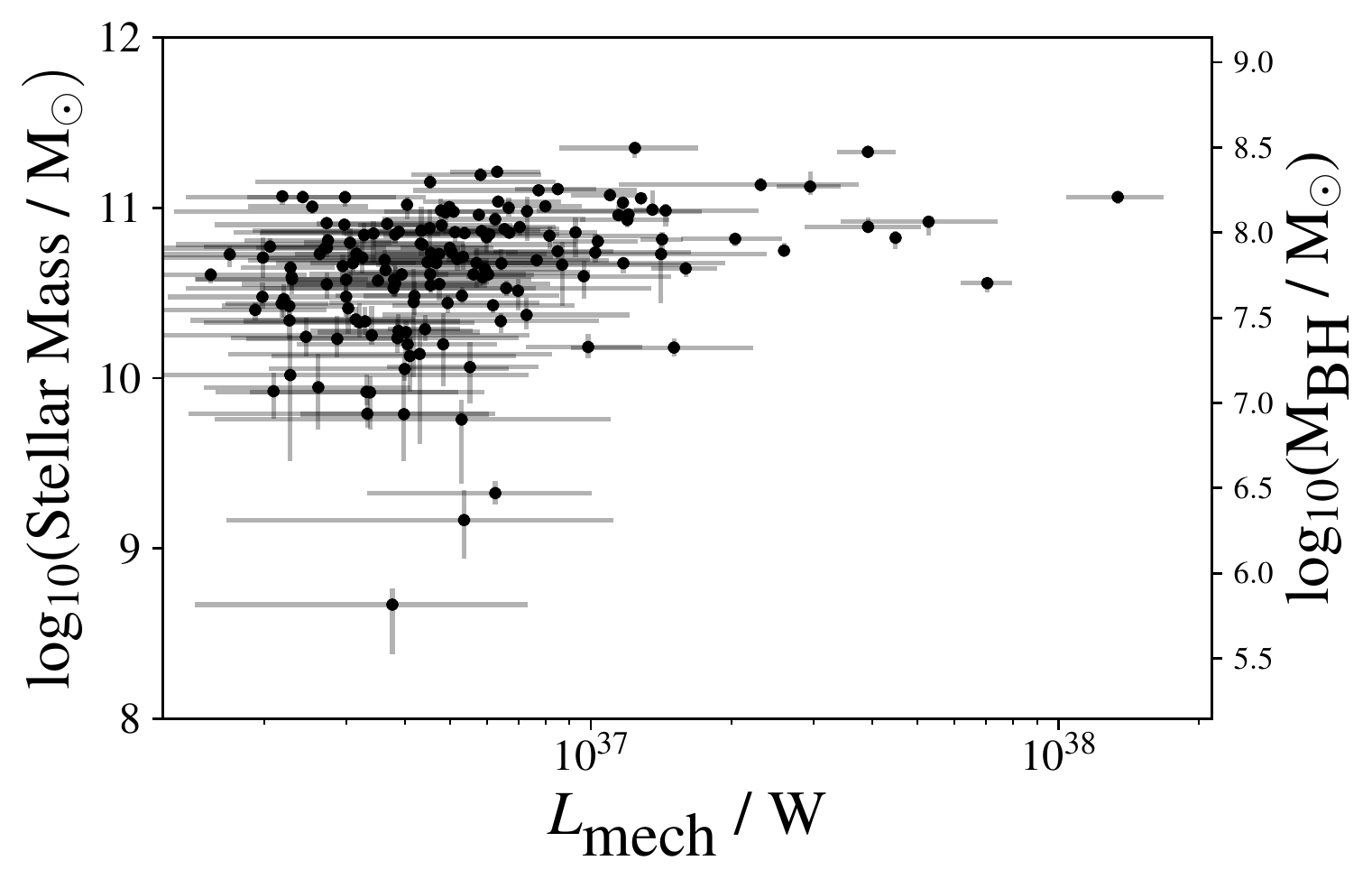}
            \includegraphics[width=6cm]{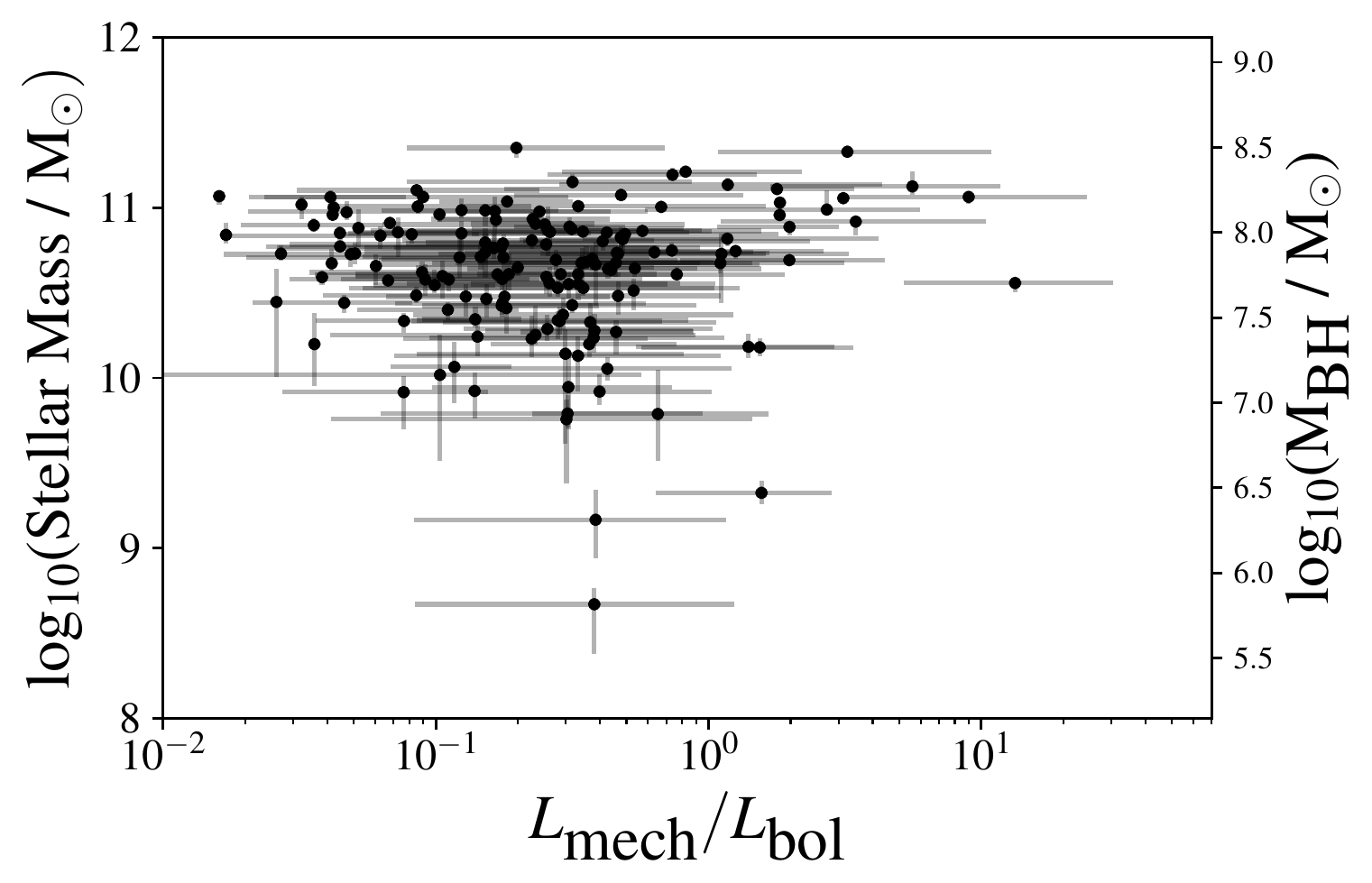}}
\centerline{\includegraphics[width=6cm]{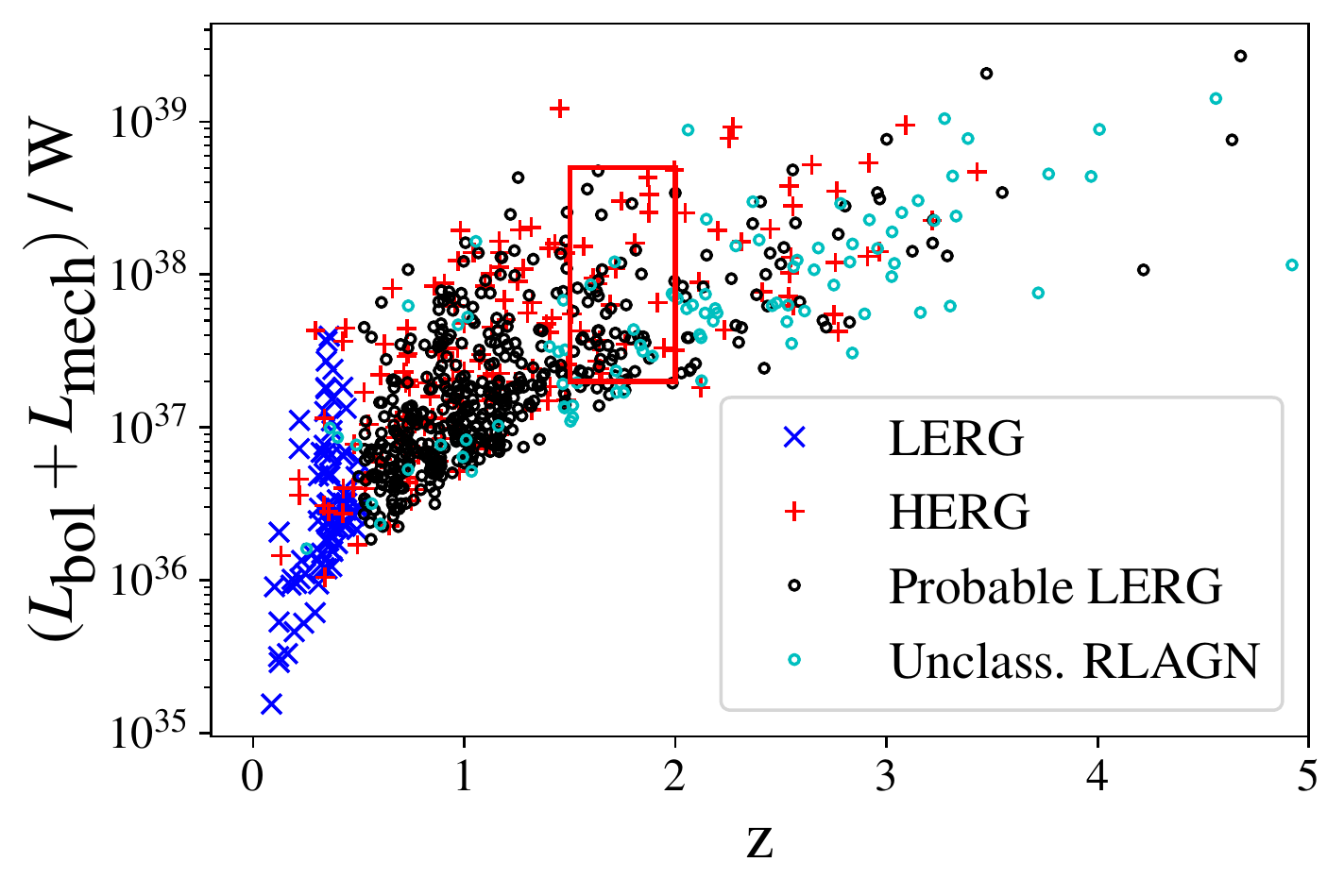}
            \includegraphics[width=6cm]{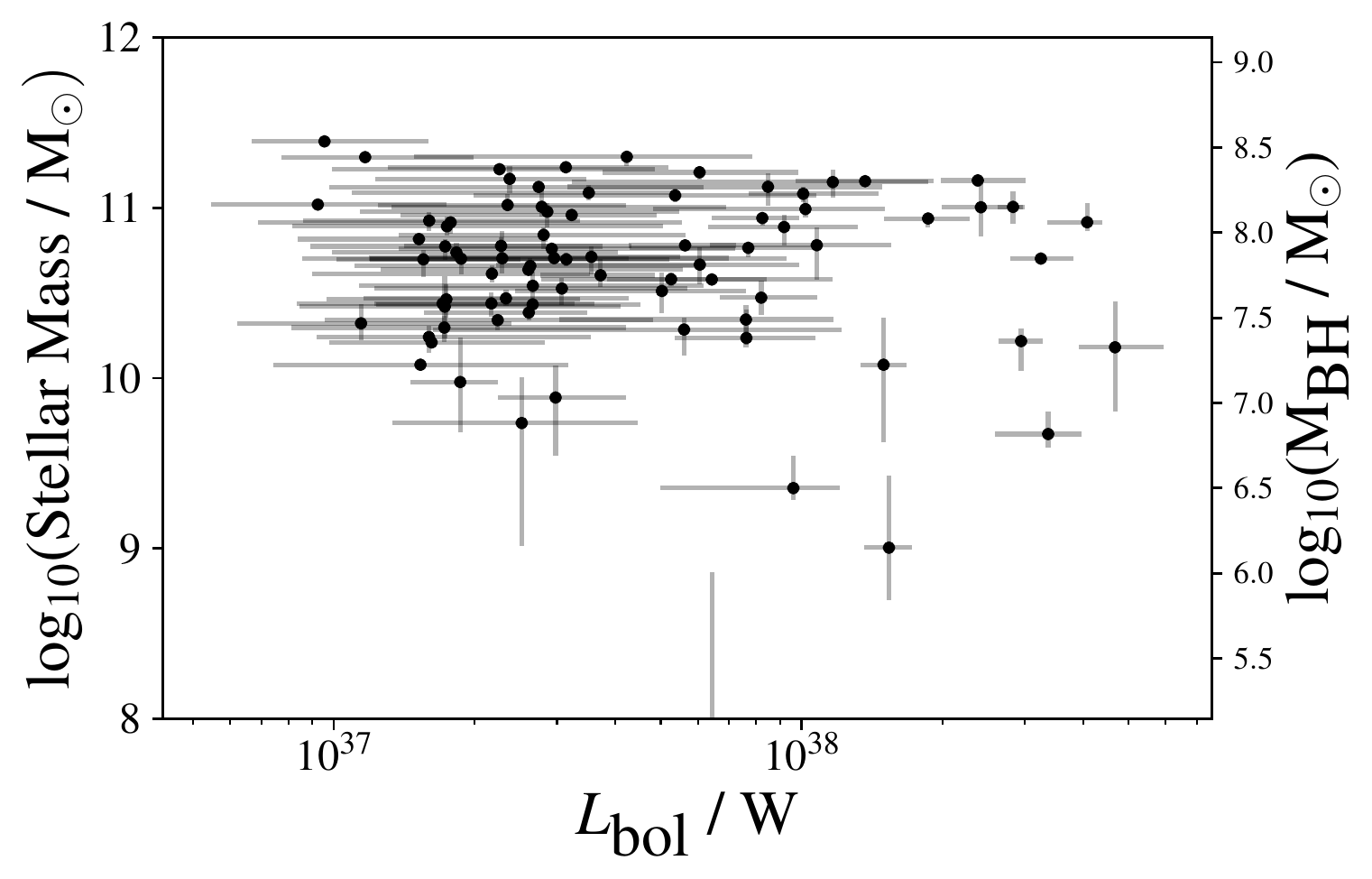}
            \includegraphics[width=6cm]{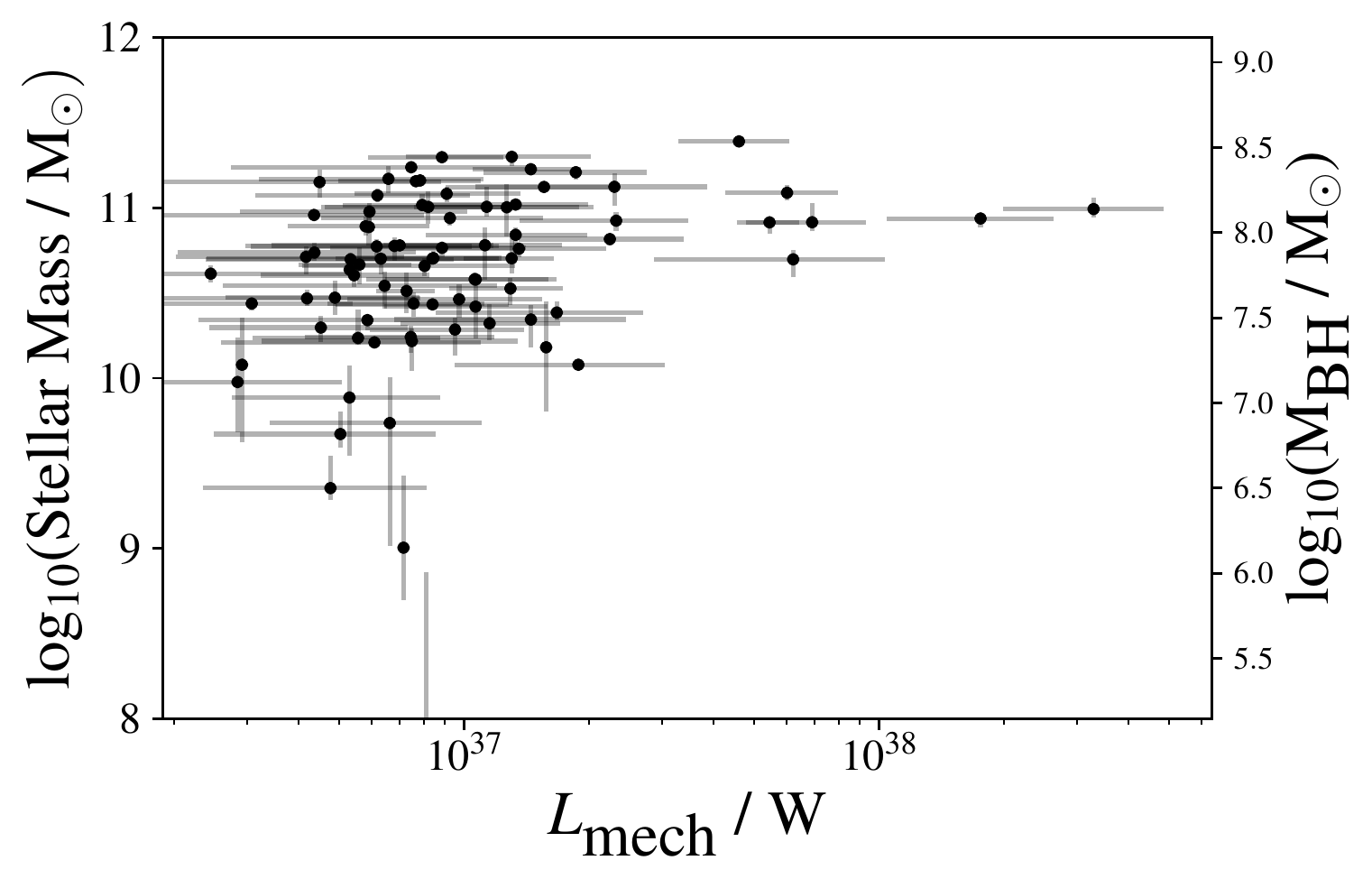}
            \includegraphics[width=6cm]{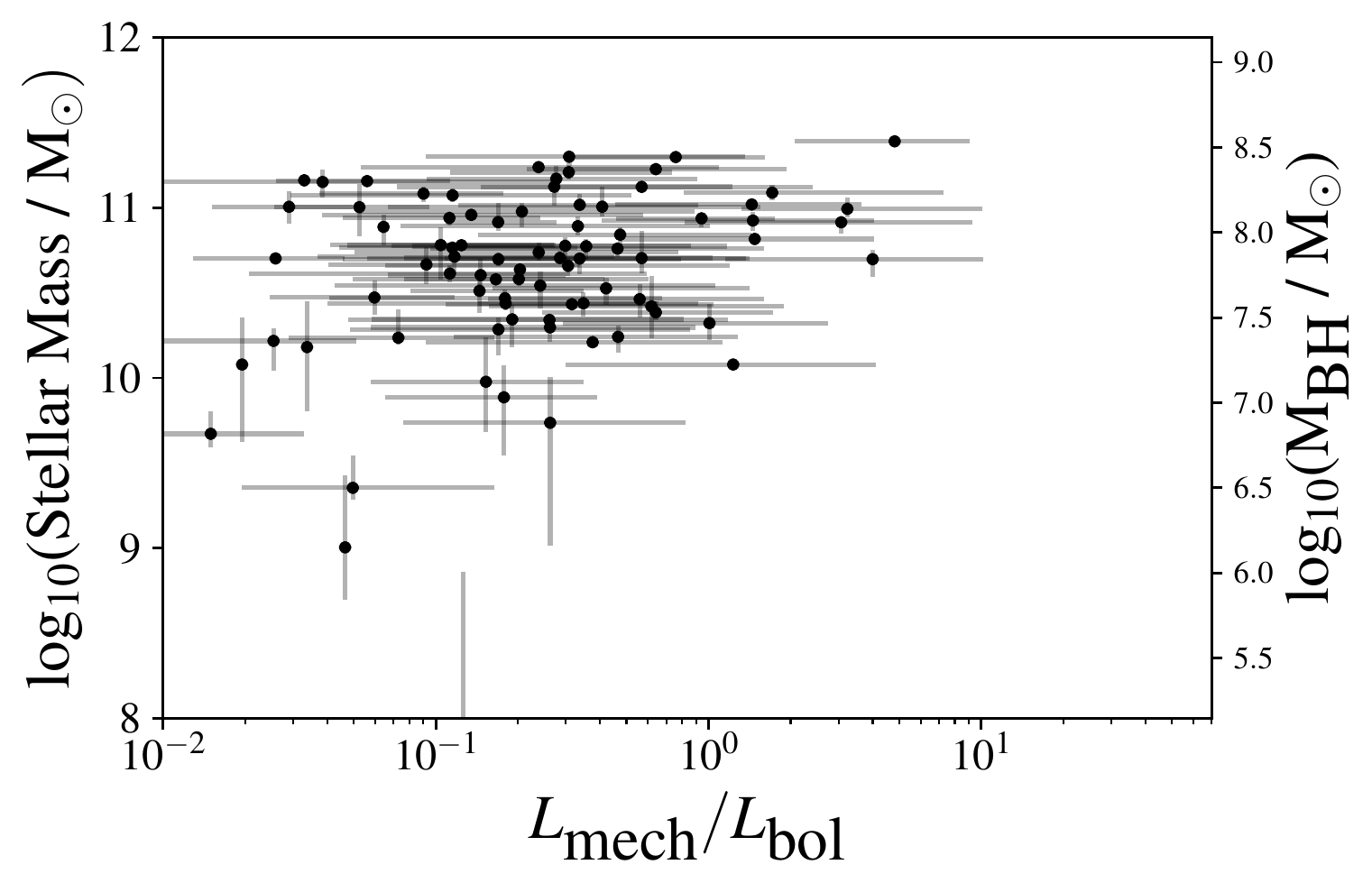}}
\caption{Each row corresponds to a redshift and accretion-rate selected sample, demonstrated by the red box in the figure in the first (left) column, which shows the total accretion rate ($L_\textrm{mech} + L_\textrm{bol}$) as a function of redshift. The top row shows sources with $0.2 < z < 0.5$ and $1.2 \times 10^{36} < (L_\textrm{mech} + L_\textrm{bol}) / \textrm{W} < 5 \times 10^{37}$. The second row shows sources with $0.5 < z < 1.0$ and $5 \times 10^{36} < (L_\textrm{mech} + L_\textrm{bol}) / \textrm{W} < 1 \times 10^{38}$. The third row shows sources with $1.0 < z < 1.5$ and $1 \times 10^{37} < (L_\textrm{mech} + L_\textrm{bol}) / \textrm{W} < 2 \times 10^{38}$. The bottom row shows sources with $1.5 < z < 2.0$ and $2 \times 10^{37} < (L_\textrm{mech} + L_\textrm{bol}) / \textrm{W} < 5 \times 10^{38}$. The second column shows stellar mass as a function of bolometric luminosity for the sources selected in this way. The third column shows stellar mass / supermassive black hole mass as a function of mechanical luminosity, and the final (right-hand) column is the ratio L$_\textrm{mech}$/L$_\textrm{bol}$.}
\label{fig:rates-discussion}
\end{sidewaysfigure*}

In order to avoid incompleteness over a large redshift range, we now split the sample into redshift slices, and for each redshift range we consider only sources with accretion rates where we know we are complete in that redshift range. This selection is demonstrated by the red box in the left hand column of Fig.~\ref{fig:rates-discussion}. The remaining columns of Fig.~\ref{fig:rates-discussion} show how the bolometric and mechanical luminosities of the sources in each redshift slice vary with stellar mass. We find no correlation between bolometric luminosity and stellar mass (the Pearson's correlation coefficient $<|0.3|$ in all redshift bins out to $z=4$), suggesting that the radiative power of these AGN is independent of stellar mass. As there is a tight correlation between a galaxy's stellar mass and the mass of the supermassive black hole, this implies that the radiative power of these radio-loud AGN is independent of the supermassive black hole mass. 

When we consider the mechanical power, as shown by the third column in Fig.~\ref{fig:rates-discussion}, a trend becomes evident. While sources with low mechanical powers are found in host galaxies with a range of stellar masses, higher powered jets are only found in galaxies with stellar masses greater than $10^{10.5} \textrm{M}_\odot$. As there is a tight correlation between a galaxy's stellar mass and the mass of the supermassive black hole, this implies that a SMBH above a certain mass is required to launch a powerful jet. Using the stellar mass - black hole mass relation of \citet{2004ApJ...604L..89H}, the stellar mass above which the galaxies in our sample appear to be able to launch a powerful radio jet corresponds to a black hole mass of $\sim 10^{7.8} \textrm{M}_\odot$. 

The jet power at which this switch between being hosted by galaxies with a range of masses, to only being hosted by the most massive galaxies, increases with redshift. In the lowest redshift bin ($0.25 < z < 0.5$) a large supermassive black hole ($M_\textrm{BH} > 10^{7.5} \textrm{M}_\odot$ is required to launch jets with mechanical powers $>10^{36}$~erg/s, while at higher redshifts ($z > 0.5$), black holes of this size are required to launch jets with mechanical powers $>10^{37}$~erg/s, while less powerful jets are found in host galaxies with a range of black hole masses. This may suggest that it is easier to launch and sustain a powerful jet at higher redshifts, likely due to the availability of gas to provide fuel for accretion. However, further data is required to confirm this, as the sample studied here does not contain similarly powerful sources at low redshifts, due to the small volume probed at these redshifts.

The final column of Fig.~\ref{fig:rates-discussion} shows the ratio between the mechanical and radiative powers of the radio-loud AGN in this sample as a function of stellar mass/black hole mass. Across the redshift range, only galaxies with supermassive black hole masses $\gtrsim 10^8 \textrm{M}_\odot$ host AGN with $L_\textrm{mech} / L_\textrm{bol} > 1$. This suggests that a large black hole mass is required to launch and maintain a radio jet that is powerful relative to the radiative output. This is discussed further in the next section.

%
\makeatletter
\let\@makecaption=\SFB@makefigurecaption
\makeatother
\setlength{\rotFPtop}{0pt plus 1fil}
\setlength{\rotFPbot}{0pt plus 1fil}

\section{Discussion}\label{section:discussion}

In the previous section we have investigated the properties of radio-loud AGN with lower radio powers and at higher redshifts than have been studied by previous work (see Fig.~\ref{fig:LzS82}). In contrast to previous studies at higher radio powers, we find very little difference in the host galaxy properties of HERGs and LERGs (Section~\ref{section:hosts}). We also find considerable overlap in the Eddington-scaled accretion rates of the HERGs and LERGs; while the small number of sources accreting very inefficiently (with Eddington-scaled accretion rates $<10^{-3}$) are all LERGs, the accretion rates of the bulk of the population of HERGs and LERGs are indistinguishable (Section~\ref{section:accretion-rates}).  This is in contrast to previous works which have found a dichotomy in the accretion rates of the two classes \citep{2012MNRAS.421.1569B,2014MNRAS.440..269M}, due to the fact that the MIGHTEE survey is probing higher redshifts and lower radio powers than previous studies, so is detecting a significant population of slowly accreting HERGs which were not detectable by previous work, together with the fact that we are not sensitive to the rare, efficiently accreting HERGs at low redshifts which dominated the previous samples.  It is possible that source misclassification could be partly responsible for the increased overlap in accretion rates observed, as it is possible that some of the LERGs and probable-LERGs in our sample with high Eddington-scaled accretion rates could be misclassified HERGs. However, source misclassification is unlikely to be responsible for inefficiently-accreting HERGs in our sample.

Taken together with the results of \citet{2018MNRAS.480..358W}, who found a difference in the accretion rates of the two classes but considerably more overlap than had been found previously, these results suggest that as we probe lower radio powers and higher redshifts, the HERG and LERG populations are becoming increasingly similar, both in terms of Eddington-scaled accretion rates and host galaxy properties. 

Although there is no significant difference between the HERGs and LERGs in our sample, there is a relationship between host galaxy properties and Eddington-scaled accretion rate; the most-efficiently accreting sources are found in galaxies with lower stellar masses and higher star-formation rates than the sources with higher Eddington-scaled accretion rates (Section~\ref{section:properties-accrate}). This fits in with the widely-accepted scenario that these sources contain a supply of cold gas, which provides the necessary supply of fuel to both accrete efficiently and form stars. What is noteable, however, is that this trend does not seem to translate into HERG/LERG classification; many of the sources with low Eddington-scaled accretion rates also have low star-formation rates suggesting that they do not have a plentiful supply cold gas, but are classified as HERGs. This means that despite accreting slowly they display at least one characteristic usually associated with efficient accretion, such as X-ray emission or the mid-infrared signature of a dusty torus. This implies that as we reach lower radio powers and higher redshifts the situation becomes less clear-cut than it is for more powerful radio galaxies in the local universe. 


Recent work by \citet{2022MNRAS.513.3742K} identified a population of LERGs hosted by star-forming (rather than quiescent) galaxies at $z>0.5$, which they suggest may be fuelled by cold gas, in a similar way to HERGs, rather than via the hot gas mechanism which is typically thought to fuel LERGs in quiescent galaxies. This again raises the question as to whether the distinction between HERGs and LERGs is meaningful for this higher-redshift and lower-powered population.

In Section~\ref{section:power-mass} we found that while AGN with relatively weak radio jets compared to the radiative emission from the nucleus (with $L_\textrm{mech}/L_\textrm{bol} < 1$)  are found in galaxies with SMBH masses across the full range found in our sample ($6< \textrm{log}_{10}(\textrm{M}_\textrm{BH} / \textrm{M}_\odot) <8.5$), all galaxies hosting jets with $L_\textrm{mech}/L_\textrm{bol} > 1$  have black hole masses $\gtrsim 10^{7.8} M_\odot$. This suggests that a large black hole mass is required to launch and sustain a powerful radio jet.  Our results show that this holds out to at least $z \sim 2$. This is in agreement with \citet{2004MNRAS.353L..45M}, who found that the radio-loudness of quasars is dependent on black hole mass, with all the genuinely radio-loud quasars in their sample having M$_\textrm{BH} \geq 10^8 \textrm{M}_\odot$. \citet{2011MNRAS.410.1360H} also found that radio luminosity scales with black hole mass. 
\citet{2018A&A...609A...1B} suggest that only when a supermassive black hole has both a high mass and high spin is it able to sustain powerful radio jets for long enough to produce an extended radio galaxy. This could explain why in our sample a large black hole mass seems to be required to sustain a powerful radio jet, but only a fraction of the radio-loud AGN which have large black hole masses produce powerful jets. 

The spin of a supermassive black hole depends on both the accretion and the merger histories of the system. Major mergers are the main mechanism for spinning-up black holes \citep{2011MNRAS.418L..84M}, while chaotic accretion reduces the spin of the central black hole. Galaxies with higher masses are more likely to have undergone more mergers, and therefore have a higher black hole spin. Supermassive black hole spin also evolves with redshift; \citet{2011MNRAS.418L..84M} showed that galaxies at $z\gtrsim 1$ are expected to have lower spins than those in the local universe, where a significant fraction will have high spins. 

Theoretical work suggests there is a connection between jet speed and black hole spin \citep{1977MNRAS.179..433B,1986bhmp.book.....T,1999ApJ...522..753M}, where black holes with higher spin values are able to produce more powerful relativistic jets. 
High spins may be required to launch a jet at all; \citet{2012JPhCS.355a2016M} use general-relativistic magnetohydrodynamic (GRMHD) simulations and gamma ray observations of blazars to argue SMBH spins $>0.5$ are required to launch a relativistic jet from a black hole. 
It has been suggested that black hole spin could therefore be responsible for the radio loud/radio quiet dichotomy, with AGN where the central SMBH is spinning faster producing radio jets, while those with lower spins result in radio quiet AGN. However, observational evidence for the connection between black hole spin and the jet is inconclusive. Several studies support the idea that spin is necessary for jet production (e.g.\ \citealt{2012MNRAS.419L..69N,2016ChA&A..40..198Z}), while other studies of both X-ray binaries \citep{2010MNRAS.406.1425F} and AGN \citep{2013A&A...557L...7V} find no evidence for a connection between black hole spin and jet power.

However while a high black hole spin may be necessary to produce a powerful radio jet, it is not sufficient. Radio-quiet AGN with high spins have been observed (e.g.\ \citealt{2014SSRv..183..277R}), showing that not all AGN with high spin have a jet. There are clearly other factors involved in launching and sustaining a powerful radio jet, such as black hole mass, accretion rate, the fuel available and magnetic fields.

It is also possible that black hole spin is one of the reasons for the difference between the properties of HERGs and LERGs observed in this study and those observed in previous studies. As previous studies are limited to high radio luminosities, and therefore powerful radio jets, they may preferentially select SMBHs with high spin. Additionally, these studies were limited to lower redshifts, where SMBHs would have higher spins in this scenario. The radio galaxies in this sample extend to lower radio luminosities and higher redshifts, so may include SMBHs with lower spins as well as those with high spins. This range of spin values could explain why we see a larger range of jet powers and accretion efficiencies for HERGs and LERGs, resulting in a larger overlap between the two populations.

\section{Conclusions}\label{section:conclusions}

In this work we have classified the radio sources in the MIGHTEE Early Science data in the COSMOS field, allowing us to probe the faint radio source population down to $S_{1.4~\textrm{GHz}} \sim 20~\muup$Jy, and used this sample to investigate the nature of radio-excess AGN at both lower radio powers and higher redshifts than previous studies. Making use of the extensive multi-wavelength data in the COSMOS field, we identify 1806 AGN and 767 SFG, with a further 2040 sources classified as `probable SFGs'. The radio-loud AGN are further classified as high-excitation and low-excitation radio galaxies, finding 249 HERGs, 150 LERGs and a further 782 `probable LERGs'. This classification catalogue is released with this work. 

We have used this sample to investigate the properties of radio-loud AGN with $10^{20} < L /\textsc{W Hz}^{-1} < 10^{27}$ out to $z \sim 5$, probing both lower radio powers and higher redshifts than previous studies of radio galaxies. Our main findings are:

\begin{itemize}
    \item We find no significant difference in the host galaxy properties of the HERGs and LERGs in the MIGHTEE sample, except for the lowest redshift bin ($z < 0.4$) where LERGs are hosted by galaxies with higher stellar masses than HERGs.
    \item There is considerable overlap in the accretion rates of the HERGs and LERGs in our sample. This is in contrast to previous works which have found a dichotomy in the accretion rates of the two classes \citep{2012MNRAS.421.1569B,2014MNRAS.440..269M}, due to the fact that the MIGHTEE survey is probing both higher redshifts and lower radio powers than previous studies.
    \item Sources with higher Eddington-scaled accretion rates tend to be hosted by galaxies with higher star-formation rates and smaller stellar masses. 
    \item As it is becoming increasingly hard to separate HERGs and LERGs into separate populations at higher redshifts and lower radio powers, we suggest that this classification may not be a helpful way to think about the radio galaxy population, and instead one should consider how properties vary with Eddington-scaled accretion rate.
    \item A black hole mass $\gtrsim 10^{7.8}~\textrm{M}_\odot$ is required to power a jet with $L_\textrm{mech}/L_\textrm{bol} > 1$; we discuss that both a high black hole mass and a high black hole spin may be necessary to launch and sustain a radio jet with a high mechanical power relative to the radiative output of the AGN.
    
\end{itemize}

\section*{Acknowledgements}

The authors thank the anonymous referee for their careful reading of the paper and insightful comments which improved the quality of the manuscript. The MeerKAT telescope is operated by the South African Radio Astronomy Observatory, which is a facility of the National Research Foundation, an agency of the Department of Science and Innovation. 
We acknowledge the use of the ilifu cloud computing facility – www.ilifu.ac.za, a partnership between the University of Cape Town, the University of the Western Cape, the University of Stellenbosch, Sol Plaatje University, the Cape Peninsula University of Technology and the South African Radio Astronomy Observatory. The Ilifu facility is supported by contributions from the Inter-University Institute for Data Intensive Astronomy (IDIA – a partnership between the University of Cape Town, the University of Pretoria, the University of the Western Cape and the South African Radio astronomy Observatory), the Computational Biology division at UCT and the Data Intensive Research Initiative of South Africa (DIRISA).
The authors acknowledge the Centre for High Performance Computing (CHPC), South Africa, for providing computational resources to this research project.
This work is based on data products from observations made with ESO Telescopes at the La Silla Paranal Observatory under ESO programme ID 179.A-2005 (Ultra-VISTA) and ID 179.A- 2006(VIDEO) and on data products produced by CALET and the Cambridge Astronomy Survey Unit on behalf of the Ultra-VISTA and VIDEO consortia.
Based on observations obtained with MegaPrime/MegaCam, a joint project of CFHT and CEA/IRFU, at the Canada-France-Hawaii Telescope (CFHT) which is operated by the National Research Coun- cil (NRC) of Canada, the Institut National des Science de l’Univers of the Centre National de la Recherche Scientifique (CNRS) of France, and the University of Hawaii. This work is based in part on data products produced at Terapix available at the Canadian Astronomy Data Centre as part of the Canada-France-Hawaii Telescope Legacy Survey, a collaborative project of NRC and CNRS.
The Hyper Suprime-Cam (HSC) collaboration includes the astronomical communities of Japan and Taiwan, and Princeton University. The HSC instrumentation and soft- ware were developed by the National Astronomical Observatory of Japan (NAOJ), the Kavli Institute for the Physics and Mathematics of the Universe (Kavli IPMU), the University of Tokyo, the High Energy Accelerator Research Organization (KEK), the Academia Sinica Institute for Astronomy and Astrophysics in Taiwan (ASIAA), and Princeton University. Funding was contributed by the FIRST program from Japanese Cabinet Office, the Ministry of Education, Culture, Sports, Science and Technology (MEXT), the Japan Society for the Promotion of Science (JSPS), Japan Science and Technology Agency (JST), the Toray Science Foundation, NAOJ, Kavli IPMU, KEK, ASIAA, and Princeton University.
This research was funded in whole, or in part, by the UK Science and Technology Facilities Council [ST/N000919/1]. For the purpose of Open Access, the author has applied a CC BY public copyright licence to any Author Accepted Manuscript version arising from this submission.
IHW, MJJ and PWH acknowledge generous support from the Hintze Family Charitable Foundation through the Oxford Hintze Centre for Astrophysical Surveys.
IH and MJJ acknowledge support from the UK Science and Technology Facilities Council [ST/N000919/1]. 
CLH acknowledges support from the Leverhulme Trust through an Early Career
Research Fellowship. 
NM acknowledges the support of the LMU Faculty of Physics.
NJA acknowledges support from the European Research Council (ERC) Advanced Investigator Grant EPOCHS (788113).
MG was partially supported by the Australian Government through the Australian Research Council's Discovery Projects funding scheme (DP210102103).
LKM and NLT were supported by the Medical Research Council [MR/T042842/1]. 
RB acknowledges support from an STFC Ernest Rutherford Fellowship [grant number ST/T003596/1].
Y.A. acknowledges financial support by NSFC grants 12173089 and 11933011.
LM, IP and MV acknowledge financial support from the Italian Ministry of Foreign Affairs and International Cooperation (MAECI Grant Number ZA18GR02) and the South African Department of Science and Innovation's National Research Foundation (DSI-NRF Grant Number 113121) under the ISARP RADIOSKY2020 Joint Research Scheme.
DJBS acknowledges support from the UK Science and Technology Facilities Council (STFC) under grant [ST/V000624/1]
JA acknowledges financial support from the Science and Technology Foundation (FCT, Portugal) through research grants PTDC/FIS-AST/29245/2017, UIDB/04434/2020 and UIDP/04434/2020.
This research has made use of NASA's Astrophysics Data System.  
This research used \textsc{Astropy}, a community-developed core Python package for Astronomy \citep{2013A&A...558A..33A,2018AJ....156..123A}. 
This work made use of the cubehelix colour scheme \citep{2011BASI...39..289G}. 
We also used \textsc{Topcat} \citep{2005ASPC..347...29T,2011ascl.soft01010T}.

\section*{Data Availability}

The release of the MIGHTEE Early Science continuum data used for this work is discussed in depth in \citet{2022MNRAS.509.2150H}, details of the data release and how to access the data are provided there. The cross-matched catalogue and its release is described in the work of Prescott et al. (in prep) and will be released in accompaniment with their work. The catalogue presented in this article is summarised in Appendix~\ref{appendix:level3-cat} and is published with this article as supplementary material.



\bibliographystyle{mnras}


%
%

\setlength{\labelwidth}{0pt}



\appendix

\section{Structure of the Level-3 source classification catalogue}
\label{appendix:level3-cat}

The Level-1 MIGHTEE Early Science catalogue of radio sources was released with \citet{2022MNRAS.509.2150H}. The Level-2 catalogue contains information about the multi-wavelength identifications for the Level-1 radio sources, and is released with Prescott et al. (in prep). The source classifications described in this work form the Level-3 catalogue, and are released here. The structure of this catalogue is described below.

\noindent \textbf{(0)}: Name: An IAU-style identifier of the form JHHMMSS.SS+/-DDMMSS.S, \textbf{based on the position of the host galaxy, as in Level-2.}

\noindent \textbf{(1)}: RA\_Radio: The J2000 Right Ascension of the radio source in degrees. If the source is multi-component radio source this is the Right Ascension brightest component, as in Level-2.

\noindent \textbf{(2)}: DEC\_Radio: The J2000 Declination of the radio source in degrees. If this is multiple component radio source it is the Declination brightest component, as in Level-2.

\noindent \textbf{(3)}: RA\_host: The J2000 Right Ascension of the object in degrees from the $K_s$-band selected multi-wavelength catalogue, as in Level-2. 

\noindent \textbf{(4)}: DEC\_host: The J2000 Declination of the object in degrees from the $K_s$-band selected multi-wavelength catalogue, as in Level-2. 

\noindent \textbf{(5)}: S\_INT14: The total 1.4-GHz flux density of the radio sources in Jy. This is scaled to 1.4-GHz from the measured frequency using the effective frequency map and assuming a spectral index of 0.7.

\noindent \textbf{(6)}: Redshift: The best available redshift value. This is a spectroscopic redshift value if available, if not it is the photometric redshift from Hatfield et al. (in prep). See Section~\ref{section:redshifts}. See column (8) for information about the origin of the redshift value.

\noindent \textbf{(7)}: Redshift\_err: Uncertainty on the best available redshift value if available. (-99 if not available -- this is the case for some spectroscopic redshifts.)

\noindent \textbf{(8)}: Redshift\_note: Note on the origin of the redshift (=`photz' if photoz is used).

\noindent \textbf{(9)}: L14: 1.4-GHz radio luminosity in W/Hz. Scaled to rest-frame 1.4~GHz assuming a spectral index of 0.7.

\noindent \textbf{(10)}: XAGN: boolean column, True if source is classified as an X-ray AGN, i.e.\ has $L_{x} > 10^{42}$ erg/s, see Section~\ref{section:XAGN_class}.

\noindent \textbf{(11)}: notXAGN: boolean column, True if source is classified as not being an X-ray AGN, i.e.\ has $L_{x} < 10^{42}$ erg/s, see Section~\ref{section:XAGN_class} (note that a source must be positively identified as having $L_{x} < 10^{42}$ erg/s, i.e.\ this is not everything False in column (10)).

\noindent \textbf{(12)}: RLAGN: boolean column, True if source is a radio-excess AGN, see Section~\ref{section:RLAGN}. 

\noindent \textbf{(13)}: notRLAGN: boolean column, True if source does not have a radio excess, see Section~\ref{section:RLAGN}. (Note that a source must be positively identified as not having a radio-excess, i.e.\ this is not everything False in column (12)).

\noindent \textbf{(14)}: midIRAGN: Donley mid-infrared AGN, boolean column, True if source is classified as a mid-infrared AGN, see Section~\ref{section:midIRAGN}.

\noindent \textbf{(15)}: notmidIRAGN: not Donley mid-infrared AGN, boolean column, True if source is classified as not being a mid-infrared AGN, see Section~\ref{section:midIRAGN} (note that this is not everything False in column (14)).

\noindent \textbf{(16)}: optAGN: optical point-like AGN, boolean column, True if source is point-like in the \emph{ACS} optical imaging, see Section~\ref{section:optAGN}.

\noindent \textbf{(17)}: notoptAGN: not optical point-like AGN, boolean column, True if source is resolved point-like in the \emph{ACS} optical imaging, see Section~\ref{section:optAGN} (note that this is not everything False in column (16)).

\noindent \textbf{(18)}: VLBAAGN: boolean column, True if source is detected in the VLBA observations, see Section~\ref{section:VLBA_AGN}. Note that everything False in this column is not a VLBA AGN.

\noindent \textbf{(19)}: AGN: overall AGN, boolean column, True if source is classified as an AGN in the overall classifications, see Section~\ref{section:overall_class}.

\noindent \textbf{(20)}: SFG: overall SFG, boolean column, True if source is classified as a SFG in the overall classifications, see Section~\ref{section:overall_class}. (Limited to sources with $z < 0.5$, see text for details.)

\noindent \textbf{(21)}: probSFG: probable SFG, boolean column, True if source is classified as a probable SFG in the overall classifications, see Section~\ref{section:overall_class}. (Note that sources with $z>0.5$ cannot be securely classified as SFG so will be classified as probable SFG if appropriate, see text for details.)

\noindent \textbf{(22)}: unclass: Unclassified sources, boolean column, True if source is not able to be classified as an AGN, SFG or probable SFG.

\noindent \textbf{(23)}: HERG: boolean column, True if source is classified as a HERG, see Section~\ref{section:overall_class}.

\noindent \textbf{(24)}: LERG: boolean column, True if source is classified as a LERG, see Section~\ref{section:overall_class}. (Limited to sources with $z < 0.5$, see text for details.)

\noindent \textbf{(25)}: probLERG: Probable LERG, boolean column, True if source is classified as a probable LERG, see Section~\ref{section:overall_class}. (Note that sources with $z \geq 0.5$ cannot be securely classified as LERGs so will be classified as probable LERGs if appropriate, see text for details.)

\noindent \textbf{(26)}: RQAGN: radio-quiet AGN, boolean column, True if source is classified as an AGN (column (19) is True) and is not radio loud (column (13) is True).

\section{Williams et al. (2018) classification scheme}\label{appendix:williams}

\begin{figure}
    \centering
    \includegraphics[width=7.9cm]{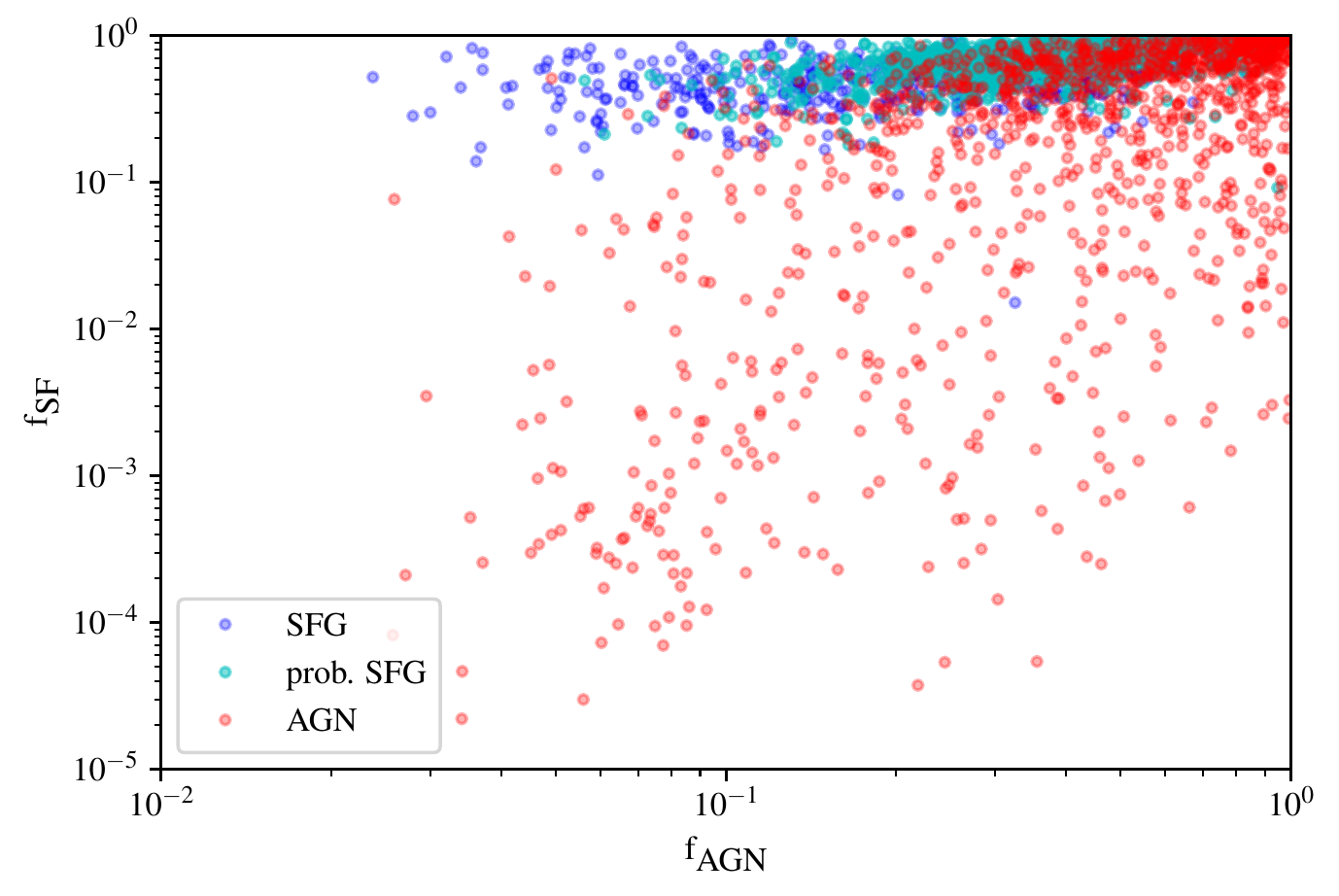}
    \includegraphics[width=7.9cm]{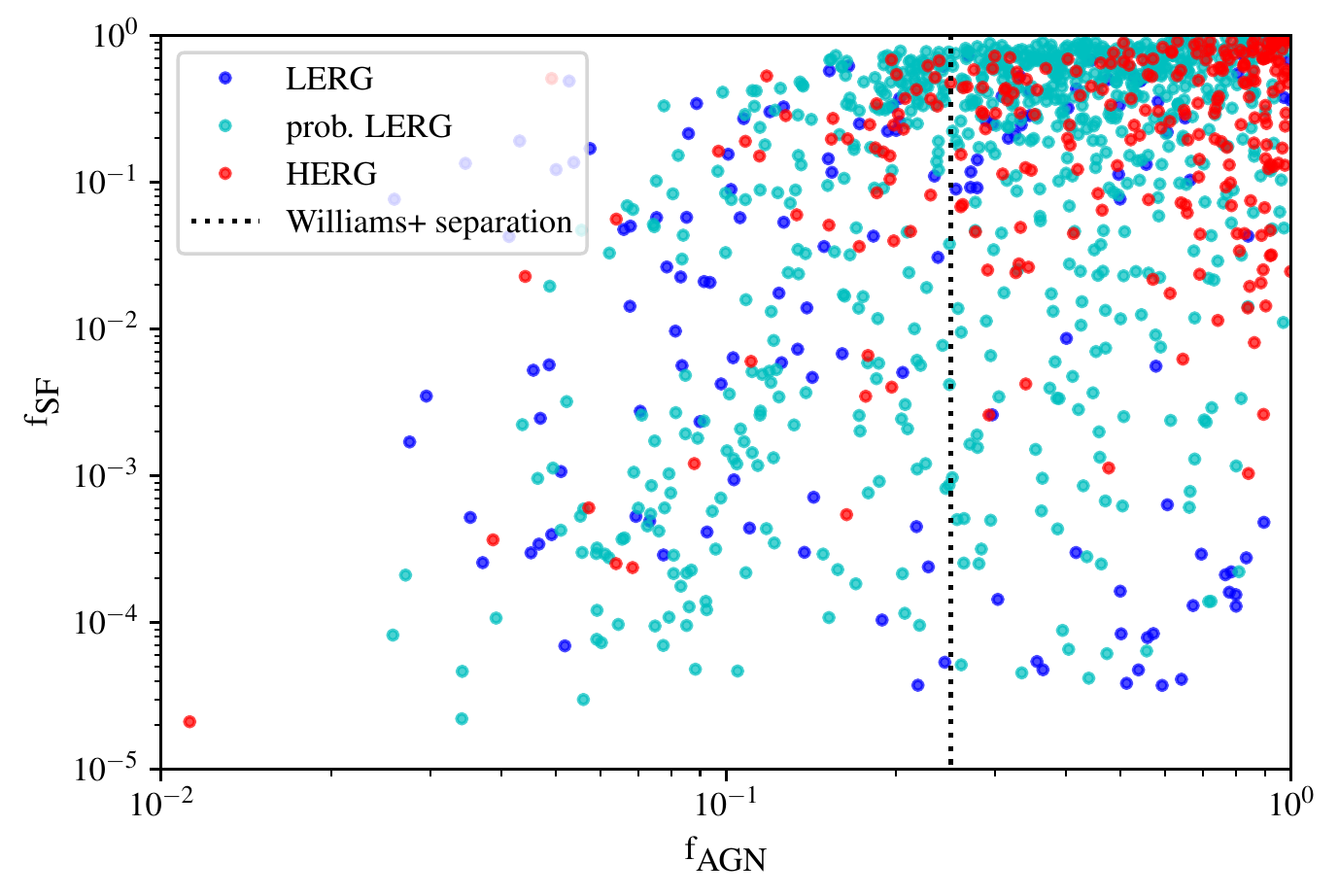}
    \includegraphics[width=7.9cm]{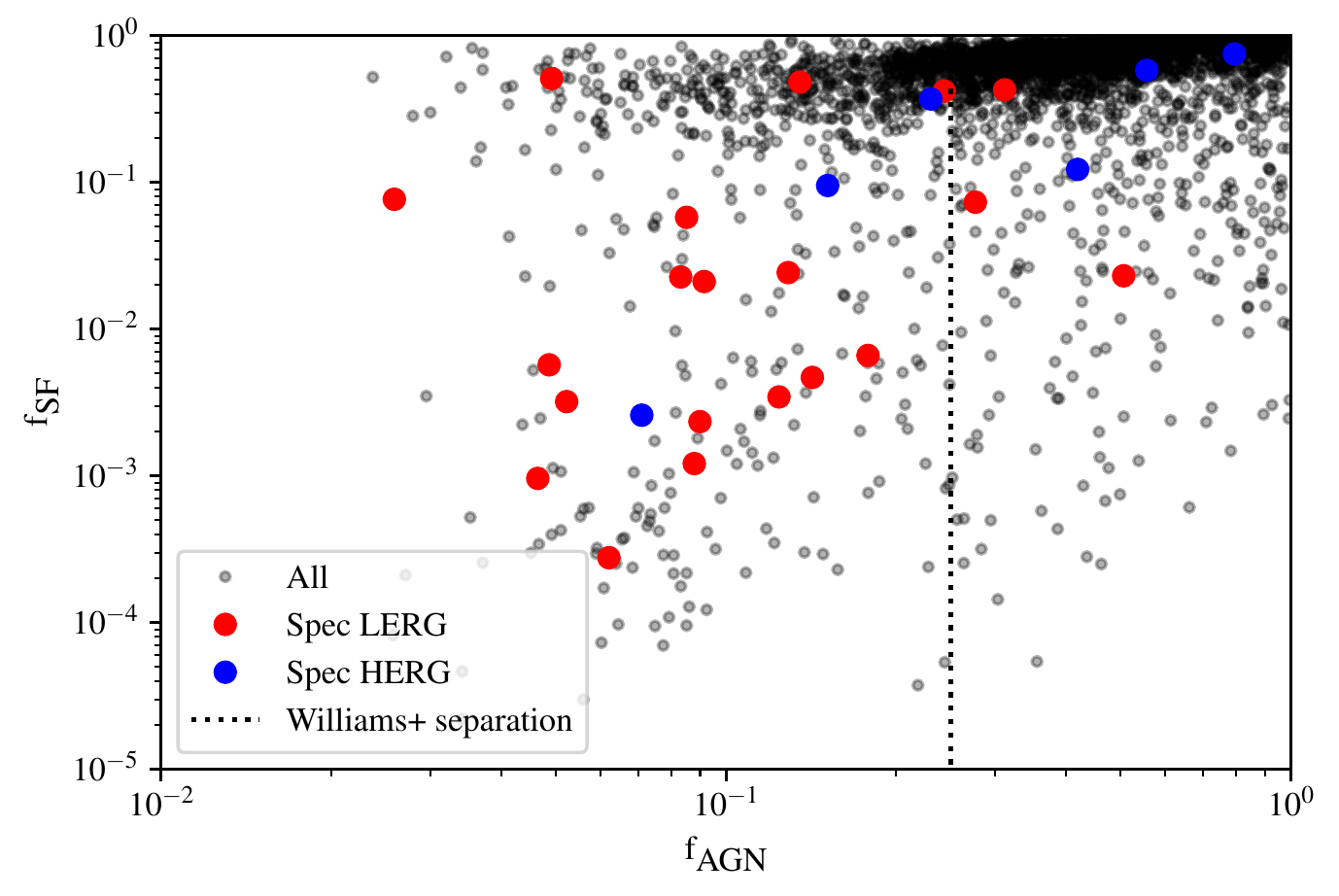}
    \caption{AGN and SF fractions derived from \textsc{AGNfitter} in the same way as \citet{2018MNRAS.475.3429W}. They classify sources with $f_\text{AGN} > 0.25$ (shown by the dotted line) as HERGs. The top panel shows sources classified as SFG and AGN using our scheme, the middle panel shows sources classified as HERGs and LERGs, the bottom panel shows sources with classifications from optical spectra, using [O\textsc{iii}] line measurements to classify HERGs and LERGs (see Section~\ref{section:opt_spectra_comparison} for details). Note that only sources with good \textsc{AGNfitter} fits (log likelihood > -20) are shown.}
    \label{fig:williams}
\end{figure}

\citet{2018MNRAS.475.3429W} use \textsc{AGNfitter} to classify sources detected in the LOFAR Boot\"{e}s field as HERGs and LERGs. Here we compare their classification scheme to that used in this work. \citeauthor{2018MNRAS.475.3429W} first select the radio-loud AGN using a criteria similar to that described in Section~\ref{section:RLAGN}. They then define $f_\text{AGN}$ as:
\begin{equation}
    f_\text{AGN} = \frac{L_\text{TO} + L_\text{BB}}{L_\text{TO} + L_\text{GA} + L_\text{BB}}
\end{equation}

\noindent where $L$ are the luminosities of the different SED components from \textsc{AGNfitter} described in Section~\ref{section:AGNfitter}; $L_\text{TO}$ is the luminosity from hot dust torus component, $L_\text{BB}$ is due to the UV/optical accretion disk, $L_\text{GA}$ is from the stellar emission and $L_\text{SB}$ is the reprocessed light from dust (we use this nomenclature to be consistent with \citealt{2016ApJ...833...98C,2018MNRAS.475.3429W}). $f_\text{AGN}$ is essentially the fraction of emission that is due to the AGN (i.e.\ the torus and accretion disk components) compared to the total emission independent of the mid-infrared star-forming component ($L_\text{SB}$). 

\citeauthor{2018MNRAS.475.3429W} also define:
\begin{equation}
    f_\text{SF} = \frac{L_\text{SB}}{L_\text{SB} + L_\text{GA}}
\end{equation}

\noindent which is the fraction due to the infrared starburst component compared to the total galaxy component. $L_\textrm{SB}$ and $L_\textrm{TO}$ are calculated over the wavelength range $1 < \lambdaup / \muup\textrm{m} < 30$, and $L_\textrm{BB}$ and $L_\textrm{GA}$ are calculated over the wavelength range $0.1 < \lambdaup / \muup\textrm{m} < 1$. \citeauthor{2018MNRAS.475.3429W} classify sources with $f_\textrm{AGN} > 0.25$ as HERGs, and sources with AGN fractions below this value as LERGs. 
The top panel of Fig.~\ref{fig:williams} shows $f_\textrm{SF}$ and $f_\textrm{AGN}$ for sources classified as AGN, SFG and probable SFG in this work. As expected, the SFG all have very high values of $f_\text{SF}$ (with a very small number of exceptions). A significant fraction of the AGN also have large $f_\text{SF}$ values; this is not necessarily unexpected, and these AGN generally also have reasonably high values of $f_\text{AGN}$. The middle panel of Fig.~\ref{fig:williams} shows HERGs, LERGs and probable LERGs classified in this work, along with the $f_\text{AGN} = 0.25$ HERG/LERG separation used by \citeauthor{2018MNRAS.475.3429W} The majority of the sources classified as HERGs in this work have $f_\text{AGN} > 0.25$, however so do a significant fraction of the LERGs and probable LERGs. This suggests that while this cut may be a reasonable method to select HERGs, the sample may be contaminated by a significant number of LERGs. The \textrm{bottom panel} of Fig.~\ref{fig:williams} show the positions of sources in our sample with spectroscopic classifications. While 17 of the 20 sources with spectroscopic classifications with $f_\text{AGN} < 0.25$ are LERGs, so are 3 out of the 6 sources with $f_\text{AGN} > 0.25$. This again suggests that while the majority of the sources with $f_\text{AGN} < 0.25$ are indeed LERGs (although 3 out of 20 are HERGs), there is some LERG contamination in the sample found above this divide. 


\bsp	
\label{lastpage}
\end{document}